\documentclass{aastex}
\usepackage{emulateapj5}

\def\degree{\hbox{$^\circ$}}

\shortauthors{Jimenez-Garate et al.}
\shorttitle{X-ray Emission of Accretion Disk Atmospheres}

\begin{document}

%% LaTeX will automatically break titles if they run longer than
%% one line. However, you may use \\ to force a line break if
%% you desire.

\title{The Structure and X-ray Recombination Emission of
a Centrally Illuminated Accretion Disk Atmosphere and Corona}

%% Use \author, \affil, and the \and command to format
%% author and affiliation information.
%% Note that \email has replaced the old \authoremail command
%% from AASTeX v4.0. You can use \email to mark an email address
%% anywhere in the paper, not just in the front matter.
%% As in the title, you can use \\ to force line breaks.

\slugcomment{Accepted to ApJ}
%\slugcomment{\date}

\author{M. A. Jimenez-Garate}
\affil{MIT Center for Space Research, 70 Vassar St.,
NE80-6009, Cambridge, MA 02139}
\email{mario@alum.mit.edu}

\author{J. C. Raymond}
\affil{Center for Astrophysics, 60 Garden St., Cambridge, MA 02138}
\email{raymond@cfa.harvard.edu}
\and
\author{D. A. Liedahl}
\affil{Lawrence Livermore National Laboratory, 
Department of Physics and Advanced Technologies, 7000 East Ave., L-41, Livermore, CA 94550}
\email{liedahl1@llnl.gov}

%% Notice that each of these authors has alternate affiliations, which
%% are identified by the \altaffilmark after each name.  Specify alternate
%% affiliation information with \altaffiltext, with one command per each
%% affiliation.

%% Mark off your abstract in the ``abstract'' environment. In the manuscript
%% style, abstract will output a Received/Accepted line after the
%% title and affiliation information. No date will appear since the author
%% does not have this information. The dates will be filled in by the
%% editorial office after submission.

% make present tense
\begin{abstract}
We model an accretion disk atmosphere and corona photoionized by a
central X-ray continuum source.  We calculate the opacity and
one-dimensional radiation transfer for an array of disk radii, to
obtain the two-dimensional structure of the disk 
and its X-ray recombination emission. 
The atmospheric structure is extremely insensitive to the
viscosity $\alpha$.  We find a feedback mechanism between the disk
structure and the central illumination, which expands the disk and
increases the solid angle subtended by the atmosphere. We apply the
model to the disk of a neutron star X-ray binary. The model
is in agreement with the $\sim 12 \degree$ disk half-angle 
measured from optical light curves.
We map the temperature, density, and ionization
structure of the disk, and we simulate high resolution
spectra expected from the {\it Chandra} and {\it XMM-Newton} grating
spectrometers. X-ray emission lines from the disk atmosphere are detectable,
especially for high-inclination binary systems.  
The grating observations of two classes of X-ray binary systems
already reveal important spectral similarities with our models.
The model spectrum is dominated by double-peaked lines of H-like and He-like
ions, plus weak Fe L. The line flux is proportional to the
luminosity and is dominated by the outer radii.  Species with a broad range of
ionization levels coexist at each radius: from \ion{Fe}{26} in the hot corona,
to \ion{C}{6} at the base of the atmosphere. 
The line spectrum is very sensitive to the temperature, ionization, and
emission measure of each atmospheric layer, and it probes the heating
mechanisms in the disk.
We assume a hydrostatic disk dominated by gas pressure, 
in thermal balance, and in ionization equilibrium.
As boundary conditions, we take a
Compton-temperature corona and an underlying Shakura-Sunyaev disk.
The choice of thermally stable solutions
strongly affects the spectrum, since a thermal instability is present
in the regime where X-ray recombination emission is most intense.
\end{abstract}

%% Keywords should appear after the \end{abstract} command. The uncommented
%% example has been keyed in ApJ style. See the instructions to authors
%% for the journal to which you are submitting your paper to determine
%% what keyword punctuation is appropriate.


%% From the front matter, we move on to the body of the paper.
%% In the first two sections, notice the use of the natbib \citep
%% and \citet commands to identify citations.  The citations are
%% tied to the reference list via symbolic KEYs. The KEY corresponds
%% to the KEY in the \bibitem in the reference list below. We have
%% chosen the first three characters of the first author's name plus
%% the last two numeral of the year of publication as our KEY for
%% each reference.

\section{Introduction}

When the infall of matter into a deep gravitational potential
is mediated by an accretion disk, gravitational energy 
is converted to the thermal radiation which powers both low mass X-ray binary
(LMXB) systems and active galactic nuclei (AGN).
Accretion disks present unique problems involving
magnetized plasma dynamics, photoionization, atomic kinetics, thermal and
ionization equilibria, general relativity, and radiation transfer.
The accretion disks in LMXBs and AGN are expected to have many common properties.
The compactness of the accretor in LMXBs and AGN and their inferred
accretion rates imply temperatures of $T> 10^7$~K
and intense X-ray emission in the inner disk region.
The inner radii of these disks, as well as the disk atmosphere as a whole,
is substantially more ionized than the case in which the accretor is a
white dwarf.
In both LMXBs and AGN, the vast energy emitted
in the inner disk region is reprocessed in the outer disk, where
the external radiative heating can dominate the local thermal emission.
The subsequent photoionization of the disk plasma radically alters its
equilibrium state, structure, and spectrum, especially in the
atmospheric and coronal disk layers, which are the subject of this study.

High resolution X-ray spectroscopy is an essential tool to study the 
physics of this "hot class" of accretion disks and the conditions near
black hole event horizons.  
In this paper, we concentrate on the outer radii of
disks in neutron star LMXBs,
since current observational constraints provide more stringent tests
for LMXBs than for AGN.
The following points support these assertions:

\begin{itemize}

\item High resolution spectra can reveal
discrete emission or absorption from atomic transitions
within the accretion disk plasma, providing 
information on the accretion disk structure,
dynamics, and physics. These spectra open a window into 
photoionized gases and their phase equilibria.

\item X-rays, and in particular discrete atomic transitions
of hydrogen- and helium-like ions, probe the regions in the disk with
the highest levels of ionization.  Regions closer to the compact
object will have the highest ionization levels, although vertical
stratification is also expected.

\item The knowledge of the accretion disk physics
directly impacts our ability to probe the physical conditions around
the compact object. For example, the Fe K emission
originating in the innermost regions of an accretion disk has been
proposed as a direct probe of the general relativistic effects
near a black hole event horizon in AGN, by virtue of the observed
characteristic line shape \citep{tanaka1995}. 
However, very little is known about the physical conditions 
in the Fe K emission region, and it is still unclear how our
ignorance of the physical processes within the accretion disk affect
the modeled Fe K line profile and flux. It is also unclear whether the
soft X-ray line features reported by \citet{brandu} are feasible. 

\item While in neutron star LMXBs the photoionizing source
must be near the neutron star surface, in AGN and galactic black hole candidates (BHC)
the location of the
ionizing source is unknown.
In AGN, various authors have assumed the ionizing source to be
located in the rotation axis of the black hole, above the disk
midplane, possibly close to the base of a jet. Alternatively, an
ionizing source might be present on the upper layers of the disk,
perhaps due to disk flares, or to Comptonization of thermal UV photons in the 
accretion disk corona (ADC).

\item LMXB systems are observed in less crowded regions than AGN. 

\item In contrast to AGN, LMXBs often have measured orbital parameters which 
constrain the geometry of the system, such as the maximum disk
radius.  LMXBs may also have a measured value of the disk
inclination, while orbital phase and eclipse phase variations provide
tomographic information.  For example, measurements of the 
optical light curve amplitudes of LMXBs have yielded estimates
of the angle subtended by the disk of $\sim$ 12$\degree$
\citep[]{dejong}. 

\item To our knowledge, the inner disk radii can only be
studied in the X-ray band.
Theoretically, the thermal emission of the inner disk radii
must peak in the X-ray band for both LMXBs and AGN. 
The optical and UV emission originates in the outer
regions of the accretion disk and further away from the
compact object, and the radio and gamma ray emission is likely
dominated by emission from jets.

\end{itemize}
Fewer physical ingredients are needed to model the outer radii of
disks than the inner radii, so the logical progression is to
successfully model the outer radii first.
The X-ray spectroscopy of neutron star LMXBs will allow us to construct a
physical picture of their accretion disks, which we can
then use to investigate the inner disk in BHC and AGN. 

To fully exploit the high energy-resolution X-ray spectra of
accretion disks, we created physical disk models and calculated synthetic spectra.
  The disk plasma, at
$10^5$ to $10^7$~K, cools
through atomic line emission that can be detected with space-borne
X-ray observatories such as {\it Chandra} and {\it XMM-Newton}.
Modeling the equilibrium state of the plasma and the 
radiation transfer within the disk allows a calculation of
the disk structure and its X-ray spectrum. A synthetic spectrum can be
compared to the data.  The model spectrum is unique in that it is
calculated purely on physical, and not just phenomenological, grounds.

We describe four fiducial disk models. Two of these models were introduced
in \citet[]{jimenez}. We use a newly developed adaptive-mesh disk structure calculation,
the \citet{ray93} photoionized plasma code, and a new X-ray emission code which uses
$HULLAC$ data (Hebrew University/Lawrence Livermore Atomic Code,
\citet[]{hullac}).
The models consist of a disk illuminated
by a pure neutron star continuum, and they contain as boundary
conditions a Compton-temperature ADC at the top of the disk and a modified
\citet[hereafter abbreviated as SS73]{ss73} disk at the bottom \citep{vrtilek}. Thus, the region of interest
has a temperature and ionization which is intermediate of these two
regions, and it emits copious X-ray radiation.

In section \ref{sub:line-lmxb}, we introduce the X-ray line observations 
prior to {\it Chandra} and {\it XMM-Newton}; in section \ref{sub:raddisk},
we introduce theoretical work on the structure of X-ray
illuminated accretion disks; in section \ref{sec:modelatm}, we describe
the disk structure calculations and the assumptions of hydrostatic, thermal and
ionization equilibrium; in section \ref{sec:instab},
we detail the effects of a thermal instability on a layer of the disk
atmosphere throughout the disk; in section
\ref{sec:adamodel}, we discuss the calculation of the high resolution
spectrum, which is done {\it a posteriori} from the structure
calculation; in section \ref{sec:diskstruct}, the disk density, temperature
and ionization structure are presented; 
in section \ref{sec:spec}, the model spectra are shown, 
assuming a full, partial or obstructed view of the neutron star region,
and we show simulated spectra utilizing the response of the {\it
XMM-Newton} reflection grating spectrometer (RGS) and the {\it Chandra} 
medium energy gratings (MEG); 
in section \ref{sec:disc},
comparisons of the model to the observed X-ray 
spectra of LMXBs are discussed briefly, and we discuss the limitations
of the model. In section \ref{sec:conclusions}, concluding remarks are
presented.

\section{LMXB accretion disks}
\subsection{X-ray line emission from LMXBs}
\label{sub:line-lmxb}

With the exception of Fe K emission in the 6.4--7.0~keV range
\citep[]{asai2000}, discerning X-ray line emission in LMXBs has
been challenging, owing to limitations in sensitivity and spectral
resolving power, as well as the difficulties associated with
attempts to extract line emission from data dominated by
intense continuum emission.
Measurements obtained with the \it Einstein \rm Objective
Grating Spectrometer \citep[]{vrtilek1991},
the \it ROSAT \rm Position Sensitive Proportional
Counter \citep[]{schultz1999}, and the \it ASCA \rm CCD
imaging detectors \citep[]{asai2000} have shown
that the spectra of a large fraction of bright LMXBs
exhibit line emission. 
X-ray lines at $\sim 1$~keV are often mixed with
various species, so that only the brightest LMXBs had
clear line identifications, as in the case of \ion{Ne}{10} Ly$\alpha$ 
in 4U1626-67 \citep[]{angelini1995}, Fe L in Sco X-1, or
\ion{O}{8} Ly$\alpha$ \& Ly$\beta$ in 4U1636-53
\citep[]{vrtilek1991}.  

The X-ray line emission arises presumably as
the result of irradiation of the disk by the X-ray continuum,
producing an extended source of reprocessed emission.
Evidence of X-ray emission from extended regions in LMXBs comes from 
the spectral variations during ingress and egress phases of eclipses,
and during rapid intensity fluctuations known as dips.
Most dips, which are observed to precede eclipses,
are thought to result from variable obscuration and attenuation
of the primary continuum by material near the outer disk
edge, which has been thickened due to impact of the
accretion stream with the disk  \citep[]{white1982,frank1987}. 
Dips that are uncorrelated with orbital phase
can be produced by orbiting clouds crossing the line
of sight, as shown in Figure \ref{fig:schem_lmxb}.
A cloud larger than $\sim 10^6$~cm can obscure the X-rays
from the neutron star.
Hard X-ray emission, presumably originating in the 
ADC, and representing a few percent of the non-eclipse flux,
remains visible during mid-eclipse in several LMXBs, implying that
the ADC is larger than the secondary star
\citep[]{whiteh1982,mcclintock1982}. LMXB
spectra during eclipses or dips may {\it harden} or {\it soften}, i.e. the
proportion between hard ($\sim 3$--10~keV) and soft ($\sim 1$--3~keV) 
X-rays changes. Most sources harden during dips
\citep[]{parmar1986}, consistent with
photoelectric absorption, but there are exceptions like the softening
of 4U1624-49 \citep[]{church1995}, and an unchanging X1755-33
\citep[]{white1984,church1993}.
Sources such as EXO0748-67, X1916-05,
and X1254-69 show evidence for an unabsorbed spectral component
during dips \citep[]{parmar1986,church1997}, revealing an
extended source of X-rays which is larger than the ADC. These soft
X-rays are likely radiation reprocessed in the accretion disk. Dip
ingress/egress times indicate ADC sizes in the $10^9$ to $5 \times
10^{10}$~cm range, a factor of a few smaller than the accretion disk
sizes calculated from typical orbital parameters \citep{church2000}.
A soft X-ray emission component distinct from the hard X-ray continuum
has also been interpreted as being due to the effects of absorption edges or line emission 
in some LMXBs \citep{parmar2000,church1998}. The unequivocal identification
of the accretion disk X-ray emission, 
requires both an energy resolution that is higher than CCD detectors,
and high throughput, plus a quantitative theoretical prediction of the
X-ray emission from the disk. We discuss the recently observed high
resolution spectra in section \ref{sub:gspectra}.

\subsection{Radiatively heated accretion disks}
\label{sub:raddisk}

\subsubsection{Radial structure}
\label{sub:radialdisk}
In LMXBs roughly half of the gravitational potential energy is
released in the vicinity of the compact object (i.e., in a
boundary layer near the neutron star surface). The disk is exposed to
this radiation, and it will be heated by it. Radiative heating
can exceed internal viscous heating in the outer region of the disk.
The temperature structure of the disk can thus be controlled by the 
X-ray field photoionizing the gas, suppressing convection,  and
increasing the scale height of the disk. 

Assuming that all the viscous heating and radiative heating from
illumination by the central source is
radiated locally as a blackbody (as in the SS73 model),
\citet[]{vrtilek} calculated the temperature
for a geometrically thin disk, with $r \gg R_1$,
where $R_1$ is the radius of the compact X-ray source:
\begin{equation}
\label{eq:locdisk}
\sigma T_{\rm phot}^{4} \simeq \frac{3 G M_1 \dot{M} }{8 \pi r^{3} }
+ \frac{(1 - \eta) L_{\rm x} \sin \theta(r)}{4 \pi r^{2}},
\end{equation}
where $T_{\rm phot}$ is the photospheric temperature, $M_1$ is the mass
of the compact X-ray source, $G$ is the gravitational constant,
$\sigma$ the Stephan-Boltzmann constant, $\theta$ is the
grazing angle of the incident X-ray flux with respect to the disk surface,
and $\eta$ is the X-ray {\it albedo} such that $(1-\eta)$ is the fraction of
X-rays absorbed at the photosphere. The albedo has been derived
from  optical observations \citep[]{dejong}.
The first term on the right-hand side of equation (\ref{eq:locdisk}) is the 
energy dissipated within the SS73 disk, and the second term is the
radiative heating.
The radiative heating term will dominate where
\begin{equation}
\label{eq:disk}
r>2.3 \times 10^8 ~\biggl(\frac{M_1}{M_{\sun}}\biggr) ~
\biggl(\frac{1 - \eta}{0.1}\biggr)^{-1}~
\biggl( \frac{\sin \theta}{0.1}\biggr)^{-1}~
\biggl(\frac{\epsilon_{\rm x}}{0.1}\biggr)^{-1} ~\rm cm,
\end{equation}
where the X-ray luminosity is written in terms of an
X-ray {\it accretion efficiency} $\epsilon_{\rm x}$, according to
$L_{\rm x}= \epsilon_{\rm x} \dot{M} c^2$. For example, accretion onto
a neutron star results in roughly 1/2 of the gravitational potential energy
being converted into X-rays, or $\epsilon_{\rm x} =GM_1/2c^2 R_1$.
The disk, therefore, is radially divided in two regions: an inner region
dominated by internal dissipation, and an outer region dominated by
external illumination.  External radiation will dominate the disk
atmosphere energetics for the outer two or three decades in radii, and
the local dissipation and magnetic flare heating, if any, will be
ignored there (see also section \ref{sub:bfield}).

\subsubsection{Vertical Structure}
The radial dependence of the disk temperature in equation (\ref{eq:locdisk})
relies on averaging physical quantities such as
the dissipation parameter $\alpha$
in the direction perpendicular to the disk plane, which is valid for
regions in the disk that are optically thick. 
However, as we will show in this article,
the radiative recombination spectrum is very sensitive to
the radial {\it and} the vertical ionization structure, including regions
with an optical depth $\tau \lesssim 1$.

To obtain a high resolution spectrum of an accretion disk, and in
particular one for which the outer (or upper) layers are X-ray photoionized,
several authors have calculated the vertical structure by
solving the radiation transfer equations, assuming hydrostatic
equilibrium. Models have been applied to AGN and LMXBs in the
high-$L_{\rm x}$ state, since in the low-state radiatively
inefficient accretion ensues, which is described by a separate family
of models (\citet[]{lowacc} and references therein).
In radiatively efficient accretion disks,
the radiative transfer is typically simplified by using an
on-the spot approximation and the escape probability formalism.
Due to photoelectric absorption and Compton scattering, the
ionization structure of the disk is stratified, and it is
approximated by a set of zones, each with a single ionization parameter.
The ionization structure of the disk can be solved by using
photoionization codes such as 
CLOUDY \citep{ferland1998} and XSTAR \citep{kallman1982},
which calculate the ionization and thermal equilibrium state
of the gas at each zone.

\citet{ko1991,ko1994} calculated the vertical structure
of an illuminated accretion disk and obtained the
recombination X-ray spectrum for individual rings on the disk.
\citet{ray93} utilized the temperatures in equation (\ref{eq:locdisk})
and calculated the vertical
structure and the UV spectrum from the entire disk.
Both assumed parameters for LMXBs, and gas pressure-dominated disks.
Later models of photoionized accretion disks focused primarily on
calculating the Fe K$\alpha$ fluorescence emission from AGN disks.

\citet{rozanska1996} and \citet{rozanska1999} modeled
semi-analytically the stratified, photoionized transition region
between the corona and the disk in AGN. They found that their approximations,
which included on-the-spot absorption, matched more accurate radiation
transfer codes for optical depths $\la 10$.
They also discussed the existence of a two-phase
medium, stopping short, however, of calculating an X-ray spectrum.
\citet{nay2000} modeled a radiation-pressure dominated
disk and showed that the vertical structure of the disk implied
significant differences in the Fe K fluorescence line spectrum compared to
that predicted by constant-density disk models
\citep{ross1993,matt1993,zycki1994}.
In addition, \citet{nay2000} also found that the gas was
thermally unstable at certain ionization parameters, which created an
ambiguity in choosing solutions and a sharp transition in temperature
in the disk. This instability is discussed in section \ref{sec:instab}.
\citet{balla} calculated
the vertical structure of disk ring as a function of radius, accretion
rate, the angle of incidence of radiation, the photon index, and the black
hole mass, albeit using a diffusion approximation.
\citet{roza2002} calculated the hydrostatic disk structure including
Compton scattering and line transfer without assuming the escape
probability approximation. \citet{roza2002} also 
calculated the structure of the optically thick part of the disk, by use of the
diffusion approximation and the local
$\alpha$-prescription (Eq. [\ref{eq:viscosity}]).
All of the above models calculate the disk structure for
one radius at a time. 

\citet{li} found the static solution that resolves the
thermal instability in the gas by considering the effect of
conduction, and they computed the X-ray recombination and resonance line
scattering spectrum for the conduction transition region that forms
between stable solutions in the disk.  With this procedure, the
unphysical, sharp transition between stable phases was eliminated. Li
et al. considered ionizing continua typical of AGN, which can yield 
stable solutions with three different temperatures for a given pressure
ionization parameter $\Xi$ (defined in section \ref{sec:ionpar}).
Up to three distinct transition layers
can form.  The reflection and recombination spectrum of the transition
regions in the 0.5--1.5~keV range was computed by considering the
vertical structure of an isobaric, optically thin region.
They found that resonant scattering can be important within the
transition region, depending on the local gravity and luminosity,
which yields a line spectrum which is different from
that of pure recombination emission.

The vertical structure of an optically thick accretion disk
can be obtained using the diffusion approximation, which assumes
that the photon mean free path $\lambda$ is much smaller than the scale of
temperature and density gradients $T/\nabla T$ and $\rho / \nabla \rho$,
respectively. Adding convective heat transfer by introducing an
adiabatic temperature gradient, \citet[]{meyer1982} have calculated
the vertical structure of an isolated accretion disk which is
dominated by convection zones.  Such techniques are used in the
standard stellar structure equations. X-ray illumination from the
central compact object suppresses convection, reduces the thermal
gradients in the disk, and has a stabilization effect in the outer
radii; but if X-ray illumination is combined with the diffusion approximation,
it also produces a convex disk that self-shadows the outer
disk regions. This contradicts the observed spectra, which show evidence
of reprocessing from the outer disk \citep[]{dub99}. A semi-analytical model using
a variable $\alpha$-viscosity prescription was used to model
AGN disks and to investigate its effects on the Lyman edge absorption
and emission
\citep[]{rozanska1999}.

The failure of the diffusion-equation models to reproduce 
a concave disk that can efficiently reprocess the central X-rays
may indicate that important effects were neglected.
First, the effect of the disk atmosphere and corona was ignored.
Second, turbulent heat transfer may produce
a vertical disk structure that is nearly isothermal. 
The strong turbulence occurring at the scale of the disk
thickness in magneto-hydrodynamic (MHD) models supports this hypothesis
\citep[]{miller2000}. A reliable calculation of the turbulent heat
transfer in an accretion disk is needed. Therefore, we prefer
to use the \citet[]{vrtilek} vertically-isothermal disk for the 
optically thick region.

The diffusion approximation is inadequate when
calculating high resolution spectra since line radiation must
originate in a region where the photon mean-free-path $\lambda$ exceeds
the scale of the temperature gradient, i.e. $\lambda \gtrsim T/\nabla T$.
Thus, just as for stellar atmospheres \citep{mihalas},
an explicit radiation transfer calculation 
without assumption of local thermodynamic equilibrium is needed.
The modeling of photoionization heating, recombination cooling,
and X-ray opacities is then required in the atmosphere.

\section{Model atmosphere}
\label{sec:modelatm}
We consider a LMXB with a $M_{*} = 1.4$ $M_{\sun}$ primary
radiating an Eddington luminosity ($L_{\rm x} = 10^{38.3}$~erg~s$^{-1}$)
bremsstrahlung continuum, with $T=8$~keV.  A set of fiducial system
parameters for a bright LMXB is used, so application to a particular
source will require using the observed X-ray continuum to improve accuracy.
The maximum radius of the centrally-illuminated disk is $10^{11}$~cm,
so the orbital period $\sim 1$ day. The minimum radius is $10^{8.5}$~cm,
below which the omitted effect of radiation pressure, in large part, 
determines the disk structure.

The vertical structure of the disk atmosphere for each annulus in the
array, is obtained by integrating the hydrostatic balance and 1-D
radiation transfer equations for a slab geometry (Fig. \ref{fig:disk_model}):
\begin{equation}
\label{eq:hydro}
\frac{\partial P}{\partial z} = - \frac{G M_{*} \rho z }{r^{3}}
\end{equation}
\begin{equation}
\label{eq:rad1}
\frac{\partial F_{\nu}}{\partial z} = 
	- \frac{\kappa_{\nu} F_{\nu} }{\sin \theta}
\end{equation}
\begin{equation}
\label{eq:rad2}
\frac{\partial F_{\nu}^{\rm d}}{\partial z} = 
	- \kappa_{\nu} F_{\nu}^{\rm d} 
\end{equation}
while satisfying local thermal equilibrium (see also eq. [\ref{eq:netheating}]):
\begin{equation}
\label{eq:thermal}
\Lambda (P, \rho,  F_{\nu}) = 0 
\end{equation}
and ionization balance (see also eq. [\ref{eq:ioneq}]):
\begin{equation}
\label{eq:ioneq2}
{\rm ion \ formation \ rate} = {\rm ion \ destruction \ rate }
\end{equation}
where $P$ is the total pressure, $\rho$ is the mass density, 
$F_{\nu}$ is the net flux of incident radiation (which is the intensity
integrated over all solid angles in erg~cm$^{-2}$~s$^{-1}$~Hz$^{-1}$), $F_{\nu}^{\rm d}$ is the
reprocessed net flux propagating down towards the
disk midplane, $z$ the vertical distance from the midplane,  $G$ the
gravitational constant, $\theta$ the grazing angle of the radiation on
the disk, $r$ the radius, $\nu$ is the frequency,
and $\kappa_{\nu}$ is the local absorption
coefficient. The rays corresponding to
$F_{\nu}$  and $F_{\nu}^{\rm d}$ 
are defined in Figure \ref{fig:column}.
Hydrostatic equilibrium is satisfied to a $\lesssim 1$\%
accuracy and thermal balance to $\lesssim 0.01$\%.

For the structure calculation only, 100 logarithmically spaced energy bins, 
in the range $1$~eV $< h\nu < 1000$~keV, were used for $F_{\nu}$ and
$F_{\nu}^{\rm d}$. The grid is coarse, and yet sufficiently broad to
accomodate a hard X-ray tail in future models.
The reprocessed radiation propagating upwards,
$F_{\nu}^{\rm u}$, is omitted to accelerate the computation. This is a
good approximation since the radiative heating is dominated by the
direct flux $F_{\nu}$. The reprocessed flux $F_{\nu}^{\rm u}$ 
is calculated {\it a posteriori} by a high resolution spectral model
(section \ref{sec:adamodel}).
The difference between cooling and heating, $\Lambda$, includes Compton
scattering, bremsstrahlung cooling, photoionization heating,
collisional line cooling, and recombination line cooling (section
\ref{sec:thermal}).
Cosmic abundances \citep{allen} are assumed.
The code from \citet[]{ray93} 
computes the net heating and ionization equilibrium, models
Compton scattering in one dimension, and calculates line scattering
using escape probabilities.  A new disk structure calculation
simultaneously integrates equations 
(\ref{eq:hydro})--(\ref{eq:rad2}) by the
Runge-Kutta method, using an adaptive step-size control routine with error
estimation, and equation (\ref{eq:thermal}) is solved by a globally convergent
Newton's method \citep[]{numrec}. At the ADC height $z_{\rm cor}$, the equilibrium 
$T$ is close to the Compton temperature $T_{\rm compton}$,
from which we begin to integrate downward 
until $T < T_{\rm phot}(r)$. The optically thick part of the disk, with
temperature $T_{\rm phot}$, is assumed to be vertically isothermal
\citep[]{vrtilek}. To get $T_{\rm phot}$, the viscous
energy and the illumination energy are assumed to be locally
(re)radiated with a blackbody spectrum. 
Thus, for $z_{\rm phot} \ll r$ and $R_{*} \ll r$,
equation (\ref{eq:locdisk}) can be used with $M_1 \equiv M_{*}$.
The height at which the integration ends is defined as the photosphere
height $z_{\rm phot}$.  Thus, we assume that 
viscous dissipation dominates heating
for $z < z_{\rm phot}$ (Fig. \ref{fig:column}).

The boundary conditions, shown schematically in
Figure \ref{fig:column}, are set at the ADC to 
$P(z_{\rm cor})= \rho_{\rm cor} k T_{\rm compton} / \mu m_{p}$,
$\int F_{\nu}(z_{\rm cor}) d\nu =L_{x}/4 \pi r^2$, and
$F_{\nu}^{\rm d}(z_{\rm cor})=0$,
where $k$ is the Boltzmann
constant, and $\mu$ is the average atomic weight of baryons
in units of the proton mass $m_{p}$. The boundary
conditions at the photospheric height
($= z_{\rm phot}$) for $F_{\nu}$ and $F_{\nu}^{\rm d}$ are set free, 
and the shooting method \citep[]{numrec} is used
with shooting parameter $\rho_{\rm cor}$, which is adjusted until 
$P(z_{\rm phot})= \rho_{\rm phot} k T_{\rm phot} / \mu m_{p}$ is satisfied
at the photosphere.
Note $\rho_{\rm phot}$ is the viscosity-dependent density 
calculated for an X-ray illuminated SS73 disk.

The shooting method consists of guessing the value of the
coronal density which matches the desired pressure
at the bottom of the gas column. 
The boundary conditions define $P_{\rm cor}$ once $\rho_{\rm cor}$
is chosen. Equations (\ref{eq:hydro})--(\ref{eq:ioneq2})
are simultaneously solved during the integration. The temperature
drops as the integration proceeds downward through
the atmosphere.
When $T$ reaches a value below $T_{\rm phot}$, the pressure at 
that point is compared to the expected pressure of the isothermal
disk at that height. If it does not match to better than $\sim$1 \%,
the integration is repeated with a new estimate of the coronal
density. While it is not clear that photoionization will cease to be
important for temperatures less than $T_{\rm phot}$, such zones
emit negligible X-ray fluxes if $r \gtrsim 10^9$~cm.
Our new structure calculation also includes the effects of physical
instabilities (section \ref{sec:instab}), and it removes the numerical
instabilities obtained by \cite{ray93}.

A novel and important feature of this model is that the incident
radiation is allowed to modify the disk atmosphere geometry, such that
the heating and expansion of the atmosphere resulting from illumination
are used to calculate the height profile of the atmosphere
as a function of radius. This feedback between the radiative heating
and the atmospheric structure is depicted in Figure \ref{fig:feed}.
The atmospheric height is
used to derive the input grazing angle of the radiation
for the next model iteration. This contrasts with calculating
the grazing angle using the pressure scale height
of the optically thick disk \citep{vrtilek}, which is in general 
much smaller than the photoionized atmosphere, and which
underestimates the grazing angle and the line intensities
by an order of magnitude.
To get $T_{\rm phot}$ self-consistently from equation (\ref{eq:locdisk}), the equation
\begin{eqnarray}
\label{eq:diskangle}
\theta(r) & \simeq & \beta - \alpha + \arctan \biggl(\frac{R_{*}}{r}\biggr) \nonumber \\
 & = & \arctan \biggl(\frac{dz_{\rm atm}}{dr}\biggr) 
  - \arctan \biggl(\frac{z_{\rm atm}}{r}\biggr) + \nonumber \\
& & \arctan \biggl(\frac{R_{*}}{r}\biggr)
\end{eqnarray}
is needed, where $z_{\rm atm}(r)$ is defined as the height where the
frequency-integrated grazing flux $\int F_{\nu}(z_{\rm atm}) ~d\nu$ is attenuated by $e^{-1}$,
and $\alpha,\beta$ are defined in Figure \ref{fig:disk_model}.
The $\arctan (R_{*}/r)$ term is neglected, which is valid for
$r \gtrsim 10^{8.5}$~cm. As discussed above, $\theta(r)$ is calculated
iteratively from equation (\ref{eq:diskangle}).
After an initial guess for $z_{\rm atm}(r)$, it is re-calculated from
the newly obtained disk structure. A power-law fit to
$z_{\rm atm}(r)$ works well to obtain $\theta(r)$. 
This iteration is performed with a limited number of radial bins (5),
to save computation time. 
The iteration is stopped after $\theta(r)$
and $T_{\rm phot}$ converge to $\lesssim 10$ \%. 
After convergence, the number of logarithmically spaced radial
bins is increased to 26. The process of convergence does
not depend on the initial choice of $\theta(r)$, and it is
shown in Figure \ref{fig:conv}.  However, convergence does depend on the
choice of $z_{\rm atm}$, which is a free parameter in the model.
Since $z_{\rm atm}$ is not physically determined, it must be defined
{\it ad hoc}, but it is bound by $z_{\rm atm}>z_{\rm phot}$.
For $z \lesssim z_{\rm phot}$, the illumination $F_{\nu } \rightarrow
0$, and the disk blackbody flux $F_{\rm bb}(T_{\rm phot})$ takes over.  
To test how sensitive is the result to the definition of $z_{\rm atm}$,
we calculate $\theta(r)$ taking
$z_{\rm atm} = z_{\rm phot}$, and we use this to estimate the systematic
errors of the 1-D radiation transfer calculation.

\subsection{The Choice of Assumptions}

The validity of the assumptions is reviewed, both for the the model
presented here and for some of the 
accretion disk models in the astrophysical literature.

For modeling X-ray line emission from the disk atmosphere, the 
commonly used assumptions of $LTE$ and the diffusion
approximation will not hold. In addition, assuming a constant density
in the vertical direction will be inadequate, since the hydrostatic
equilibration time is small or comparable to other relevant
timescales, and the line emission is highly sensitive to the vertical
ionization structure. The recombination emission is especially
sensitive to this structure, since each ion emits clearly resolvable
line energies. Also, fluorescence emission can be reprocessed
by a Compton-thick, fully ionized gas above it  \citep[]{nay2000}.

Thermal equilibrium and ionization equilibrium
are reasonable assumptions in a time averaged sense.  
Hydrostatic equilibrium is also assumed, to avoid explicit computation
of the plasma dynamics with radiative transfer.
Deviations from hydrostatic equilibrium are smoothed in the timescale
\begin{equation}
t_{\rm hydro} = z_{\rm atm}/c_{\rm s} \sim 2.7  \biggl( \frac{z_{\rm atm}}{10^7~{\rm cm}} \biggr) 
T_5^{-1/2} \ \ \rm sec
\end{equation}
where $c_{\rm s}$ is the sound
speed. Material from the disk moves radially within
the viscous timescale $t_{\rm visc} \sim r^2 / ( \alpha z_{\rm atm} c_{\rm s} ) > t_{\rm hydro}$
\citep[]{fkr1985}, so the gas can reach hydrostatic
equilibrium before it flows inward (i.e, the radial accretion velocity
is always subsonic). However, the Keplerian orbital velocity
$v_{\rm k} \gg c_{\rm s}$ is highly supersonic. If the gas
flow in the corotating frame of the gas is also supersonic, then
shocks would collisionally ionize and heat the gas.
In such a case, the observed spectrum of the disk would
significantly deviate from a photoionized gas in ionization and
thermal equilibrium.
The thermalization and ionization equilibrium timescale of the
atmosphere is driven by the recombination timescale
\begin{equation}
t_{\rm th} = t_{\rm rec} \sim 0.3 \biggl( \frac{ T_{5}^{1/2} }{ n_{14} Z^2 } \biggr) \ \ \rm sec
\end{equation}
which can be derived from equation (\ref{eq:recrate}), and
where $T_{5}$ is the temperature in units of $10^5$~K, 
$n_{14}$ is the density in units of $10^{14}$~cm$^{-3}$,
and $Z$ is the atomic number \citep[]{rey1995}. 
The photoionization timescale is shorter than $t_{\rm rec}$
where the gas is fully stripped; otherwise both
timescales are comparable for the relevant ions having similar 
abundances.
The Coulomb collision relaxation timescale between electrons and ions
$t_{\rm ep}$ is slower than between identical particles, and is \citep{spitzer}
\begin{equation}
t_{\rm ep} \simeq 3 \times 10^{-6} \frac{ T_{5}^{3/2}}{n_{14}} \ \ {\rm sec}.
\end{equation}
Electron-electron relaxation is $\sim m_{\rm p}/m_{\rm e}$ times
faster, and proton-proton relaxation is $\sim \sqrt{  m_{\rm p}/m_{\rm e} }$
times faster.
For the hot corona at the outer disk
at $T \sim 10^7$~K and $n_{\rm e} \sim 10^{11}$~cm$^{-3}$ (from the coronal
structure in section \ref{sec:diskstruct}), the relaxation timescale is 
$t_{\rm ep} \sim 3$~sec.
The fully ionized gas in the corona, which is near the Compton temperature,
has a thermal timescale of 
$t_{\rm th} = t_{\rm compton} = 10 r_{10}^2 L_{38}^{-1}$~sec, where
$r_{10}$ is the disk radius in units of $10^{10}$~cm,
and $L_{38}$ is the X-ray luminosity in units of $10^{38}$~ergs$^{-1}$
\citep[]{rey1995}. Thus, thermalization in the
disk atmosphere and corona is driven by the ionization timescales,
since the Coulomb relaxation times are comparatively fast due
to the large density.
Thermal and ionization equilibrium occur faster
than hydrostatic equilibrium, for length scales 
$z_{\rm atm} \gtrsim 10^6$~cm.
If $t_{\rm th} < t_{\rm hydro}$, 
luminosity fluctuations with timescales $t_{\rm flux}$ such that
$t_{\rm th} < t_{\rm flux} < t_{\rm hydro}$ will take the gas outside
hydrostatic equilibrium, but not out of thermal equilibrium.
Integrated spectral observations on timescales
$t \gg t_{\rm flux}$ cannot observe this effect.

The radiation transfer is complex, and the assumptions used to
simplify calculations could be problematic.  In particular, by
dividing the disk atmosphere into annular zones with a given vertical
gas column, our 1-D radiation transfer calculation assumes that 1) the
primary continuum is not absorbed before reaching the top of the
column, 2) the radiation in the column propagates from top to bottom
at a given grazing angle, and 3) there is no significant radiative
coupling from one disk annulus to another, which is used to justify
the slab approximation.  The above assumptions are inadequate if the
column height is comparable to the disk radius, or if the photon mean
free path in the gas column is many times the local radius. Thus,
future 2-D calculations will result in better bookkeeping of photons,
a more accurate structure, and a more reliable X-ray line
spectrum.

The correct calculation of line transfer in the gas is also a concern,
since the disk atmosphere is optically thick in the lines (section \ref{sec:diskstruct}). Line
transfer is complicated by the Keplerian velocity shear, which has to
be taken into account for a given viewing angle
\citep{murray_chiang1997}.  The escape-probability approximation
used to calculate line transfer in the disk may also be inadequate
because of the large optical depths.

Our calculations show that the proper treatment of a thermal
instability \citep{field1965,kmt} and conduction
affect the spectrum significantly 
\citep{zel,li} (section \ref{sec:instab}).
A two-phase gas could form, with clouds of an unknown size
distribution and with undetermined dynamics of evaporation and
condensation \citep{bemc}, with each phase having distinct
ionization parameter and opacity. The instability is sensitive to 1)
the metal abundances, 2) the continuum shape \citep[]{hess}, and 3)
the atomic kinetics \citep{savin1999}. 

The local viscous energy dissipation rate per unit volume in the disk
atmosphere may be included in equation (\ref{eq:thermal}) with the form
\citep[SS73]{czerny}
\begin{equation}
\label{eq:viscosity}
Q_{\rm visc} = \frac{3}{2} \Omega \alpha P
\end{equation}
where $\Omega$ is the Keplerian angular velocity, $\alpha$ is the
viscosity parameter, and $P$ is the local pressure.
Equation (\ref{eq:viscosity}) is an extension of the $\alpha$-disk model
(where the viscous dissipation is vertically averaged),
and it assumes the local validity of the $\alpha$
prescription, which is untested. Fortunately, our numerical modeling
indicates that the viscosity term is negligible in most regions of the
disk atmosphere except for the inner disk and $\alpha \sim 1$ (in particular,
near the Compton-temperature corona). This viscosity term
enhances a thermal instability between $10^6$ and $10^7$~K.  Vertically
stratified MHD models \citep[]{miller2000}, although inconclusive, owing to
the uncertain effect of boundary conditions, show that
the viscous dissipation drops rapidly at $\ga 2$ pressure scale
heights away from the disk midplane, providing evidence against
equation (\ref{eq:viscosity}). Since the disk atmosphere is always a few
scale heights above the midplane, we choose not to include
equation (\ref{eq:viscosity}) in our models.  
Equation (\ref{eq:viscosity}) has been applied in
the optically thick regions of the disk by \citet{dub99} and in the disk
atmosphere by \citet{rozanska1996}. Other forms for the local
dissipation that reduce to the $\alpha$-disk have been used
\citep{meyer1982}.

\subsection{Ionization balance}
 
In steady state the equation of
ionization balance for each ion $Z^{+i}$ is
\begin{eqnarray}
\label{eq:ioneq}
\frac{\partial n_{z,i}}{\partial t} & = &
 n_{z,i+1} n_{\rm e} \alpha_{z,i+1} 
+ n_{z,i-1} ( \beta_{z,i-1} 
+  n_{\rm e} C_{z,i-1} ) \nonumber\\
 & & - n_{z,i} ( 
 \beta_{z,i} 
+ n_{\rm e} \alpha_{z,i} 
+ n_{\rm e} C_{z,i} ) \nonumber\\
& & + n_{z,i-2} \beta_{z,i-2}^{\rm K} B_{z,i-1}
 = 0 ,
\end{eqnarray}
where $\beta_{z,i}$ is the photoionization rate (s$^{-1}$) of $Z^{+i}$,
$C_{z,i}$ is the collisional ionization rate coefficient ($\rm{cm^3 ~s^{-1}}$)
 of $Z^{+i}$, $\alpha_{z,i+1}$ is the recombination rate
coefficient ($\rm{cm^3 ~s^{-1}}$) of ion $Z^{+(i+1)}$, and so on.
The code also includes K-shell photoionization followed by
Auger ionization in the last term of equation (\ref{eq:ioneq}). The 
Auger branching fraction is $B_{z,i-1} = 1 - Y_{z,i-1}$, where
$Y_{z,i-1}$ is the
fluorescence yield.
Multiple Auger decays are ignored in equation (\ref{eq:ioneq}).
The terms with $\alpha_{z,i+1}$ and $\alpha_{z,i}$ account for all
two-body recombination processes. The coefficients $C_{z,i}$ and
$\alpha_{z,i}$ depend on the electron temperature $T$ for any ion
$Z^{+i}$.  Given the photoionization cross-section $\sigma_{PE,z,i}$ 
and ionization threshold energy $\chi_{z,i}$ of the ion $Z^{+i}$,
the photoionization rate for a point source of ionizing continuum is
\begin{equation}
\label{eq:photorate}
\beta_{z,i} =\frac{L_{\rm x}}{r^2}~ \int_{\chi_{z,i}}^{\infty}dE~\frac{S_E(E)}{4\pi 
E}~\sigma_{PE,z,i}(E),
\end{equation}
where $S_E$ is the spectral shape function, normalized on a suitable
energy interval. 
For the accretion disk atmospheres orbiting neutron stars, which are
of interest here, collisional ionization rates are negligible compared
to photoionization rates.

\subsection{The ionization parameter}
\label{sec:ionpar}

Let $\xi=L_{\rm x}/n_{\rm p} r^2$ (in erg~s$^{-1}$~cm)
be the {\it ionization parameter},
where $n_{\rm p}$ is the proton number density \citep{tarter}. 
The ionization parameter $\xi$ is factored out of equation (\ref{eq:photorate}), and
together with the spectral shape function $S_E$, it defines uniquely
the charge state distribution in an optically thin photoionized gas
(eq. [\ref{eq:ioneq}] does not include three-body recombination, which is important
at $n_e \gtrsim 10^{16}$~cm$^{-3}$).

Another, dimensionless ionization parameter is constructed with
the radiation pressure $P_{rad}$ and the proton gas pressure $P_{gas}$
\citep[]{kmt}. This ionization parameter, $\Xi$, is defined as
$\Xi \equiv P_{rad}/P_{gas}$, where $P_{rad} = \int F_{\nu} d\nu / c$.
Note $\Xi = \xi/4 \pi c k T$.  The new parameter $\Xi$ is useful when
the local pressure can be defined. For an optically
thin gas, an isobar has constant $\Xi$. 

\subsection{Thermal Equilibrium}
\label{sec:thermal}

We review the terms in the thermal equilibrium equation, which
we solve with the \citet{ray93} photoionized plasma code. 
Thermal equilibrium is enforced at each zone in the disk atmosphere.
The explicit form of the thermal equilibrium condition,
equation (\ref{eq:thermal}), is
\begin{eqnarray}
\nonumber
 & {\rm Compton \ net \ heating } + {\rm photoionization \ heating } = & \\
 & {\rm bremsstrahlung \ cooling } + {\rm recombination \ cooling} & \nonumber \\
 & + {\rm collisional \ cooling} , &
\end{eqnarray}
which corresponds to \citep{halpern} 
\begin{eqnarray}
\label{eq:netheating}
\int F_\nu \biggl( n_{\rm e}  \Gamma^{{\rm com}}_{\nu} + \sum_{z,i} n_{z,i} 
\Gamma_{\nu,z,i}^{{\rm phot}} \biggr) d\nu =  \nonumber \\
 n_{\rm e} \sum_{z,i}  n_{z,i} (
\Lambda_{z,i}^{{\rm brem}} + 
 \Lambda_{z,i}^{{\rm rec}} + \Lambda_{z,i}^{{\rm col}})
\end{eqnarray}
in units of erg~s$^{-1}$~cm$^{-3}$, where the 
rates for each process and other dependencies are included
in the group of rate coefficients 
$(\Gamma^{{\rm com}}_\nu, \Gamma_{\nu,z,i}^{{\rm phot}}, 
\Lambda_{z}^{{\rm brem}}, \Lambda_{z,i}^{{\rm rec}}, \Lambda_{z,i}^{{\rm col}}$),
$n_{\rm e}$ is the electron number density, $n_{z,i}$ is the
$Z^{+i}$ ion density, and
the sums are performed over all abundant ions. 
The radiative heating is directly proportional
the net flux $F_\nu$ (in units of erg~s$^{-1}$~cm$^{-2}$~Hz$^{-1}$),
and the density.
Cooling processes, which originate from electron-ion interactions,
are proportional to the square of the density and are in
general dependent on the electron temperature $T$. The 
coefficients in equation (\ref{eq:netheating}) can be obtained from \citet{halpern}, and some,
such as the recombination coefficients, are very dependent on the 
available atomic data. A list of the data used for the coefficients in
the model, and a list of the processes and transitions included in the
calculations can be found in \citet{ray93}. Recombination cooling
includes both radiative recombination and dielectronic recombination.
Collisional cooling includes cooling due to 
line emission and collisional ionization. 
Photon trapping and subsequent collisional
de-excitation reduces the cooling rate compared to the
optically thin case, but this affects the UV lines
more than the X-ray lines because the resonant scattering
opacity is larger in the UV.

The ions of  H, He, C, N, O, Ne, Mg, Si, S, Ar, Ca, and Fe 
are included in the thermal equilibrium equation (\ref{eq:netheating}),
and the ionization balance equation (\ref{eq:ioneq}).

For a fully ionized gas, such as the hot corona above the accretion disk,
Compton heating and inverse-Compton cooling dominate equation (\ref{eq:netheating}).
The Compton net heating term may be positive or negative,
since the transfer of energy between the photons and the electron gas
depends on the electron temperature and the shape of
the ionizing spectrum. In such cases, and in the non-relativistic case,
the equilibrium temperature is
\begin{equation}
\label{eq:comptemp}
T_{\rm compton} = \frac{h \int \nu F_\nu d\nu }
					{4 k \int  F_\nu d\nu }
\end{equation}
where $k$ and $h$ are the Boltzmann and the Planck constants, respectively.
The Compton temperature $T_{\rm compton}$ is determined uniquely by
the shape of the ionizing continuum. For an 8~keV bremsstrahlung spectrum,
$T_{\rm compton} \sim 2 \times 10^7$~K.

\section{Thermal Instability in Photoionized Gases}
\label{sec:instab}

Irradiated gas is subject to thermal instabilities for temperatures in
the $10^5$--$10^7$~K range \citep{buff,field1969}, such that
X-ray line emission at those temperatures may be suppressed.  The
\citet{field1965} stability criterion,
together with plasma equilibrium calculations
\cite[]{david1979,kallman1982,ray93,ferland1998},
indicate that a photoionized gas becomes thermally unstable when
recombination cooling of H- and He-like ions is important.  
Consequently, the disk atmosphere structure has a thermally unstable region,
which modifies the X-ray spectrum.

To clarify the nature of the thermal instability, consider the 
calculated net heating (Fig. \ref{fig:instab}).
The gas is externally heated, and the net heating depends on the state
variables and ionization of the gas, making this a peculiar system. 
The thermal balance locus, where the net heating is zero,
is denoted as the {\it S-curve} and 
is displayed in Figure \ref{fig:instab}.
The region to the left of the S-curve undergoes net
cooling, and the one to the right has net heating. 

To test stability, consider small $T$-perturbations starting from the
S-curve.  A vertical displacement from any point in Figure
\ref{fig:instab} represents an isobaric 
perturbation in $T$. From this, we find that the points in the
S-curve with positive slope
are stable, while those with negative slope are
unstable \citep[]{field1965}. This splits the
S-curve into three branches.  
The shape of the S-curve determines 
which range of $\Xi$ are unstable
(section \ref{sec:ionpar}). At such $\Xi$, thermal balance is
achieved by three distinct $T$ on the S-curve, two 
stable $T$ and one unstable $T$.
On the unstable branch, isobaric
$T$-perturbations cause a thermal runaway to one of the stable
branches.

The shape of the S-curve depends on the metal abundances and the ionizing
spectrum \citep[]{hess}. The S-curve is subject to uncertainties
in the atomic data \citep{savin1999}, and its shape may vary (albeit
not dramatically) from one plasma code to another.  Our calculated
S-curve for the disk atmosphere is shown in Figure \ref{fig:fig1}.

Most spectral studies of heated accretion disks in LMXBs and AGN
have either used unstable solutions, or just selected a subset of the
stable solutions.  \citet{roza2002} chose a monotonic density, which
is equivalent to selecting all points on the S-curve. This results in
a pressure which oscillates with height and a transition region 
which is {\it not} in hydrostatic equilibrium.  \citet{ko1994} and \citet{nay2001}
selected the hot branch of solutions, which produce a condensing
atmosphere biased towards high-ionization species. \citet{balla} do not
specify how the choice of solutions within the instability was made, but
they acknowledge the effects of the instability,
which are seen in the sharp temperature transition
obtained with their models.

The instability implies a large $\nabla T$
as the gas is forced to move between stable branches, requiring
the formation of a transition region whose size may be determined by
electron heat conduction, convection, or turbulence,
depending on which dominates the heat transfer.
For simplicity, calculations of emission from the transition region are omitted
in this article.
Upon calculation of the Field length $\lambda_F$, 
the lengthscale below which conduction dominates
thermal equilibrium \citep{bemc}, we estimate that conduction 
forms a transition layer $\sim 10^{-2}$ times thinner than the
size of the X-ray emitting zones. Nevertheless, X-ray line emission
from the neglected conduction region may not be negligible in
all cases \citep{li}. The $\xi$ values present in the 
transition region are absent in other regions,
which may allow the transition region to have observable spectral signatures.
Resonant scattering from the transition region may be observable 
in some situations \citep{li}.  The importance of the transition
region also depends on the shape of the S-curve and on the local gravity.

Conduction tips the balance of stability at sufficiently
small spatial scales. If the gas is not in static equilibrium, conduction can drive
phase transitions, and produce dynamic condensing or evaporating
fronts.  In static equilibrium, conduction quenches the instability
and produces a transition layer at $\Xi_{\rm stat}$ (stretching
the S-curve in Fig. \ref{fig:fig1}). This transition layer \citep{zel}
connects the low-$T$ stable branch at $\Xi < \Xi_{\rm stat}$ with the high-$T$
stable branch at $\Xi > \Xi_{\rm stat}$.
In the dynamic case, if the transition layer is located away from $\Xi_{\rm stat}$,
it will dynamically approach $\Xi_{\rm stat}$ by a conduction driven mass
flow, as shown in Figure \ref{fig:massflow}.  A transition layer with
$\Xi_{\rm stat} < \Xi < \Xi_{\rm evap}$ produces an evaporating front, while
a transition layer with $\Xi_{\rm cond} < \Xi < \Xi_{\rm stat}$ 
produces a condensing front \citep[]{zel,li}.

The disk structure for both condensing
and evaporating solutions is computed. We assume a steady state,
condensing or evaporating mass flow through the transition layer at
$\Xi_{\rm cond}$ or $\Xi_{\rm evap}$, respectively. 
The static equilibrium solution is an
intermediate case of the latter extreme cases. 
A single-valued $T(\Xi)$ is used, since a two-phase solution may be
buoyantly unstable, making the denser (colder) gas sink. 
The evaporating disk corresponds to the low-$T$ branch, while
the condensing disk corresponds to the high-$T$ branch (Figure \ref{fig:fig1}).
This introduces spectral differences (section \ref{sec:spec}). 

We do not know from first principles whether
the disk atmosphere is evaporating, condensing, or static.
However, a Compton-heated wind might be expected in the corona 
for large radii \citep{bms1983}.
The speed of the conduction mass flow is estimated to be $v_{\rm cond} = 2
\kappa T / 3 P_{\rm gas} \lambda_F$, by using the characteristic conduction
time at the Field length, where $\kappa$ is the \citet{spitzer}
conductivity \citep{mckee}.
A conduction mass flow speed $v_{\rm cond}=$ 1--2$\times
10^{-2}$ times the local sound speed is obtained. Thus, the phase dynamics will
depend on the subsonic ($v \gtrsim v_{\rm cond}$) flow patterns in the
disk atmosphere, and these flows will in part determine the evaporation
or condensation rates, together with the boundary conditions on mass flow. 

If the disk is in a steady state of evaporation or condensation,
the implied mass flow can have an effect on the global mass budget,
due to mass conservation. Steady state evaporation implies mass loss
or a disk wind, while condensation implies a mass gain \citep{zel,li}. 

A thermal instability due to Compton heating
and bremsstrahlung cooling can ensue between $10^6$ and $10^7$~K
if the ionizing spectrum extends well above $\sim 10$~keV \citep{kmt}.
For a 8~keV bremsstrahlung spectrum, this 
additional instability regime is suppressed \citep{hess}. Nevertheless,
some LMXBs have harder spectra \citep{white1995}. 
A double S-curve results from the hard spectra in AGN \citep{nay2001},
which allows a three-phase gas.

As mentioned above, gas dynamics which are not included in the model
can have an impact on the gas phase. The only physical mechanism known to
transport the necessary angular momentum for disk accretion
involves a magneto-rotational instability (MRI) which drives turbulent
flow in the disk \citep[]{bh98}. These turbulent flows are nearly
supersonic in the disk midplane region, where most of the mass is
accreted. Enhanced heat transfer rates due to this turbulent
flow could quench the thermal instability and affect the disk
structure. Turbulent heat transfer rates can be orders of magnitude
larger than the saturated conduction heat transfer rate.
However, it is not known whether such turbulent motions will also
be present in the disk atmosphere, which is several scale heights above
the disk midplane and has a density which is orders of magnitude smaller 
(Section \ref{sec:diskstruct}).  A decline in the viscous $\alpha$
parameter with vertical disk height was obtained with local MHD
models, and an enhanced ratio of the magnetic pressure to the gas
pressure with increasing height \citep{miller2000}.  The MRI also
favors the assumption of vertical isothermality in the optically thick
disk. In section \ref{sub:bfield}, we discuss the effects of magnetic fields
in the atmosphere and corona. 

\section{Spectral modeling}
\label{sec:adamodel}

With the disk structure $\rho(r,z)$, $T(r,z)$ 
and ion abundances $f_{z,i}(r,z)$,
the X-ray line emission from
the disk atmosphere is calculated using $HULLAC$ data
\citep[]{hullac}.
The code calculates the atomic structure and transition rates of
radiative recombination (RR) and the ensuing radiative cascade,
which can produce both line photons and
radiative recombination continuum (RRC) photons.
We include the H-like and He-like ions of C, N, O, Ne, Mg, Si, S, Ar, Ca, and 
Fe, as well as the Fe L shell ions.  Fluorescence emission, which is prominent 
for high-$Z$ ions such as those of Fe, is omitted in these calculations,
as well as resonant scattering, an additional source of line emission.
The recombination emissivities, and the opacities 
in this model, are
calculated as described in appendices \ref{sec:radrec} and \ref{sec:opacity},
respectively. 

The spectrum for each of the 26 annuli was added to obtain the disk
spectrum.  Each annulus consists of a grid of zones in the vertical
$\hat{z}$ direction, and $T$, $\rho$ and $f_{z,i+1}$ for each zone
are used to calculate the RR and RRC emissivities. The radiation 
is propagated outwards at inclination angle $i$, including
the continuum opacity of all zones above, thus accounting
for the optical depth of the atmosphere.
Compton scattering of the irradiating continuum is included
in the disk structure calculation but it is omitted in the synthetic spectrum.
The latter scattering adds a weak continuum component with the 
spectral shape of the neutron star emission.
The spectrum is Doppler broadened by the projected
local Keplerian velocity, assuming azimuthal symmetry.

\section{Disk structure}
\label{sec:diskstruct}

Once the atmosphere and corona are accounted for, 
the disk is thicker than would be expected from
the local pressure scale height ($=Z_{\rm P}$) alone. 
We found  $Z_{\rm P} < z_{\rm phot} < z_{\rm atm}$. 
To quantify the disk geometry, the calculated height
of the photosphere and atmosphere, $z_{\rm phot}$ and
$z_{\rm atm}$, are both fitted with $z = C(r/1~{\rm cm})^n$, with fit parameters $C$ and $n$. The fitted
parameters are $C_{\rm phot} = (2.4_{-1.9}^{+0.4}) \times 10^{-3}$ 
cm, $n_{\rm phot} = 1.14^{+0.06}_{-0.01}$,
$C_{\rm atm} = (1.0^{+0.2}_{-0.1}) \times 10^{-3}$~cm,
$n_{\rm atm} = 1.21 \pm 0.01$.
The above fits imply $z_{\rm phot} \sim 3Z_{\rm P}$ to $4Z_{\rm P}$ (depending
on radius) and 
$z_{\rm atm} \sim (7^{+2}_{-0})Z_{\rm P}$ to $(8^{+2}_{-0})Z_{\rm P}$.
We account for statistical errors and estimated systematics.
\citet{vrtilek} estimated $n_{\rm atm} = 9/7=1.29$, but in spite of
the steeper radial dependence, the Vrtilek et al. disk is thinner, and
it assumes $z_{\rm atm} = Z_{\rm P}$ for $r > 10^{10}$~cm.
The disk thickness derived from the
optical light curve observations of LMXBs relies on the
large fraction of X-rays from the neutron star which are shielded from
the companion by the disk. This \it de facto \rm disk boundary should be taken to be
$\sim z_{\rm atm}$, since, by definition, a fraction $1/e$ of the central
X-rays are absorbed there. 
In Figure \ref{fig:omega}, we
compare $Z_{\rm P}$ with $z_{\rm atm}$.
We find that previous theoretical studies severely
underestimated the size of the disk atmosphere. 

The X-ray continuum opacity of the
atmosphere is $\tau \ll 1$ for most lines of sight, except
for rays originating on the neutron star which
are incident at a small grazing angle $\theta(r)$, such that 
they are nearly parallel to the disk plane.
The atmosphere's ($z > z_{\rm phot}$)
maximum photoelectric opacity is always $\tau \ll 1$ in the vertical direction,
although $\tau / \sin \theta(r) \gtrsim 1$.
Illumination heating dominates at $r \gtrsim$ $10^{10}$~cm
(eq. [\ref{eq:locdisk}]), where only
$\int F_{\nu}(z_{\rm phot}) d\nu/ \int F_{\nu}(z_{\rm cor}) d\nu \sim 0.12$ 
of the incident photons reach $z_{\rm phot}$ directly, while
$\int F_{\nu}^{\rm d}(z_{\rm phot}) d\nu/ ( \int F_{\nu}(z_{\rm cor}) d\nu \sin \theta)
 \sim 0.35$ reach $z_{\rm phot}$
after reprocessing in the atmosphere. Thus, the atmospheric albedo is
$\sim 0.5$. Both the photosphere and atmosphere
contribute significantly to the disk albedo.
The total disk albedo deduced by \citet[]{dejong} 
from optical observations in LMXBs is $\eta \sim 0.9$.
We have found that its high value is partially explained by the atmospheric contribution.
Once the latter is taken into account, the photosphere's albedo
becomes $\sim 0.8$, since 0.5+0.5(.8)=0.9, which is closer to physical
expectations.

At any fixed radius, the vertical disk structure 
has a marked boundary between the optically thick, colder disk
and an optically thin, hotter atmosphere, as
shown in Figures \ref{fig:disk_temp} and \ref{fig:disk_density}.
At the largest scales, the vertical structure has two distinct
zones: a hot corona, in which Compton heating and cooling dominates, and
an atmosphere or warm corona, where photoionization heating and 
recombination cooling are most
important. Three regions are discernible in Figure \ref{fig:heatreg}, which
shows the vertical structure of the outer radius of the disk (other
radii show a similar pattern, aside from changes in scale).

The structure of the underlying atmosphere is better discerned by
plotting the height of the atmosphere above the photosphere, 
$z-z_{\rm phot}$. This reveals the presence of fine structure, in
particular a region emitting lines from low-Z He-like ions at
$T \sim 5 \times 10^4$~K 
(Fig. \ref{fig:atm_temp} and \ref{fig:atm_density}). The evaporating
and condensing disk model solutions (section \ref{sec:instab}) are  shown in
Figures \ref{fig:disk_temp} through \ref{fig:atm_density}.
This low-Z He-like ion region is small due to the rapid increase in continuum
opacity with decreasing temperature, but is resolved by the
adaptive step-size integration. A more extended, $T \sim 10^6$~K region
emits predominantly H-like ion and mid-Z He-like ion RR lines.
Both H-like and He-like ion emission regions can be identified by the
abundance distribution of the fully ionized and H-like ions,
which recombine to produce the H-like and He-like ion emission,
respectively (Fig. \ref{fig:lowzdist} and \ref{fig:midzdist}).  The
recombination line luminosity for a $u \to l$
transition in ion $Z^{+i}$ is
$dL_{u \to l} \propto n_{\rm e} n_{z,i+1} T^{-\gamma} dV$
(eq. [\ref{eq:emme}] and eq. [\ref{eq:lumdem}]).
The highest emissivities will be produced
at low temperatures and high densities. This implies
the region of origin of the emission will track the abundances
from Figures  \ref{fig:lowzdist} and \ref{fig:midzdist}, 
with an added bias towards the lower
range of heights, which are denser and colder.

The spatial distribution of K and L-shell Fe ions shows that the
structure calculation included all the intermediate
ionization states (Fig. \ref{fig:irondist}). The ionization 
parameter $\Xi$ varied continuously with atmospheric height
from full ionization at $\Xi \sim 10^3$, down
to the thermal instability regime at $\Xi \sim 10$,
where a break occurs.

The presence of the instability has a large effect on the
luminosity of He-like ion lines from mid-Z elements, 
as can be seen from Figure \ref{fig:midzdist}. In particular,
the Mg$^{+11}$ abundance is never allowed to peak, such
that the model predicts a dim \ion{Mg}{11} line.
A similar effect occurs with \ion{Si}{13} and \ion{Ne}{9}.
Thus, in the context of this disk model,
the brightness of these three lines will determine
whether the instability is operating as modeled.

The discontinuity in the density, temperature, and ionization state
is a result of enforcing the thermal stability of the chosen
solutions, since a range of temperatures from $\sim 6 \times 10^4$~K
to $\sim 7 \times 10^5$~K is unstable (see section \ref{sec:instab}). The
discontinuity is unphysical, of course, and can be smoothed in future
models by the inclusion of conduction, or any other heat transport
mechanisms in the disk that might be present, such as those due to
turbulence or convection. 

A comparison of the spatial ion distribution from the condensing disk 
(Fig. \ref{fig:lowzdist_cond}) and the evaporating disk (Fig. \ref{fig:lowzdist})
shows that the differences in the synthetic spectra
can be attributed to differences in the vertical disk structure. The low-Z He-like ion
line producing region shrinks, while the H-like ion line and mid-Z He-like
ion line producing region expands in the condensing disk model, as compared
to the evaporating case. 
The condensing solution shows a more extended Fe L emission
region, with particularly large ion emission measure for 
Fe$^{+19}$, which recombines to Fe$^{+18}$ and produces strong \ion{Fe}{19} lines
(Fig. \ref{fig:irondist}[b]), while the evaporating disk shows a larger
emission measure for 
Fe$^{+17}$ at lower temperatures (Fig. \ref{fig:irondist}[c]), which recombines to Fe$^{+16}$ and
emits the \ion{Fe}{17} lines more efficiently,
as will be shown in section \ref{sec:spec}.
This behavior traces back to the
choice of solutions from the stability curve in section \ref{sec:instab}.

\section{Spectroscopy}
\label{sec:spec}

In this section we delineate the circumstances under which the disk emission
is rendered observable, and we describe
the disk spectroscopy and its diagnostics.

The LMXB photon net flux (photon~cm$^{-2}$~s$^{-1}$~keV$^{-1}$) is modeled by:
\begin{equation}
\label{eq:spec}
F_{E}^{\rm tot} = e^{- \sigma_{E} N_{H}^{*}} F_{E}^{*} + 
e^{- \sigma_{E} N_{H}^{\rm disk}} F_{E}^{\rm disk}
\end{equation}
where $F_{E}^{*}$ is the neutron star continuum, 
$F_{E}^{\rm disk}$ is the RR line and RRC modeled flux (Fig. \ref{fig:purelines}), 
$N_{H}^{*}$ and $N_{H}^{\rm disk}$ are the neutral hydrogen
absorption column densities, $E$ is the photon energy, and $\sigma_{E}$ are the 
\citet[]{abun} absorption cross sections. The system is assumed to be
$d = 10$~kpc away.

\begin{figure}
\epsscale{.8}
\plotone{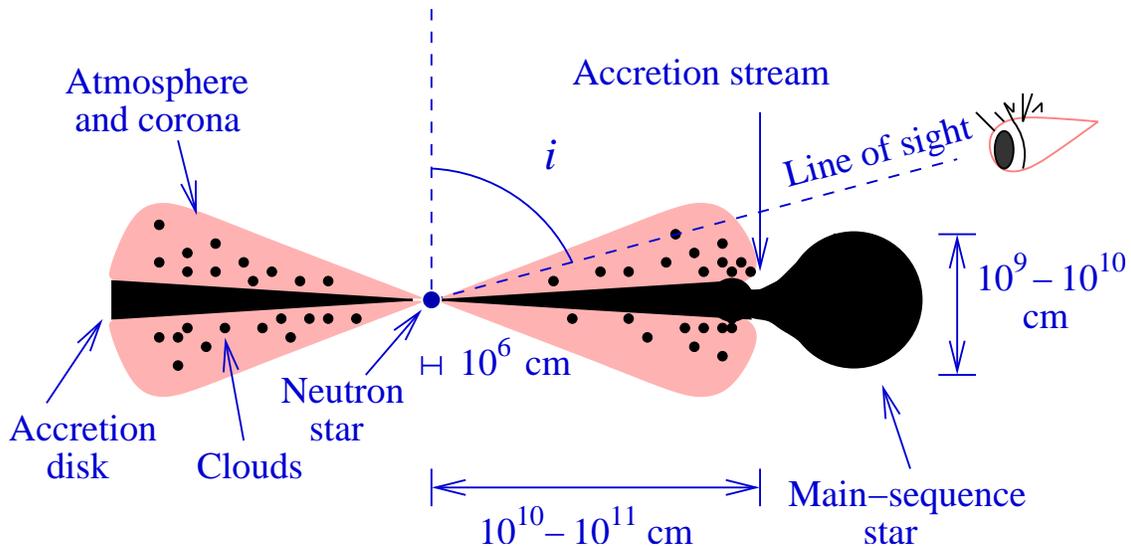}
\caption{\label{fig:schem_lmxb}
			Schematic of a low-mass X-ray binary (LMXB) system
			with a neutron star primary. Cross section view.
			The inclination angle $i$ is defined. }
\end{figure}

\begin{figure}
\plotone{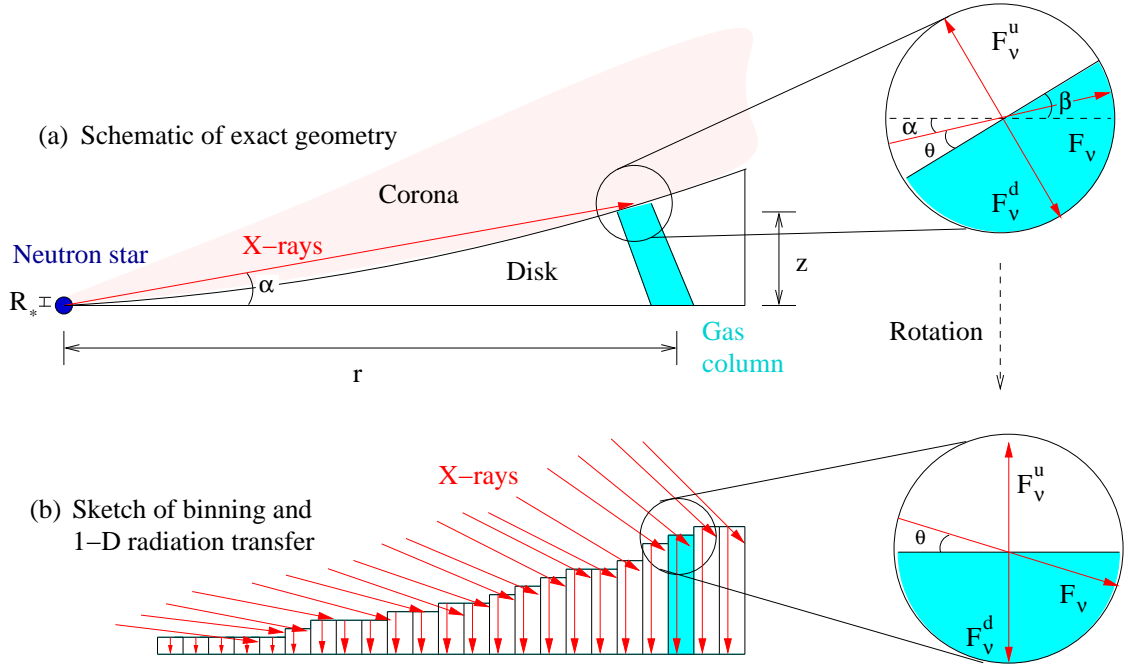}
\caption{ \label{fig:disk_model}
			Side view schematic of an illuminated accretion disk and the model
			geometry, assuming an extended corona above the disk.
			To compute the ionization structure of the disk,
			the disk geometry (a), can be approximated by
			a series of gas columns which are illuminated from the top
			(b).  }
\end{figure}

We find that the lines are swamped by the continuum 
for cases where the 
neutral column densities for the neutron star
and the disk are set equal, or $N_{H}^{*} = N_{H}^{\rm disk}$ 
in eq. [\ref{eq:spec}] (see the spectrum in Fig. \ref{fig:lineunobs}[a]).
This situation is most likely to occur in LMXBs with 
inclination in the range $i=0\degree$--$60\degree$ \cite[]{frank1987}.
The inclination angle $i$ is defined in Figure \ref{fig:schem_lmxb}.
Moreover, the continuum X-ray emission from
the inner disk ($r < 10^{8.5}$~cm)
has been neglected here, which 
according to a model by \citet{stella}, will soften the continuum below
$\sim 10$ \AA, and this will further reduce the equivalent widths of
the X-ray emission lines from the outer disk. In the inner disk,
radiation pressure dominates and the SS73 viscosity prescription 
must be modified \citep{stella}.
Thus, low inclination neutron star LMXBs are unlikely to have detectable
X-ray lines from the disk. 

Thus, consider 
$N_{H}^{*} = 5 \times 10^{22}$~cm$^{-2}$ and $N_{H}^{\rm disk} = 10^{21}$
cm$^{-2}$ with $i=75^{\circ}$, where an
obscuring medium absorbs half of the continuum flux from the neutron star.
Such a medium is compact enough to leave the disk almost unobscured (see 
Fig. \ref{fig:lineunobs}[b]). 
This situation should arise in LMXBs which exhibit flux dips,
for example, where either the disk rim or
small clouds obscure the central continuum
periodically, as explained in section \ref{sub:line-lmxb}. 
These LMXBs have inclinations in the $60\degree$ to
$80\degree$ range \citep{frank1987}. 
With a partially obscured central continuum, disk evaporation has an
observable spectral signature. We simulated 50~ks observations with the \it XMM-Newton \rm
RGS~1 and the \it Chandra \rm MEG (Fig. \ref{fig:compevap} and \ref{fig:fig3}).
Some bright lines are listed on Table \ref{tbl-1}.
The evaporating and condensing disks
have contrasting \ion{O}{7}/\ion{O}{8} and \ion{Ne}{9}/\ion{Ne}{10} 
line ratios. The evaporating
disk contains gas at $T \sim (7$--$10) \times 10^4$~K, unlike the condensing
disk. The H-like ion line intensities are higher for the condensing disk
since it has more gas at $T \sim 10^6$~K.
The spectral differences stem from the distinct differential emission measure
distributions $d(EM)/d\log \Xi$ and from the \ion{O}{7} recombination
rate $\alpha_{RR} \propto T^{-\gamma}$, where
$\gamma = 0.7$--$0.8$, and $EM=\int n_{z,i} n_{e} dV$ is the emission measure 
(appendix \ref{sec:radrec}).

In the case the central continuum is completely occulted, 
the model predicts that numerous hard X-ray lines will become detectable
with {\it Chandra} (Fig. \ref{fig:lineunobs}[c]),
such that evaporating and condensing
disk models are distinguishable.
Figure \ref{fig:megevap} contrasts the {\it Chandra} MEG~+1 simulations of
the evaporating and condensing disk models.
The column density for the neutron star continuum is taken as 
$N_{H}^{*} = 10^{24}$~cm$^{-2}$, and $N_{H}^{\rm disk} = 10^{21}$~cm$^{-2}$
for the disk.
Notably, the He-like to H-like ion line ratios still serve to
differentiate the models at larger $Z$, but with the
reverse effect. The He-like/H-like ion line ratios 
for Ar, S, and Si are larger for the hotter, condensing disk.
The \ion{Mg}{11}/\ion{Mg}{12} ratio is roughly the same for either model.
The $ Q \equiv $(\ion{Fe}{25} + \ion{Fe}{26})/\ion{Si}{14} line ratio
is 50 \% larger for the evaporating disk model. Since
\ion{Fe}{25} and \ion{Fe}{26} lines originate near or at
the hot corona, where the instability in question does not operate,
no difference in their line fluxes is observed.
Since the hot atmosphere (or ``warm corona'') of the condensing disk
is larger than the evaporating case, more \ion{Si}{14} line emission
is produced, and the $Q$ ratio is smaller.
This shows that the way the thermal instability is treated in the
models (in this case, whether we pick the evaporation or condensation
solutions), has a dramatic effect on all the line ratios which are
sensitive to the ionization distribution.

The He$\alpha$ line triplets can be used as
density diagnostics, but they may be affected by photoexcitation
by the UV field from the accretion disk \citep[]{gabriel1969,blum}.
The forbidden line $f$ of \ion{O}{7} at $22.097$  
\AA \ is suppressed due to
collisional depopulation at high density, since 
$n_e \gtrsim 10^{14}$~cm$^{-3}$
(see the density profiles in Fig. \ref{fig:atm_density}
and the O$^{+7}$ relative abundance 
distribution in Fig. \ref{fig:lowzdist}).
The \ion{O}{7} intercombination $i$ to resonance $r$ line  
ratio is $g=(f+i)/r \simeq i/r \gtrsim 4$, 
indicating a purely photoionized plasma \citep[]{porquet}.
The \ion{O}{7} and \ion{N}{6} He$\alpha$ line ratios are included
in the model at their high-density limit, while the He$\alpha$ line ratios of other ions
such as \ion{Si}{13} have not been modeled yet, since the line ratios
will start to be a function of position in the atmosphere, adding
complexity. The depopulation of the forbidden line
in many He-like ions was attributed to resonant photoexcitation
in Hercules X-1 \citep{herx1}. The intense UV fields in LMXBs imply
the same effect will operate \citep[]{lied1992}. In the context
of the present model, the
region where He-like ions are abundant is very close 
(less than $10^7$~cm away) to the
photospheric surface, such that the UV energy density is 
as high as in the photospheric surface. This implies that
the He-like diagnostics will be degenerate to high density
and UV field effects \citep[]{mauche}.

The optical depth of the atmosphere may be probed by comparing
the observed spectra to the model, which assumes the lines
are optically thin. In particular, the \ion{O}{7} $r$ line
may differ from the modeled value due to resonant scattering
of continuum photons. Whether $r$ gets enhanced or absorbed depends on
geometry and the relative placement of emissivity and opacity. The $i$
and $f$ lines should be optically thin, so optical depth can modify
the $g$ ratio substantially from $g \sim 4.2$, the value expected for a
photoionization-dominated, optically thin gas. The Ly$\alpha$ line in
hydrogenic ions also has a large scattering cross section.
Thus, the He$\alpha$ $r$ and/or Ly$\alpha$ lines are good indicators
of optical depth.

The RRC can be used for temperature diagnostics, and also for
probing the behavior of the thermal instability.
The local RRC width $\propto T$ \citep[]{lied}.
The \ion{O}{7} RRC broadening is Doppler-dominated,
resembling the RR lines, for both the evaporating and
condensing cases. 
The \ion{O}{8} RRC shape varies noticeably from
the evaporating to the condensing condition.
In the evaporating case, the \ion{O}{8} RRC
has two temperature components, one with a $FWHM \sim 2$
\AA \ that is produced at $T \sim 10^6$~K, and another
narrow component with $T \sim 10^5$~K. 
The two components are distinguishable because intermediate
temperatures are thermally unstable. Thus, the RRC profile
in this case provides evidence for the existence of
this thermal instability.
In the condensing case, the RRC has a single, broad temperature
component at $T \sim 10^6$~K, since the disk atmosphere
temperature suddenly drops from the latter value to
one where the X-ray emission is negligible.
The broadening of this RRC is also peculiar, since RRC are
usually narrow for all photoionized gases, given that
the kinetic energy of the gas particles is generally much smaller
than the recombining photon energy \citep{lied}.

\subsection{Luminosity Dependence}

The disk model was run with a lower central luminosity $L = 0.1 L_{\rm Edd}$,
for comparison to the $L_{\rm Edd}$ case explored in the previous sections, to
investigate the structural and spectral changes of the disk.

The atmospheric radiative recombination
luminosity was $\sim 20$ times lower than the
$L = L_{Edd}$ case, indicating a nearly
linear dependence of the disk luminosity to the central luminosity.
Otherwise, the low-luminosity spectrum shown in Figure \ref{fig:lowlumspec}
shows much resemblance to its $L_{\rm Edd}$ counterpart.
The dependence of the recombination luminosity can be
approximated by
\begin{equation}
\label{eq:lineprop}
L_{u \rightarrow l} \propto L_{\rm x} \frac{\Omega}{4 \pi}
\end{equation}
where $\Omega$ is the solid angle subtended by the disk atmosphere.
This relationship is then modified by the changing density, opacity
and thickness of the atmosphere.
To explain the recombination luminosity behavior,
we will first describe the calculated atmosphere structure.

The photospheric and atmospheric boundaries were fitted by power laws.
The fit parameters defined in section \ref{sec:diskstruct} were
$C_{\rm phot} = (1.8 \pm 0.3) \times 10^{-3}$ 
cm, $n_{\rm phot} = 1.14 \pm 0.01$,
$C_{\rm atm} = (1.2 \pm 0.2) \times 10^{-3}$~cm, and
$n_{\rm atm} = 1.18 \pm 0.01$. 
The fit errors are shown, while systematics are expected to follow the
same trends as in section \ref{sec:diskstruct}.
The solid angle subtended by the entire disk $\Omega/4 \pi \simeq 0.1$,
while for the Eddington Luminosity case $\Omega/4 \pi \simeq 0.2$, as
can be seen by comparing Figure \ref{fig:omega} with Figure \ref{fig:omega_low}.
This implies little variation of the disk
shape with luminosity.  With a factor of ten reduction
in luminosity, the radiative energy incident on the disk is 
20 times smaller, which coincides with 
the observed reduction in the recombination emission.

The accretion disk structure for $0.1 L_{\rm Edd}$, shown in
Figures \ref{fig:disk_temp}, \ref{fig:disk_density},
\ref{fig:atm_temp}, and \ref{fig:atm_density}, yields a density
$\sim 4.5$ times
smaller than the $L_{\rm Edd}$ case. 
Aside from the density change, the ionization structure
remains quite similar to the $L_{\rm Edd}$ case (see Fig. \ref{fig:irondist}[a]
and \ref{fig:irondist}[b]).
Naively, a factor of 10 decrease in the density would be expected to
keep $\xi$ constant.  However, since the atmospheric volume shows little change,
the observed density change implies a factor of $\sim 20$
decrease in the ion emission measure, verifying the
consistency of the density with the modeled recombination flux.

We reconcile $\xi$ with the larger-than-expected density of the atmosphere 
by accounting for a decrease in the atmospheric opacity. Considering $\xi$ to
be constant, such that
\begin{equation}
\xi \propto \frac{e^{- \tau(n)} L_{\rm x}}{ n } = 
            \frac{e^{- \tau(n^{\prime})} L_{\rm x}^{\prime}}{ n^{\prime} }  
\end{equation}
where $\tau(n) \propto n$,
implies that a decrease in density decreases the opacity, 
which increases the local flux in the atmosphere.
Thus, the overdensity of the atmosphere is explained by
a factor of $\sim 2$ decrease in $e^{-\tau}$, which was verified in the models. 
For the $L = 0.1 L_{\rm Edd}$ case, the atmosphere transmits 34 \%
of the incident flux directly for a disk annulus with $r=10^{11}$~cm, i.e.
$\int F_{\nu}(z_{\rm phot}) d\nu / \int F_{\nu}(z_{\rm cor}) d\nu \sim 0.34$, 
compared to 12 \% in the $L = L_{\rm Edd}$ case (see section \ref{sec:diskstruct}).
The atmospheric opacity is a function of radius, but the observed spectra
are weighted towards the largest radii.

The argument above holds for 
plane parallel atmospheres that are photoionized 
by radiation incident at a small grazing angle $\theta \ll 1$, such
that $\tau \sim 1$ for a grazing ray, but $\tau \ll 1$
for any other ray. Therefore, the recombination
luminosity will be proportional to the total emission measure for
most viewing angles.

As more of the X-ray
continuum is transmitted onto the disk photosphere, the heating
of the optically thick disk increases.
This effect is taken into account to derive a
self-consistent midplane disk temperature, to 10 \%.
The input grazing angle and the grazing angle calculated from
$z_{\rm atm}$ are consistent at the 25 \% level.

\subsection{A weak coupling between the disk and its atmosphere}

There is a negligible change in the atmospheric
structure by varying $\alpha$ from 0.1 to 1 in the disk model. 
As explained below, this can be understood as a decoupling which exists between
the optically thick disk and the photoionized atmosphere.

The viscosity parameter $\alpha$ 
has no effect on the disk temperature, but
it does on the density, since $\rho \propto \alpha^{-1}$ (SS73).
Assuming vertical isothermality in
the optically thick disk (below the photosphere),
its pressure can be obtained from equation (\ref{eq:hydro}) and is given by:
\begin{equation}
P(z) = P_{\rm o} e^{- z^2 / 2 Z_{\rm P}^2 }
\end{equation}
with $P_{\rm o} \propto \alpha^{-1}$.
The photoionized atmosphere is placed on top of this disk,
and satisfies two boundary conditions: matching
temperature and pressure at the photospheric boundary.
Assume the base of the atmosphere is at some
pressure $P_{\rm phot}$, and let a change in viscosity
from $\alpha$ to $\alpha^{\prime}$
produce a change in the subphotospheric disk 
pressure from $P_{\rm o}$ to $P_{\rm o}^{\prime}$.
To match the boundary conditions, the
atmosphere has to shift in height, such that
\begin{equation}
P_{\rm phot} = P_{\rm o} e^{- z^2 / 2 Z_{\rm P}^2 }
 = P_{\rm o}^{\prime} e^{- (z^{\prime})^2 / 2 Z_{\rm P}^2 }
\end{equation}
which implies that the height shift in the
atmosphere, $\Delta z = z^{\prime} - z$, is given by:
\begin{equation}
\frac{ \Delta z}{z} \simeq \frac{Z_{\rm P}^2}{z^2} \ln \Biggr( \frac{P_{\rm o}^{\prime}}{P_{\rm o}} \Biggl)
= \frac{Z_{\rm P}^2}{z^2} \ln \Biggr( \frac{\alpha}{\alpha^{\prime}} \Biggl)
\end{equation}
where the approximation $z^2 \gg Z_{\rm P}^2$ was used, which is valid
since $z_{\rm phot} > 3 Z_{\rm P}$ in the models for $\alpha < 1$. 
In such a case, it follows that $\Delta z / z \ll 1$, so that
very large changes in viscosity only produce minute shifts on the
atmospheric height and negligible effects on the atmospheric emission
(as already pointed out by \citet{nay2000}).

There are, however, other situations where the role of dissipation
in the atmosphere has to be reassessed: 
1) if the dissipated energy in the disk is no
longer negligible compared to the exterior illumination energy, as
might be the case for disks around black hole candidates, and if energy
is transported to the atmosphere via magnetic flares, for example, and
2) if there is negligible dissipation, but it is sufficient to enhance
mixing and, therefore, change the atmospheric structure. 

\citet{roza2002} noted that the structure of the optically-thick
disk has non-negligible effects in the corona when the disk scale
height is comparable to or larger than the coronal scale height (in our
notation, when $Z_{\rm P} \gtrsim z_{\rm cor}$).  This does not apply to
our case, because for a neutron star LMXB disk with $r > 10^{8.5}$~cm, we get
$z_{\rm cor} \gg Z_{\rm P}$.
\citet{roza2002} assert that
the disk and corona are coupled in BHC for $r \sim 10~R_{\rm G}$, where 
$R_{\rm G}$ is the Schwarzschild radius; while for AGN, the situation 
depends on the accretion rate. By this measure, the disk-corona
coupling may also be significant in
the outer radii of BHC disks, if the reduction of the
central illumination flux shrinks the size of the corona.

\subsection{Emission line profiles}
Synthetic profiles were produced for all the emission lines.
For simplicity, the calculation assumes that line scattering
is negligible, which may not be a valid assumption for resonance lines.
Here we select the profiles of the brightest lines 
that are not contaminated by other ions.
The line profiles are also calculated for a disk within the
$10^{8.5} < r < 10^{10}$~cm radius range, for comparison with
the $10^{8.5} < r < 10^{11}$~cm radius range shown above.
The emission line profiles have more broadly separated
peaks for smaller disks, as shown in Figure \ref{fig:lineprof}.

No variation in line broadening is obtained as a function of
charge state.
The \ion{N}{6} Ly$\alpha$ line has the same velocity profile
as the corresponding lines in \ion{O}{8},
\ion{Ne}{10}, \ion{Si}{14}, and \ion{Fe}{26}. We attribute this
to the vertical stratification of the atmosphere, which allows
the full range of ionization parameters 
detectable in the X-ray band
to exist in every annulus.
Future two-dimensional models taking into account
radiation transfer in the radial direction may exhibit
a trend for the line widths, since
the radial optical depth is not negligible.

\section{Discussion}
\label{sec:disc}

\subsection{Neutron star LMXB spectra observed with {\it Chandra} and {\it
XMM-Newton}}
\label{sub:gspectra}

The LMXB spectra observed with the {\it Chandra} 
High Energy Transmission Grating (HETG) and the {\it
XMM-Newton} RGS already provide stringent tests for the models in
this article.  The high resolution spectra of LMXBs are generally
dominated by continuum
emission, which is sometimes punctuated by emission or absorption lines. Only a fraction
of the sample of observed neutron star LMXBs show prominent lines.
The accreting pulsar 4U1626-67 shows double-peaked and broad
emission lines \cite[]{schulz2001}. The eclipsing dipper EXO0748-67
has broad emission lines \cite[]{exo0748}.
The accretion disk corona source 4U1822-37 has narrow emission lines
\cite[]{4u1822}. The dipper source 4U1624-49 
shows narrow absorption lines \cite[]{parmar2002}. Her X-1, an intermediate-mass
X-ray binary with a precessing accretion disk, has narrow emission lines
during its low- and intermediate-flux states \cite[]{herx1}.
The line emission spectra are dominated by H-like and He-like ions.
P~Cygni profiles are observed in 
Circinus X-1 \cite[]{cirx1}, a unique LMXB with
high-velocity outflows.
Most LMXBs do not exhibit 
discrete spectral features aside from interstellar absorption,
such as X0614+09 \cite[]{paerels}.
The emerging pattern 
implies that at least three classes of 
neutron star LMXBs exist which produce detectable X-ray emission lines: 1)
high inclination LMXBs with $i \gtrsim 70^\circ$, 2)
accreting X-ray pulsars, and 3) LMXBs with high velocity
winds. 

Our models are proving to be of great relevance to the
interpretation of the spectra of LMXBs with
high inclination and LMXBs with an X-ray pulsar.
Most importantly, all LMXB spectra validate the basic assumption in our model:
the plasma is photoionized. 
Evidence for other heating mechanisms, such as shocks, is
not observed in LMXB spectra.
Shocks are predicted by the \cite{miller2000} MHD disk models, but it is not clear
whether X-ray emission from such shocks would be observable.
The observed signatures of a plasma heated primarily by
photoionization are the RRC, the
peculiar He$\alpha$ line ratios, and the weakness of Fe L line emission
relative to that of low-Z and mid-Z elements \cite[]{lied99}.
The spectra have prominent line emission from H-like and He-like ions.
These properties are shared by all the spectra shown in this article.
Furthermore, the line velocity broadening observed in two LMXBs
(4U1626-67 and EXO0748-67) provides kinematic evidence for
accretion disk atmospheric and coronal emission. 

By contrast, our model is inadequate for
the interpretation of the spectra of LMXBs with high-velocity winds.
These spectra show a photoionized gas with significant optical depth.
Circinus X-1 is rare because of the extreme gas dynamics and the sizable
line optical depths which are evident in its spectrum.
The density $n_e \sim 10^{13}$~cm$^{-3}$ deduced from
the observed P Cygni profiles in Circinus X-1 \cite[]{cirx1} 
is consistent with our model calculations, 
but the wind dynamics rules out our assumption of hydrostatic equilibrium.
The modeling of an LMXB disk wind spectrum
requires a wind acceleration mechanism and a revised disk
structure \cite[]{proga}.

Initially, our model can be used as a tool for the identification
of the discrete X-ray spectral signatures from the
accretion disk atmosphere and corona.
Our model provides a quantitative expectation of the
X-ray line fluxes produced by the entire disk. Our
physical calculation of the line profiles (and therefore
the line emissivity as a function of $r$), can be used to measure the
maximum disk radius and the radial ionization distribution. 
Furthermore, we have modeled 
the density at which each
ion is produced, and this is testable 
with plasma diagnostics. The ionization distribution
(measured by the flux of the emission from the high-ionization species
relative to that of the low-ionization species)
is a fingerprint of the disk atmosphere and corona.
The temperature structure of the disk atmosphere can also be probed
with the temperatures measured with the RRC of various ions.
A full investigation of the models 
as compared with spectral data will be performed in a future paper.

\subsection{Limitations of the model}
\label{sub:limit}
The disk structure can be improved by relaxing the 
assumptions made in the radiation transfer calculations.
The spectral model calls for the inclusion
of additional lines. The observed spectra
will also allow us to investigate
additional physics in the disk which may be
missing in our current model. 

A 2-D or 3-D transfer calculation is needed to improve
the coronal structure model. 
To simplify the radiation transfer calculations, we split the disk
into a set of nested cylindrical shells, and we use 1-D transfer
to calculate the structure of each shell.
By defining $z_{\rm atm}(r)$ and iterating on $\theta(r)$ in each shell 
until self-consistency is obtained
(see eq. [\ref{eq:diskangle}] and Fig. \ref{fig:conv}), we
produce a pseudo-2-D transfer calculation.
However, the 1-D transfer approximation starts to break down at the
largest radii, since $z_{\rm cor} \sim r$ at $r \gtrsim 10^{10}$~cm
(see Fig. \ref{fig:disk_temp}). 

The UV emission is included in the structure
model but remains to be added to the high resolution spectral model.
However, we do not expect substantial differences between the UV spectrum obtained
with our disk model and the results by \citet{ray93}.
The optical depth of UV lines such as \ion{C}{4} 
is $\sim$100 in the latter model.
The UV and optical lines originate just above the
photosphere ($z_{\rm phot}$), at densities of $10^{13} < n_e <
10^{14}$~cm$^{-3}$. 
The structural difference between
our models and those by \citet{ray93} occurs at the
X-ray emitting atmosphere and corona.
In contrast, the $z_{\rm phot}(r)$
in our models agrees well with those by \citet{ray93}.

Fluorescence and resonant scattering
will need to be added in the spectral model.
The fluorescence line flux can be of the same order as
the recombination line flux.
The $r$-dependence of the Fe K fluorescence flux
should be distinct from the $r$-dependence of the recombination emission.
A 6.4~keV Fe K emission line is produced by M-shell charge states of
Fe localized at the base of the atmosphere (see Fig. \ref{fig:irondist}). 
The Fe K fluorescence flux will scale with the hard X-ray 
transmittance of the atmosphere and corona.

The structure model indicates that 
the optical depth of resonance lines is large. However, a realistic
treatment of resonant scattering in our spectral models is complicated by
the velocity shear within the disk.
Resonant scattering and Fe K fluorescence emission were only included in
the low-resolution 1-D transfer calculations to obtain the disk structure.
The propagation of resonance line photons is highly anisotropic, and it depends
on the viewing angle because of the Keplerian velocity
shear and the geometrical thickness of the atmosphere.
LMXBs with strong emission lines do not
exhibit detectable features from resonant scattering of continuum photons, with
the exception of Circinus X-1 (section \ref{sub:gspectra}).

Magnetic fields, which are not included
in our model, may affect
the structure of some regions of the disk
(see section \ref{sub:bfield}). 

\subsection{A strongly magnetized corona?}
\label{sub:bfield}

A strong magnetic field may affect the
coronal structure and the X-ray spectrum. 
The corona under consideration
is located at the outer radii of a centrally illuminated disk. 
We believe $B$ fields play a secondary role
in this type of corona, because the energy budget of the corona
is dominated by X-ray irradiation.
In detail, the role of $B$ fields
cannot be discounted, because phenomena such as magnetic flare
heating may dominate over photoionization
within localized regions of the disk.
Recent MHD models predict $B$ fields larger than the
virial value in the disk corona. However, the
applicability of these MHD models
to the illuminated corona is dependent on the effects of
radiative heating and magnetic reconnection.

MHD models of {\it radiationless} accretion disks show that above a few
scale heights, the magnetic pressure is larger than $P_{\rm gas}$
\cite[]{miller2000}. 
The gas dynamics in the disc
is dominated by the $MRI$. 
\cite{miller2000}
found that the $MRI$ produces $B$
fields which buoyantly rise to the atmosphere and corona.
In their model, 25\% of the magnetic energy generated by the
$MRI$ rises to the corona, representing 60\% of the local
heating, but $\lesssim 4$\% of the dissipative heating in the
disk. 
A 3D MHD disk model by \cite{machida}, with an
initial toroidal configuration, shows that the strong $B$
fields in the corona are confined in filaments, with a filling
factor of a few percent. Another 3D MHD model by
\cite{3disk} confirms the presence of large $B$ fields in the corona.

By contrast, when X-ray illumination is present, the $B$ field
plays a relatively minor role in the overall energetics of the
disk corona.
This is true at least in a spatially-averaged sense.
The magnetic energy produced by the $MRI$ can be no larger than
the energy dissipated in the disk.  
The maximum energy available
for the $B$ field scales as $r^{-3}$, and it is given by
the first term on the right side of equation (\ref{eq:locdisk}).  Assuming
all of the accretion energy is contained in the $B$ field, equation
(\ref{eq:disk}) indicates that the illumination energy is larger than
the magnetic energy for $r \gtrsim 10^{10}$~cm (per unit disk area).
Since MHD disk models typically assume an isothermal or adiabatic disk, 
they may not apply to an extended hot corona dominated by photoionization.

Magnetic reconnection has the effect
of decreasing the magnitude of the disk-coronal $B$ field.
Coronal flares resulting from magnetic
reconnection convert magnetic energy into kinetic energy
through particle acceleration. Most of the X-rays in black
hole accretion disks may be produced by magnetic flares
\cite[]{matteo}. In the \citet{liu} model, 
reconnection events in disk flares reduce the $B$ field
by a factor of $\sim 30$. 
Thus, the buoyant $B$ field may be dissipated in a flaring region.

\section{Conclusions}
\label{sec:conclusions}

We have calculated the hydrostatic structure
of a photoionized accretion disk atmosphere which is
in thermal equilibrium and ionization balance.
We also determined the atmosphere's thermal stability and 
its observable high resolution X-ray recombination emission 
spectrum.
\begin{itemize}
\item {\it A feedback mechanism between illumination and atmospheric
structure enlarges the atmosphere}.
The disk atmosphere is orders of magnitude less dense than the
disk midplane. The atmosphere extends for a few tens of disk pressure scale
heights (if the pressure scale height is calculated using the disk
midplane temperature). Illumination heats and expands the disk
atmosphere, increasing the number of absorbed photons in the
atmosphere and heating it further, producing further expansion of the
atmosphere, and so on. The expansion stops because the atmosphere
becomes optically thin, cooling and contracting. The inclusion of the
feedback mechanism increases the size and the line emission flux of the
atmosphere by an order of magnitude. The atmospheric
thickness is much larger than the standard $\alpha$-disk model
thickness, and
it is consistent with the $\sim 12 \degree$ subtended semi-angle
deduced from optical modulations in LMXBs. The disk atmosphere
thickness also explains the under-abundance of eclipsing LMXBs.

\item {\it The disk atmosphere subtends a large solid angle 
$0.07 \lesssim \Omega/4\pi \lesssim 0.2$}. If the inclination is 
$i \gtrsim 80 \degree$, the disk photosphere (which
subtends $0.04 \lesssim \Omega/4\pi \lesssim 0.08$)
may shield the neutron star flux, producing
an ADC source with partial eclipses or without eclipses altogether.
The disk $\Omega$ depends weakly
on the neutron star luminosity, but $\Omega$ scales linearly with disk radius.
The disk recombination luminosity scales linearly with $\Omega$.

\item {\it The atmospheric structure is independent of the viscosity
parameter $\alpha$}. The viscosity changes the density in the
optically-thick part of the disk, producing a small shift
in atmospheric height, but this has no effect on the X-ray spectrum.

\item {\it The X-ray spectra are dominated by lines from H-like and He-like
ions} of abundant elements from C to Fe, as well as RRC and weak
Fe L lines. The line ratios are a sensitive probe of the atmospheric and
coronal structure.

\item {\it Clear spectral signatures of photoionization are present, as well as
temperature, density, and radiation field diagnostics.}
An intercombination to resonance line ratio of $\sim 4$ is modeled
for low-Z He-like ion line triplets. RRC are unequivocal signposts 
of photoionization.
The density diagnostics
from He$\alpha$ lines of low-Z and intermediate-Z elements are degenerate with
the presence of an intense UV radiation field from the disk itself,
so the $R$ ratio may not give conclusive signatures of high-density in
LMXBs.
Much of the disk atmosphere is close to the photosphere, such that
the dilution factor of the UV field is small.
The He$\alpha$ density diagnostics could operate at the densities predicted by the
disk atmosphere model in LMXBs, of $10^{13} \lesssim n_e \lesssim 10^{15}$
cm$^{-3}$.

\item {\it The line fluxes are nearly proportional to the X-ray continuum luminosity}.
The disk line fluxes decreased by a factor of 20 
when the system luminosity was decreased by a factor of 10
 (to $L=10^{37.3}$~erg~s$^{-1}$).
The atmospheric density
was reduced by a factor of $\sim 5$, its optical depth 
was reduced, and the atmosphere was
$\sim 2$ times less extended than in the high-luminosity ($L=10^{38.3}$~erg~s$^{-1}$) case.

\item {\it The line equivalent widths depend strongly on 
inclination}.  The relative obscuration of the neutron star
affects the equivalent width and detectability of the disk 
X-ray emission. The modeled disk emission is almost undetectable when
the neutron star continuum is also in the line of sight. As such, 
high inclination systems, or systems with dips or ADC,
are more likely to show X-ray lines due to enhanced contrast.
This expected trend has been largely confirmed
by {\it Chandra} and {\it XMM-Newton} observations.
We have demonstrated that
for a highly absorbed neutron star continuum in our fiducial
system, the disk X-ray lines are detectable with both the {\it Chandra}
and {\it XMM-Newton} grating spectrometers.

\item {\it Double-peaked X-ray lines can be detected for $r=10^{10}$~cm disks,
but larger $r=10^{11}$~cm disks may appear blended in a single peak 
in the grating spectra.} The emission line region spans several orders
of magnitude in disk radius.  The modeled line profiles are needed to
deduce the outer disk radius.  Line emission from the outer regions of
the disk dominates.  The emission increases with disk radius,
and the Doppler broadening of the lines decreases for larger $r$. The
wings of the broadest lines are lost in the continuum, decreasing
their apparent equivalent width. 

\item {\it The resonance line optical depths can be measured.} If the $r$ line in He-like
ions has a value which differs from the model calculations, it may
be due to resonant scattering of continuum photons. The line ratios in the Lyman series
can also work as optical depth diagnostics. We have not included these effects on the current
version of the model, but our results for ionic column densities show that appreciable line
optical depths are present, and hence this process will be included in future versions of the code.

\item {\it The continuum optical depth of the atmosphere is generally small ($\tau \ll 1$),
except for photons which propagate nearly parallel to the disk plane.}
The atmosphere is optically thick to X-ray continuum photons from the neutron
star. However, most of the recombination line emission is not appreciably affected by
the continuum opacity.

\item {\it The spectrum is sensitive to a thermal instability
present in photoionized gases.} By forcing all the chosen solutions
to be thermally stable, a break in the temperature, density, and
ionization structure is created. Measurably different X-ray spectra are
obtained depending on the resolution of this instability. The shape of 
RRC profiles, which in the models show multiple temperature components, 
and the relative intensity of lines such as \ion{Mg}{11},
are useful diagnostics of the stable temperature regime.

\end{itemize}

The spectra obtained with the {\it Chandra} HETG and the {\it XMM-Newton} RGS
show that the plasmas in LMXBs are photoionized, as our model assumes.
Two LMXBs (4U1626-67 and EXO0748-67) show kinematic signatures of 
accretion disk emission \cite[]{schulz2001,exo0748}.
The line fluxes, line profiles, the ionization distribution,
density, and RRC temperatures, provide a wealth of diagnostic capability
for the identification of accretion disk atmospheres and their properties.
The spectral comparisons with the data are promising, and they will be addressed in
future work.

%% Use the figure environment and \plotone or \plottwo to include 
%% figures and captions in your electronic submission.

\begin{figure}
\plotone{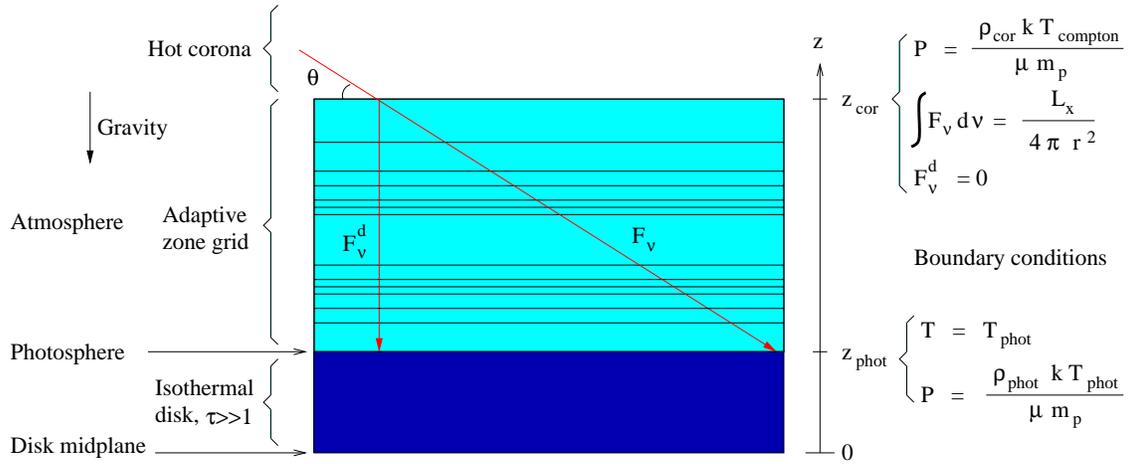}
\caption{ \label{fig:column}
			Gas column geometry for each disk annulus. The 
			boundary conditions on the atmospheric
			structure are shown.}
\end{figure}

\begin{figure}
\epsscale{.7}
\plotone{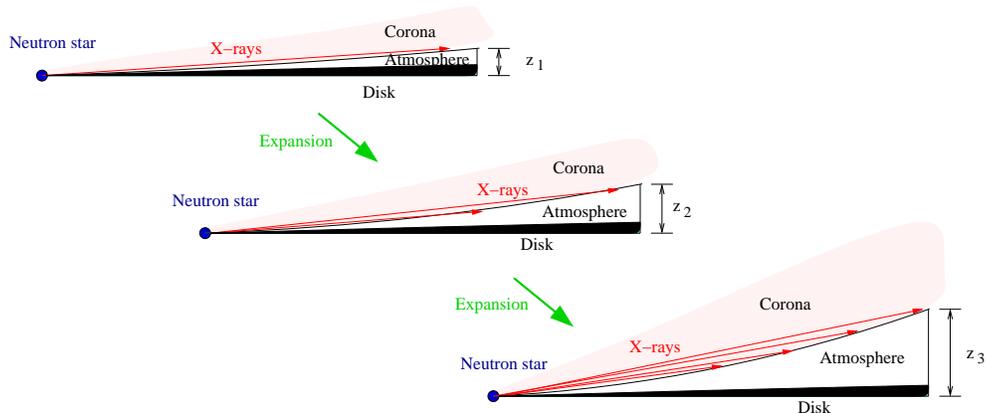}
\caption{ \label{fig:feed}
			Schematic of the feedback between radiative heating and
			disk geometry. The heated atmosphere expands and collects
			more radiation, reaching equilibrium 
			at $\sim 10$ times its initial volume.}
\end{figure}

\begin{figure}
\plotone{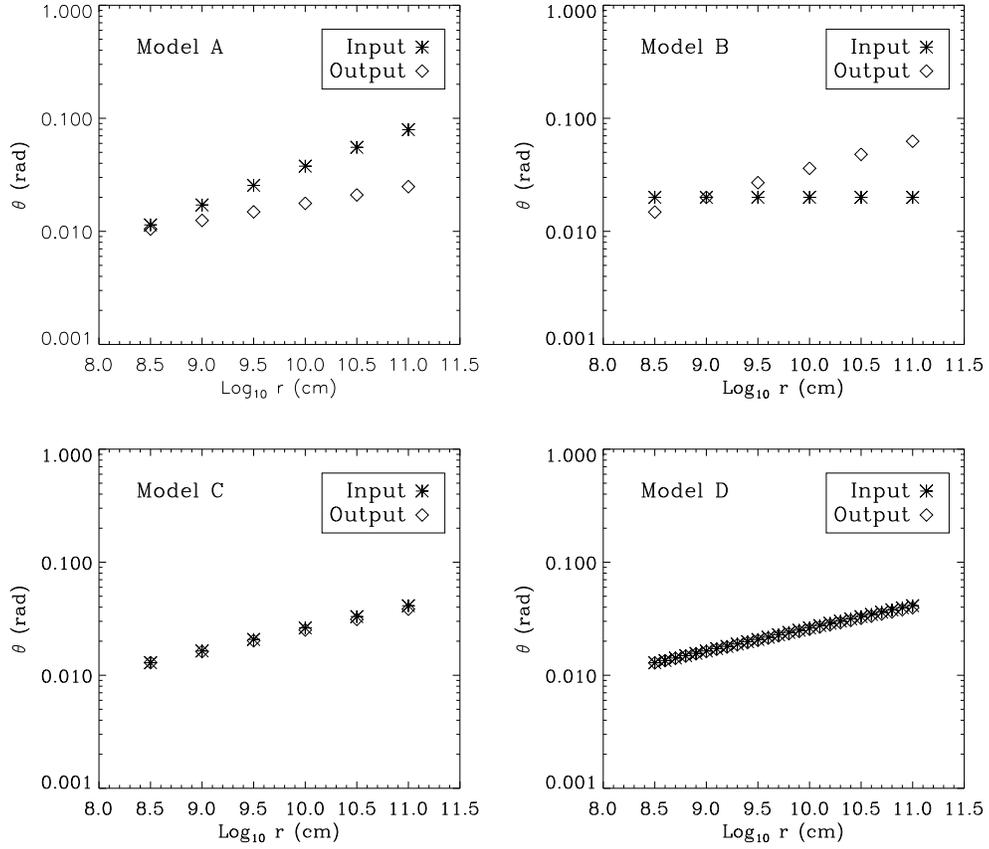}
\caption{ \label{fig:conv}
			Grazing angles $\theta(r)$ for the radiation impinging on the disk,
			at successive model iterations (A to D). The $\theta(r)$ 
			input to the model are compared to the $\theta(r)$
			extracted from the output disk structure.
			The resulting model~D is self-consistent and
			has a finer grid.}
\end{figure}

\begin{figure}
\plotone{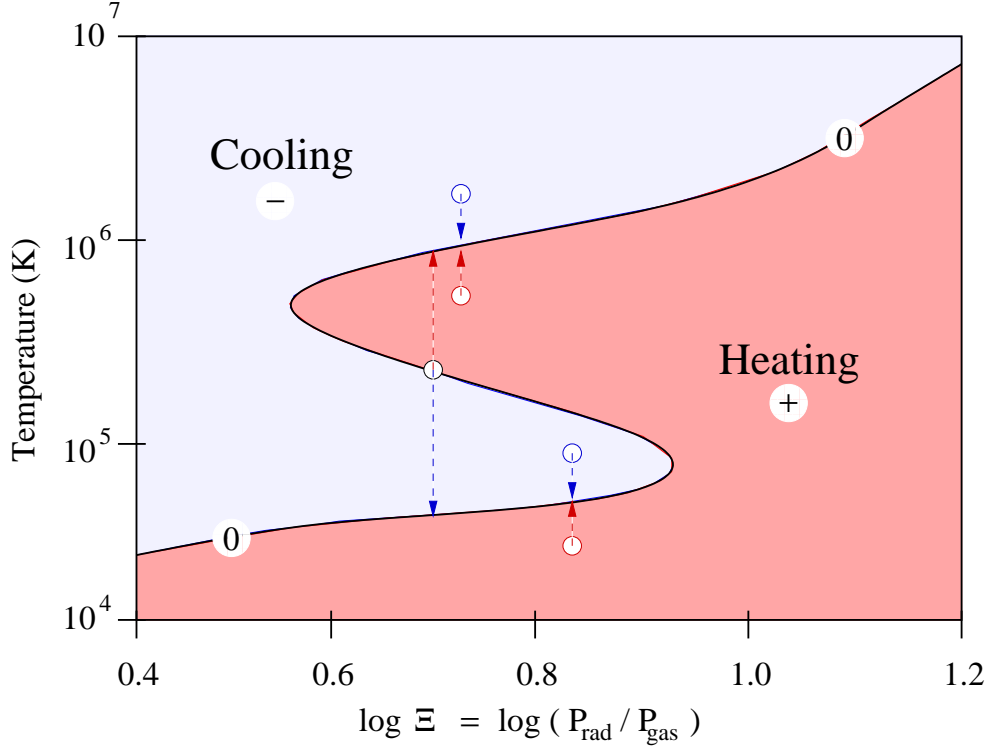}
\caption{ \label{fig:instab}
			Map of net heating in the gas,
			in the temperature $T$ vs. ionization parameter $\Xi$
			plane.
			The thermal equilibrium S-curve is labeled with ``0''. The dashed
			arrows depict the thermodynamics of the gas after
			an isobaric temperature perturbation, starting
			from points on the S-curve. 
			}
\end{figure}

\begin{figure}
\plotone{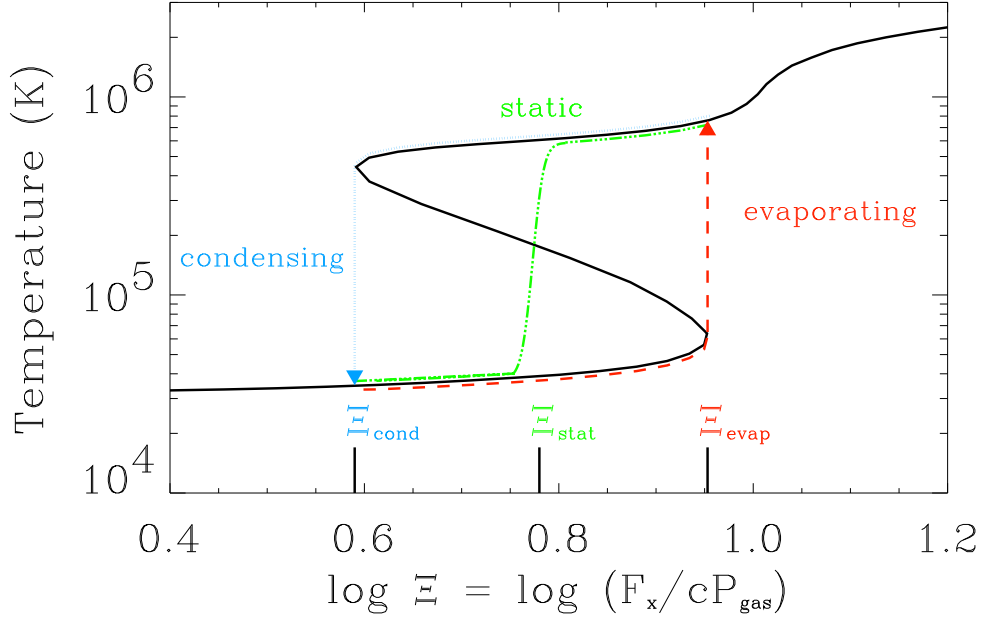}
\caption{ \label{fig:fig1}
Ionization parameter $\Xi$ vs. temperature assuming an 8~keV
bremsstrahlung continuum. The calculated S-curve corresponds to the locus of
solutions in thermal balance and ionization equilibrium. 
The branch of the S-curve with a negative slope is thermally unstable.
We over-plot three additional solutions which,
due to the addition of a conduction term in the energy
equation, avoid the unstable branch. The family of solutions 
which include conduction produce a small transition region.
We model the two extreme cases, the evaporating and
condensing disks, which include the lower and upper stable branches
of the curve, respectively. A third, static solution, 
produces a transition at an intermediate $\Xi$.
Schematics of the static transition region at
$\Xi_{\rm stat}$ and the dynamic transition regions at $\Xi_{\rm
cond}$ and $\Xi_{\rm evap}$ are shown.  For $\log \Xi > 1.2$, $T$
increases up to $T_{\rm compton} \sim 10^7$~K (not shown).}
\end{figure}

\begin{figure}
\plotone{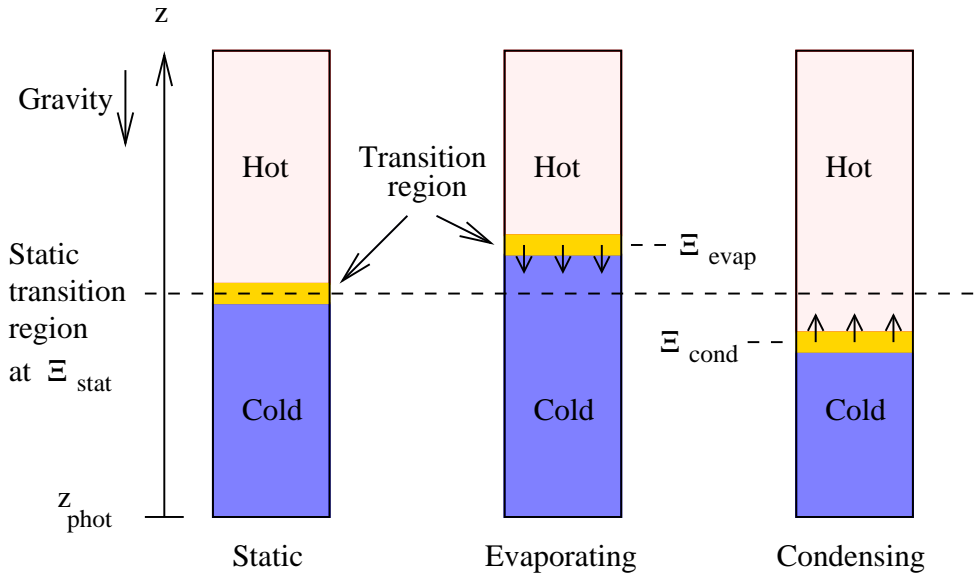}
\caption{ \label{fig:massflow}
     		Phase dynamics in a photoionized, stratified gas column.
			One static and two dynamic cases are shown.
			A conduction-driven mass flow moves the transition region
			towards $\Xi_{\rm stat}$ in the condensing and evaporating cases.
			Due to the boundary conditions at $z_{\rm phot}$,
			both the pressure and the ionization parameter $\Xi$ 
			are roughly constant at a given height $z$ among the three
			cases. 
			}
\end{figure}

\begin{figure}
\plotone{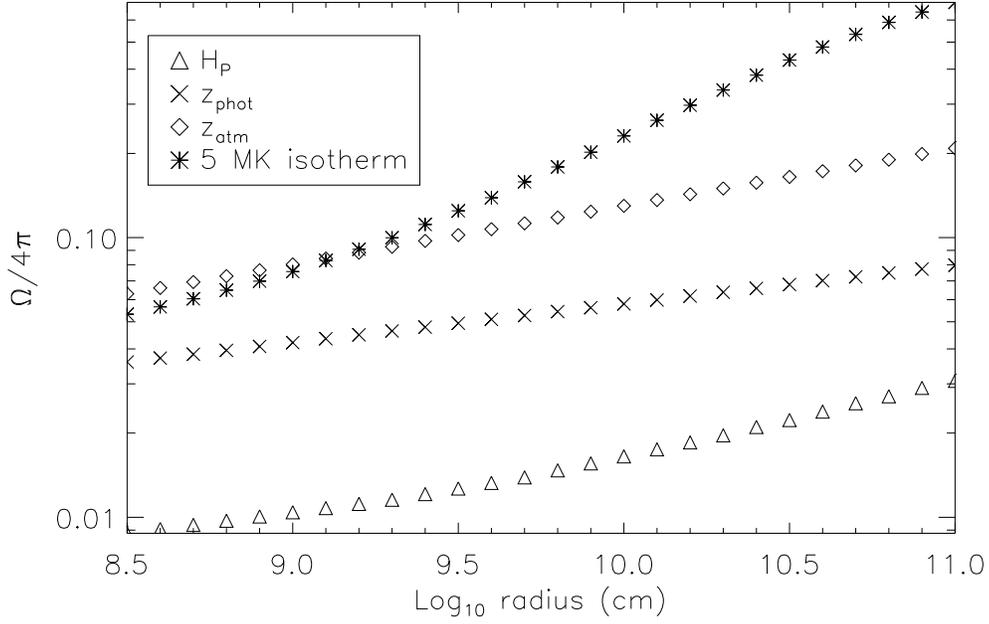}
\caption{ \label{fig:omega}
			Total solid angle subtended by the disk photosphere ($z_{\rm phot}$),
			atmosphere ($z_{\rm atm}$), and corona ($T = 5 \times 10^6$~K isotherm).
			The atmosphere subtends a much larger solid angle than would be
			expected by just considering the local pressure scale
			height at the photospheric temperature 
			($H_{\rm P} = Z_{\rm P}$). }
\end{figure}

\begin{figure}
\plotone{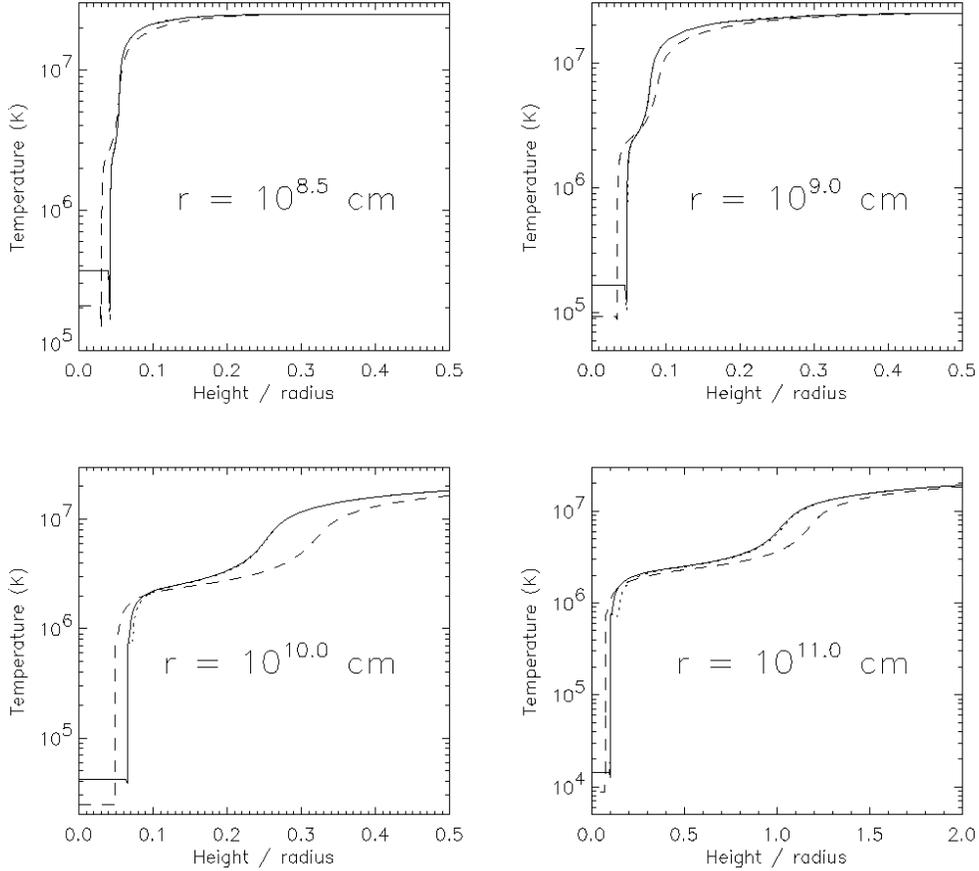}
\epsscale{.6}
\caption{ \label{fig:disk_temp}
			Modeled temperature ($T$) vs. normalized disk height ($z/r$) for various
			radii.  The diagrams can be directly compared to the 
			\citet{ray93} result.  Our new model atmosphere is
			$10 \times$ more extended. The chosen stable solutions 
			correspond to a disk under 1) steady evaporation (solid line), 
			2) steady condensation (dotted line), and 3) steady evaporation
			at low luminosity $L = 0.1 L_{\rm Edd}$ (dashed line). 
			The large-scale
			features of the first two cases are nearly identical.}
\end{figure}

\clearpage 

\begin{figure}
\plotone{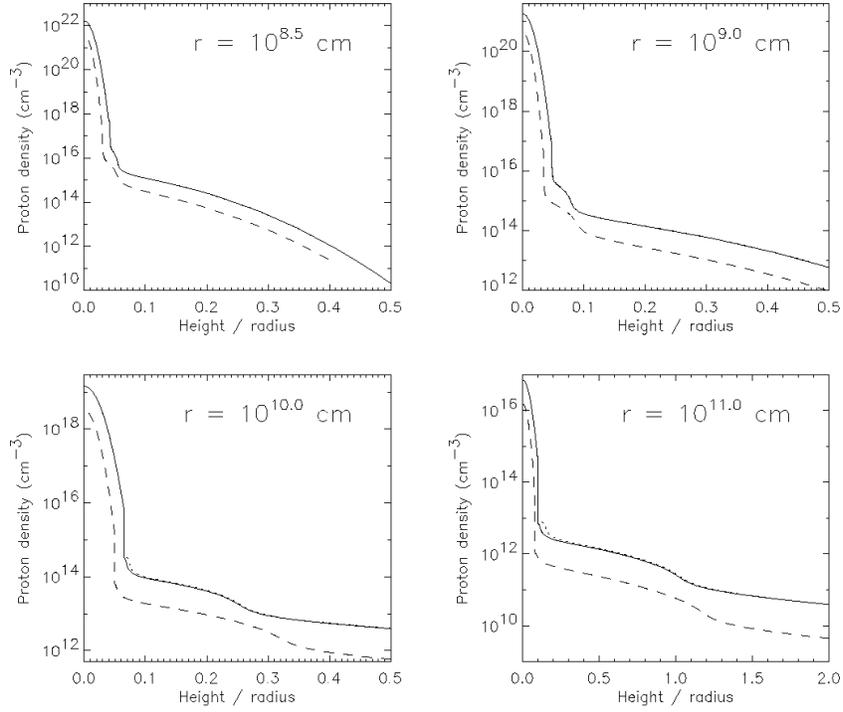}
\caption{ \label{fig:disk_density}
			Modeled proton density ($n_p$) vs. disk height ($z$) for
			various radii.  Evaporating (solid line), condensing (dotted
			line), and low-luminosity (dashed line) disk models, as in
			Fig. \ref{fig:disk_temp}.  The height is
			normalized to the local radius.} 
\end{figure}

\begin{figure}
\plotone{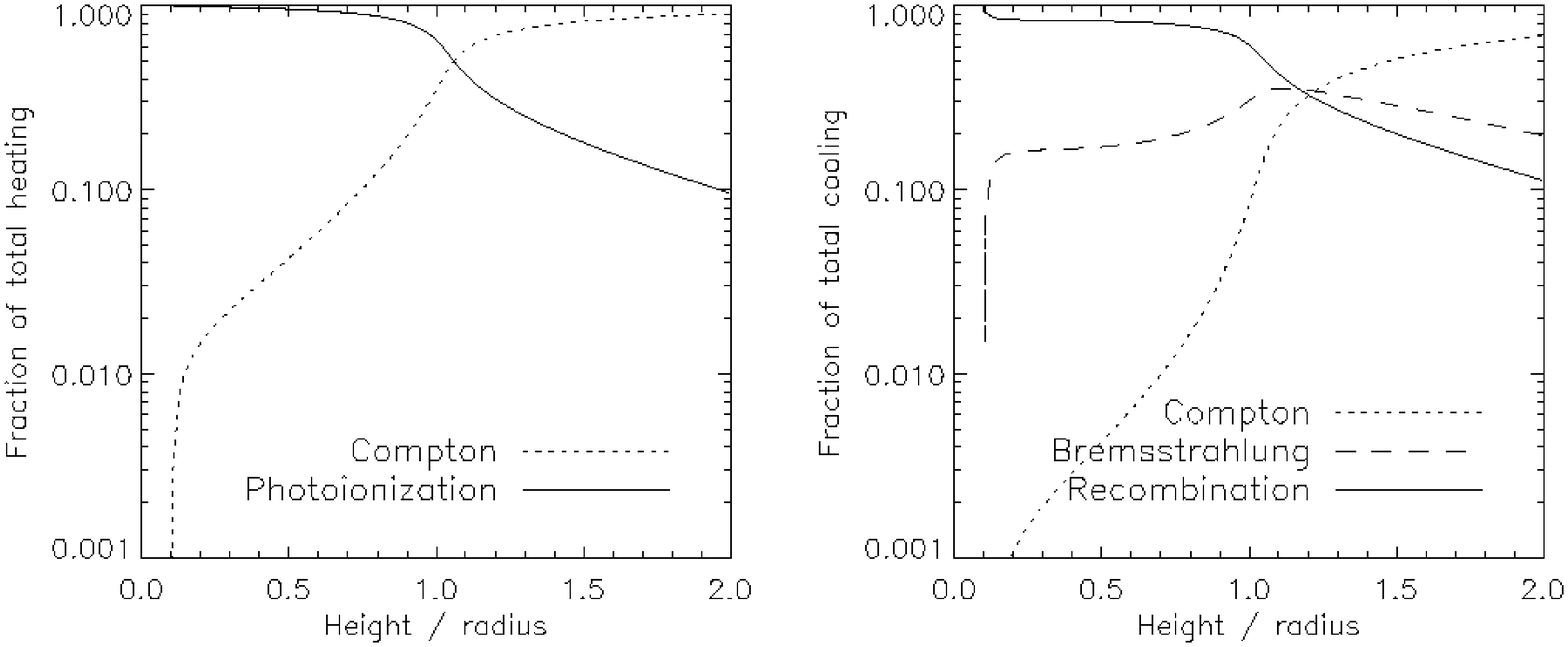}
\caption{ \label{fig:heatreg}
			Dominant heating and cooling mechanisms versus
			the vertical height of the atmosphere,
			for the disk annulus with $r=10^{11}$~cm.
			A Compton-heated corona and a recombining atmosphere
			can be discerned.}
\end{figure}

\begin{figure}
\plotone{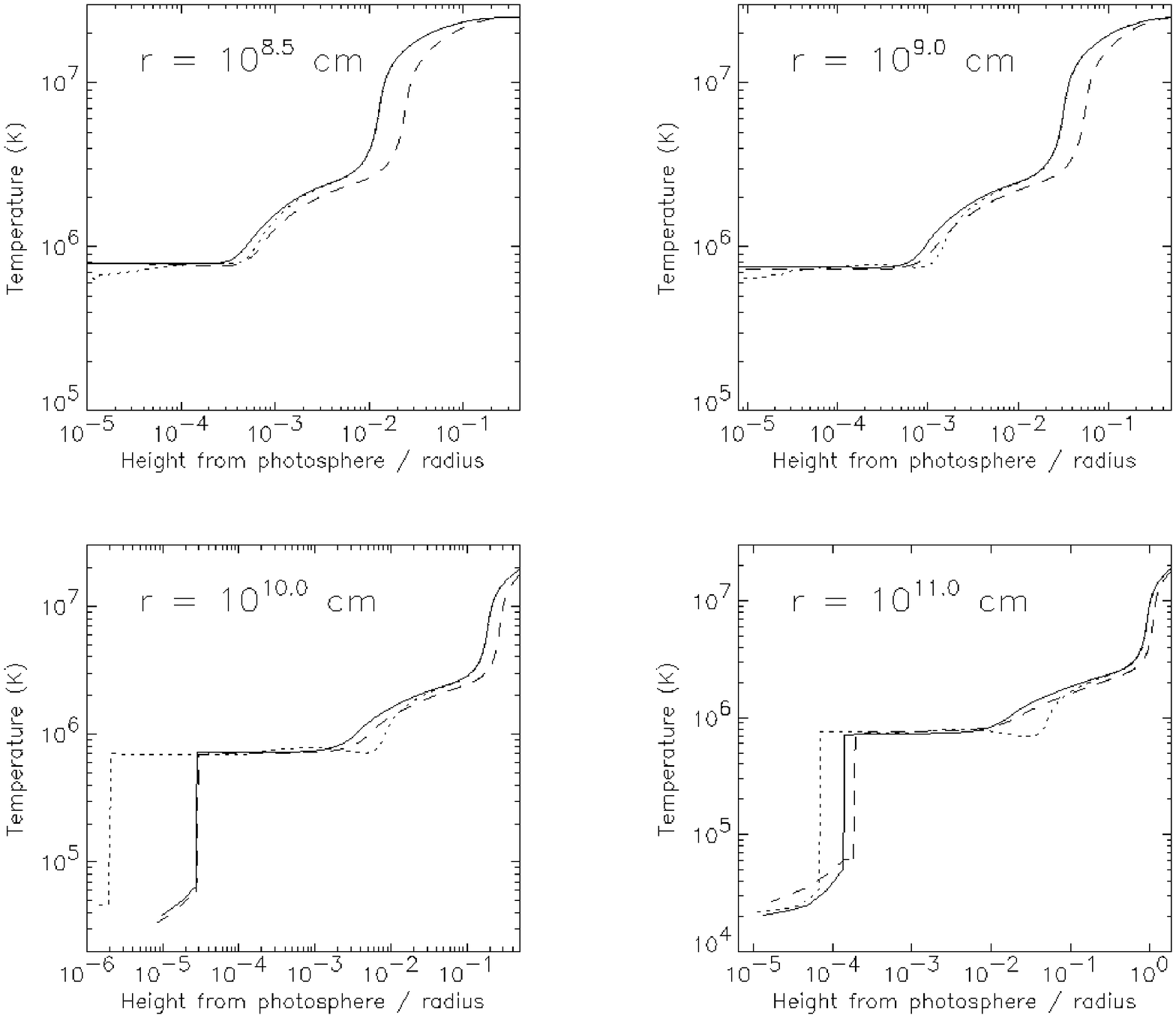}
\caption{ \label{fig:atm_temp}
			Modeled temperature ($T$) vs. disk atmosphere height
			above the photosphere
			($z_{\rm atm}-z_{\rm phot}$) for various radii.
			Evaporating (solid line), condensing (dotted line),
			and low-luminosity (dashed line) disk models.
			The height is normalized to the local radius.} 
\end{figure}

\begin{figure}
\plotone{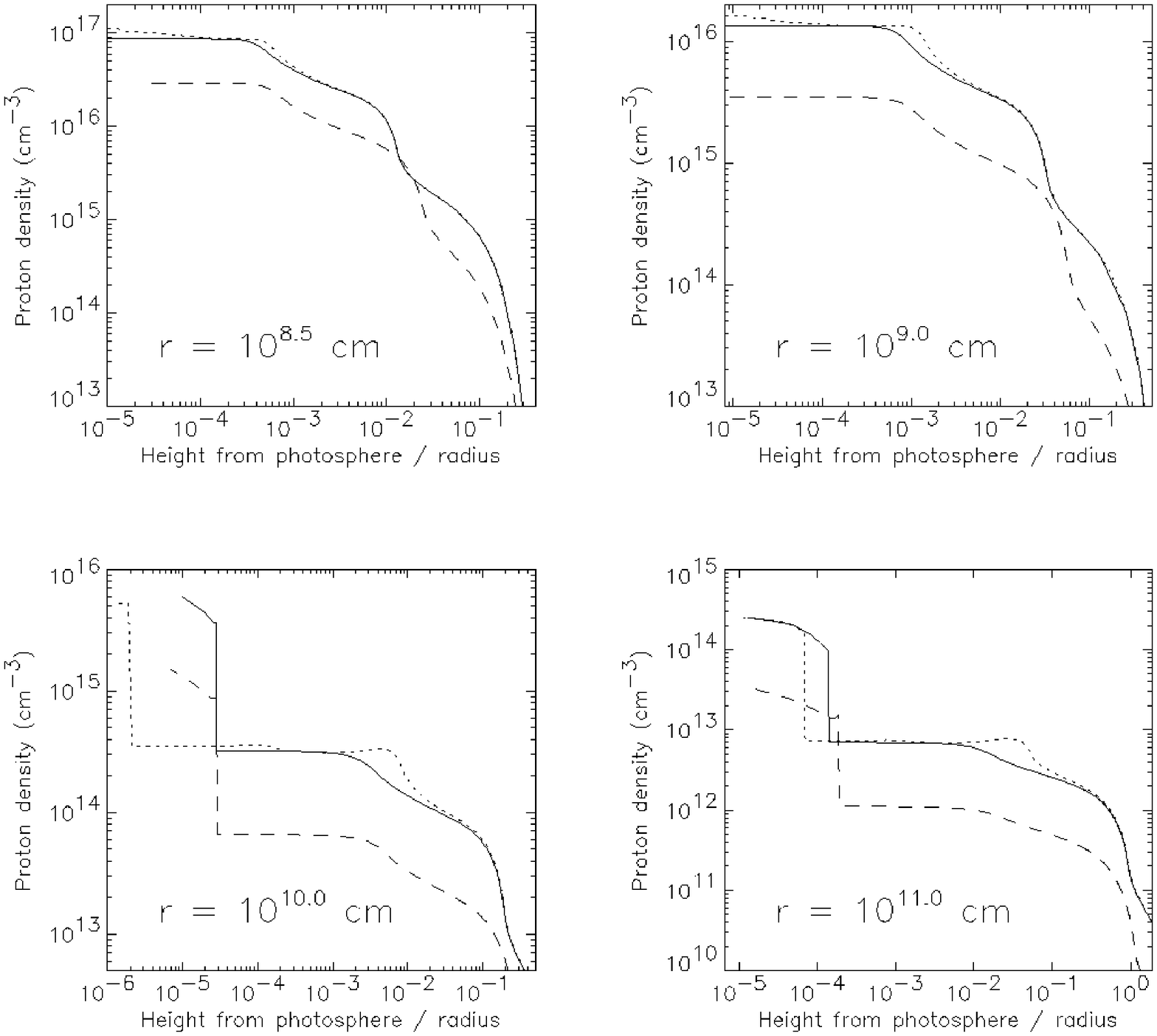}
\caption{ \label{fig:atm_density}
			Modeled proton density ($n_{p}$) vs. disk atmosphere height
			above the photosphere
			($z_{\rm atm}-z_{\rm phot}$) for various radii.
			Evaporating (solid line), condensing (dotted line),
			and low-luminosity (dashed line) disk models.
			The height is normalized to the local radius.} 
\end{figure}

\begin{figure}
\plotone{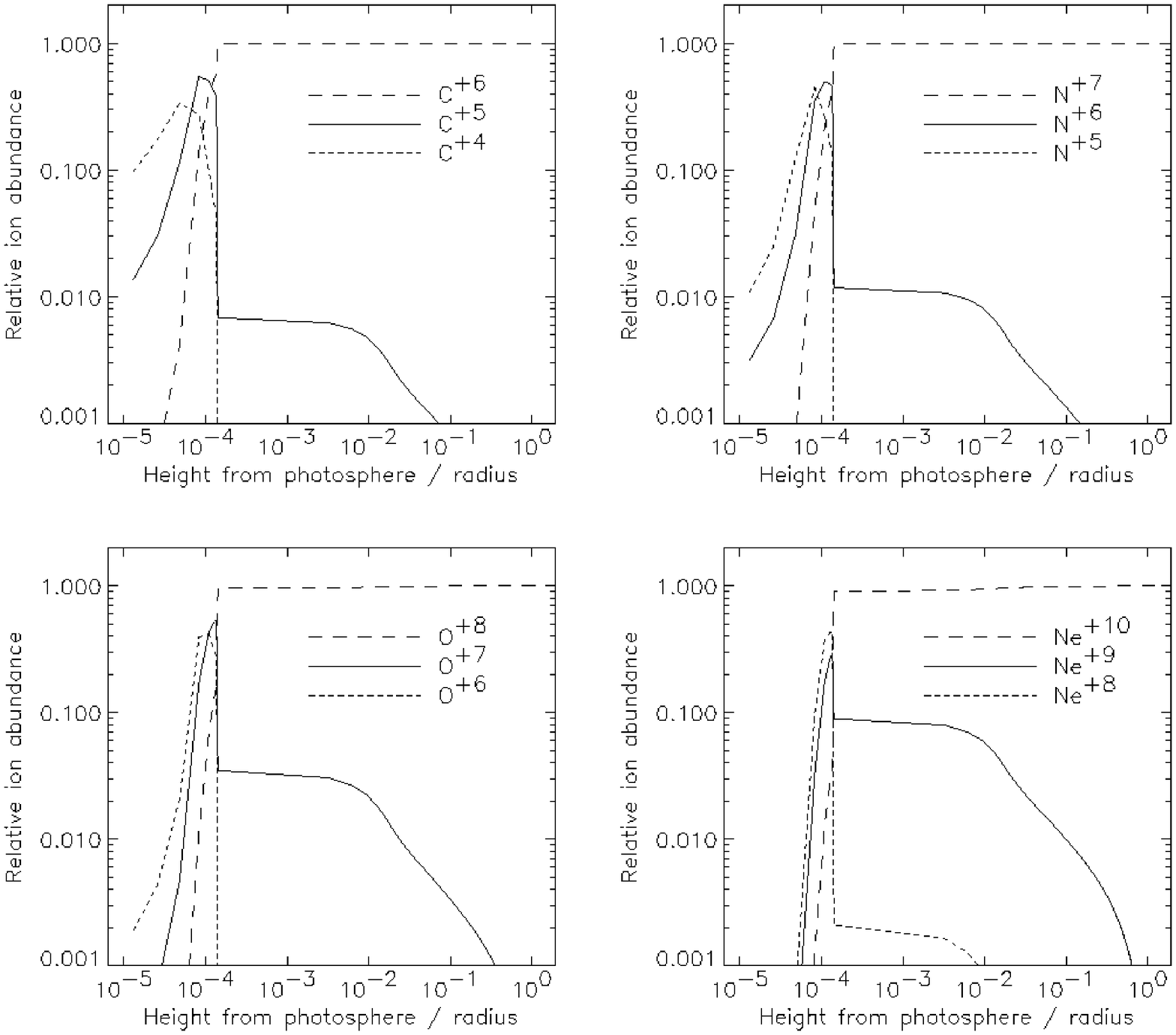}
\caption{ \label{fig:lowzdist}
			Vertical distribution of low-$Z$ ions for the evaporating
			disk model. Low charge
			states are omitted for clarity. The relative ion 
			abundances are plotted against 
			$(z_{\rm atm}-z_{\rm phot})/r$, the vertical height of the atmosphere,
			for the disk annulus with $r=10^{11}$~cm,
			which dominates line emission.
			}
\end{figure}

\begin{figure}
\plotone{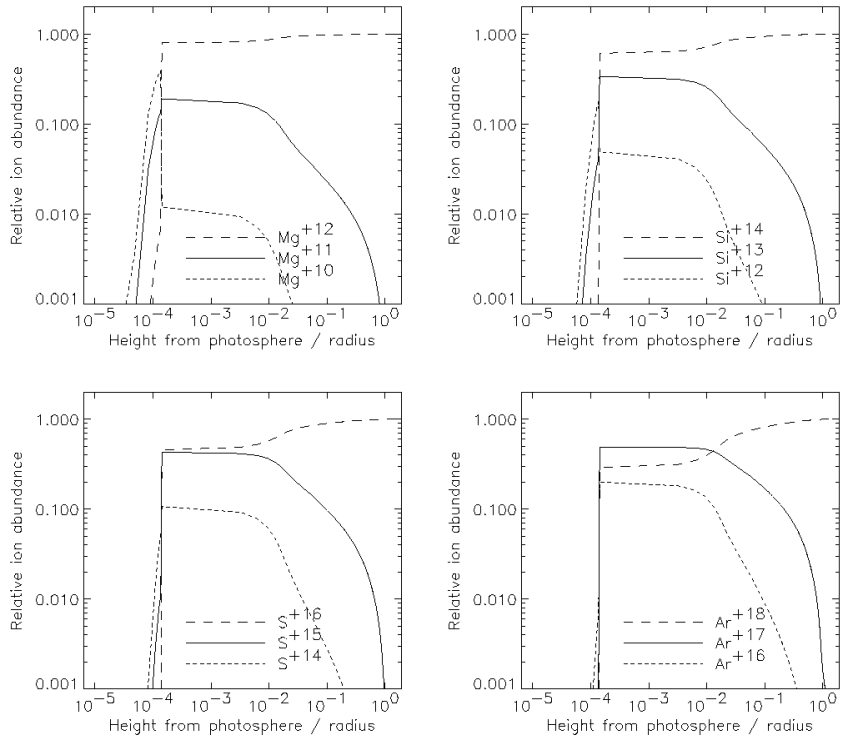}
\caption{ \label{fig:midzdist}
			Vertical distribution of mid-$Z$ ions. Low charge
			states are omitted for clarity. The relative ion 
			abundances are plotted against 
			the vertical height of the atmosphere,
			for the disk annulus with $r=10^{11}$~cm.
			The thermal instability suppresses the 
			He-like ion lines, even for the evaporating disk model,
			where they are strongest.
			}
\end{figure}

\begin{figure}
\plotone{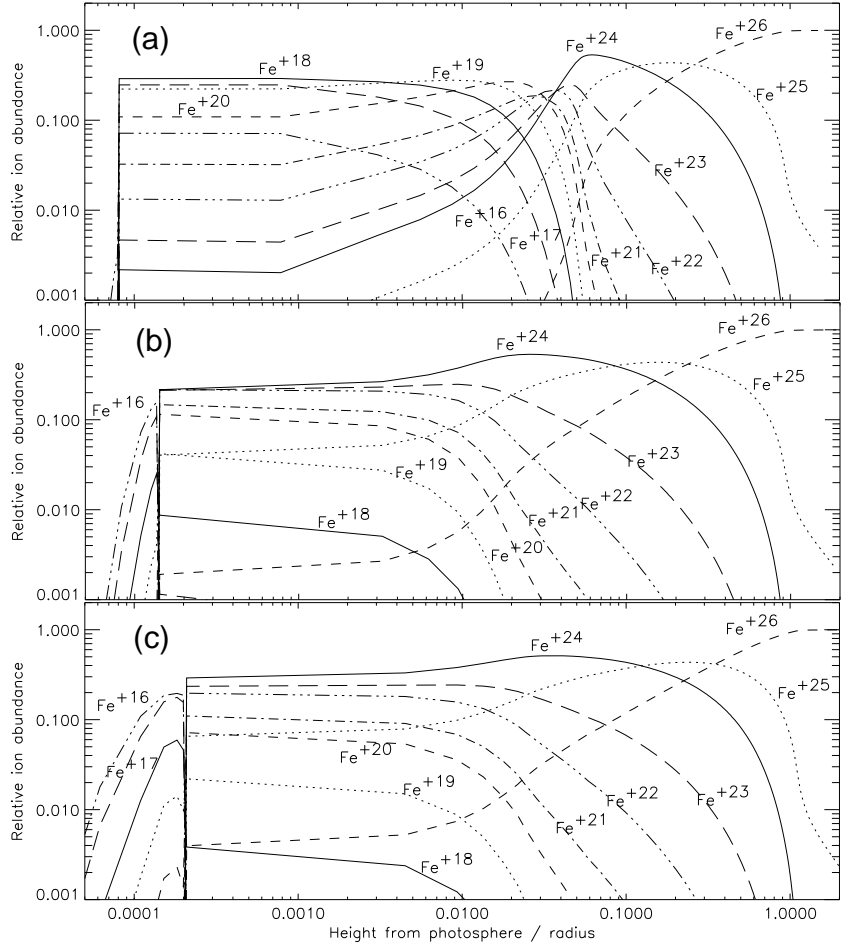}
\caption{ \label{fig:irondist}
			Vertical distribution of K and L-shell ions of Fe,
			plotted for: (a) the condensing disk with $L = L_{\rm Edd}$, 
			(b) the evaporating disk with $L = L_{\rm Edd}$, and 
			(c) the evaporating disk with low-luminosity ($0.1 L_{\rm Edd}$).
			The relative abundances are plotted vs. 
			the height $(z_{\rm atm}-z_{\rm phot})/r$, for the disk annulus with
			$r=10^{11}$~cm. 
			The plateau and break at
			$(z_{\rm atm}-z_{\rm phot})/r=0.0001$--$0.003$ correspond to
			the instability regime. }
\end{figure}

\begin{figure}
\plotone{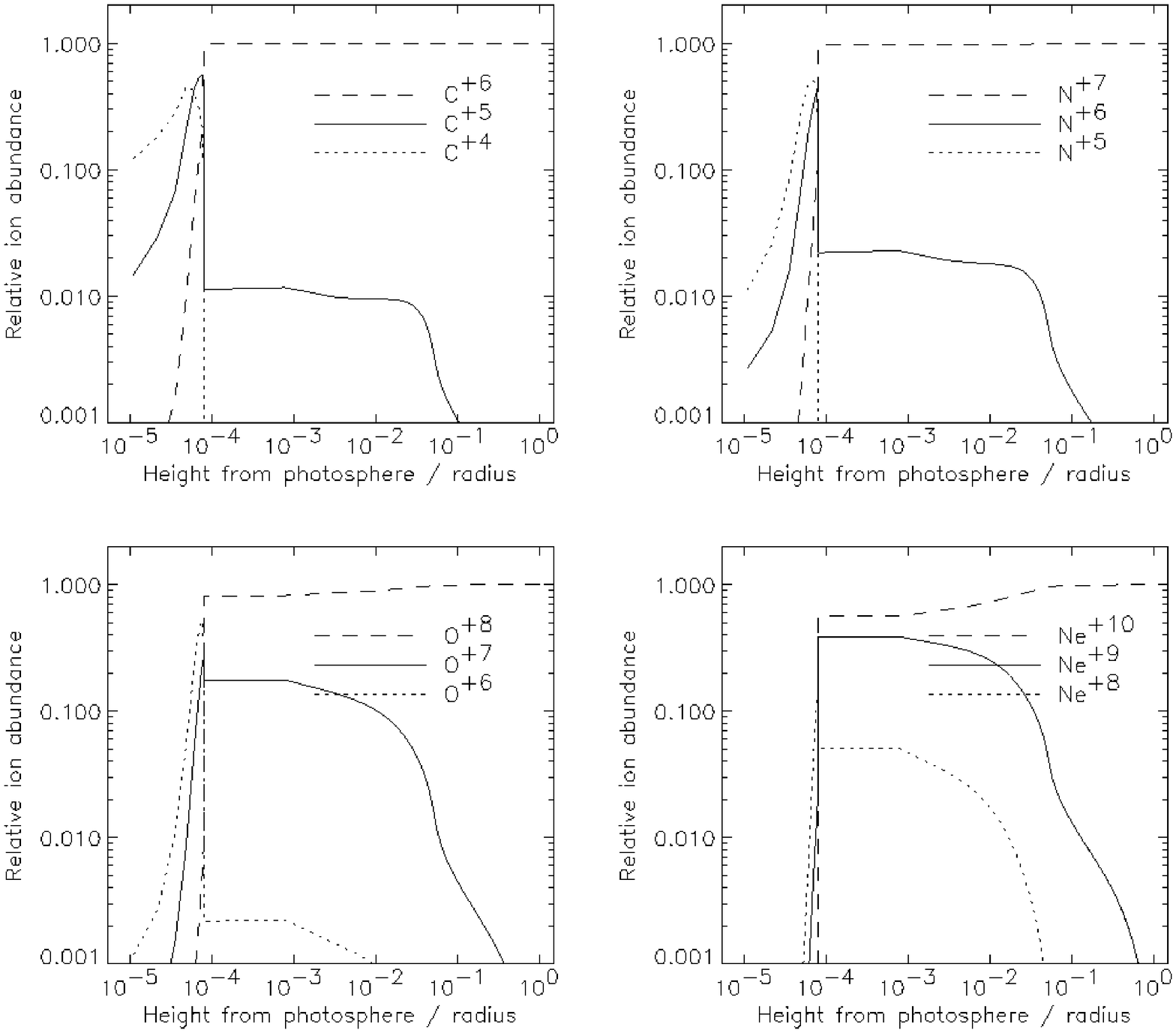}
\caption{ \label{fig:lowzdist_cond}
			Vertical distribution of low-$Z$ ions for the condensing disk
			model. Low charge
			states are omitted for clarity. The relative ion 
			abundances are plotted against 
			$(z_{\rm atm}-z_{\rm phot})/r$, the vertical height of the atmosphere,
			for the disk annulus with $r=10^{11}$~cm,
			which dominates line emission.  }
\end{figure}

\begin{figure}
\plotone{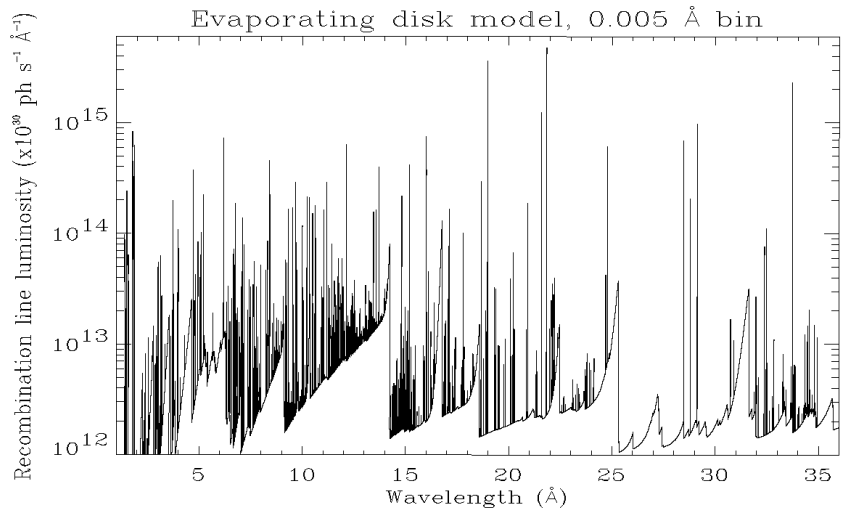}
\caption{ \label{fig:purelines}
			Modeled recombination emission from the disk atmosphere
			($L_{\nu}^{\rm disk} \equiv d^2 E^2 F_{E}^{\rm disk}$). Neither
			the ionizing continuum, interstellar absorption,
			nor Doppler broadening are included. }
\end{figure}

\begin{figure}
\epsscale{.9}
% \plotone{ada/xrb_fatdisk_50_mod_cont_ev2.eps}
\plotone{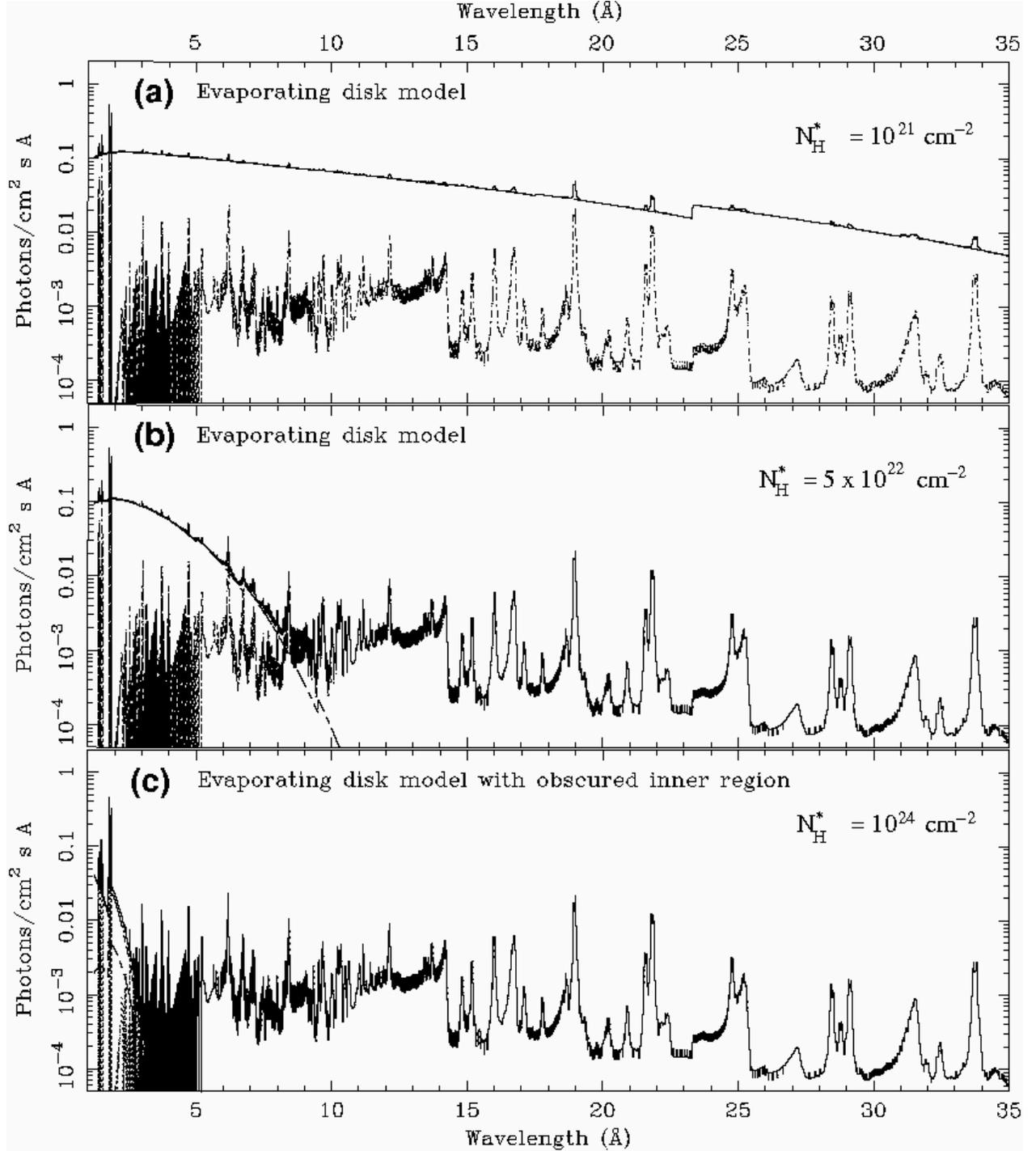}
\caption{ \label{fig:lineunobs}	
			Model spectrum of an LMXB, with $\Delta E_{\rm x} = 2$~eV bins.
			The line luminosity from Fig. \ref{fig:purelines}
			is Doppler shifted by the projected local orbital velocity,
			at an inclination of $75 \degree$. For the disk lines,
			we let $N_{H}^{\rm disk} = 10^{21}$~cm$^{-2}$ be constant. For the neutron
			star continuum, we let
			(a) $N_{H}^{*} = 10^{21}$~cm$^{-2}$,
			(b) $N_{H}^{*} = 5 \times 10^{22}$~cm$^{-2}$,
			and (c) $N_{H}^{*} = 10^{24}$~cm$^{-2}$.
			Each term in eq. (\ref{eq:spec}) is shown, as well as their
			sum (evaporating disk model).
			}
\end{figure}

% \plotone{ada/xrb_fatdisk_50_mod_ev2b.ps}
% \caption{ \label{fig:lineobs}	

\begin{figure}
\plotone{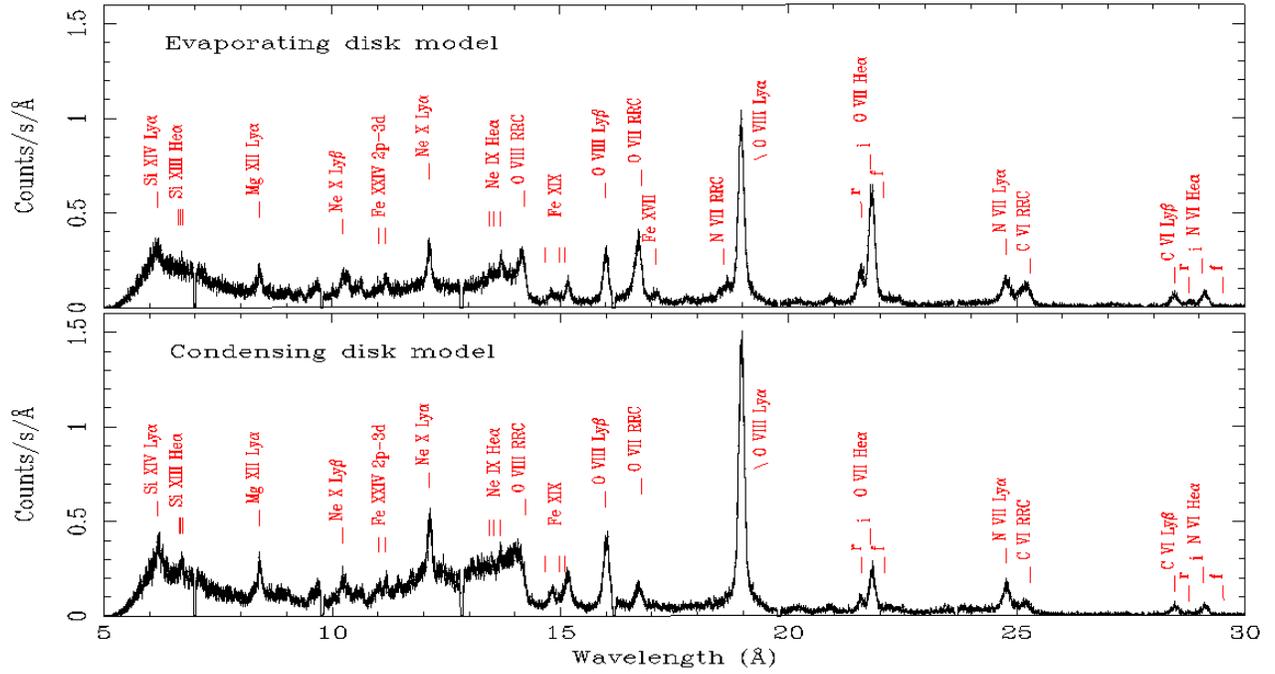}
\caption{ \label{fig:compevap}
Spectra for evaporating and condensing accretion
disks. Simulated 50~ks observation with \it XMM-Newton \rm RGS~1. 
The evaporating disk spectrum here corresponds to 
Fig. \ref{fig:lineunobs}(b).
The continuum emission from the inner ($r < 10^{8.5}$~cm)
disk is not included, and the obscuration of the ionizing continuum
from the compact object is assumed.  }
\end{figure}

\begin{figure}
\plotone{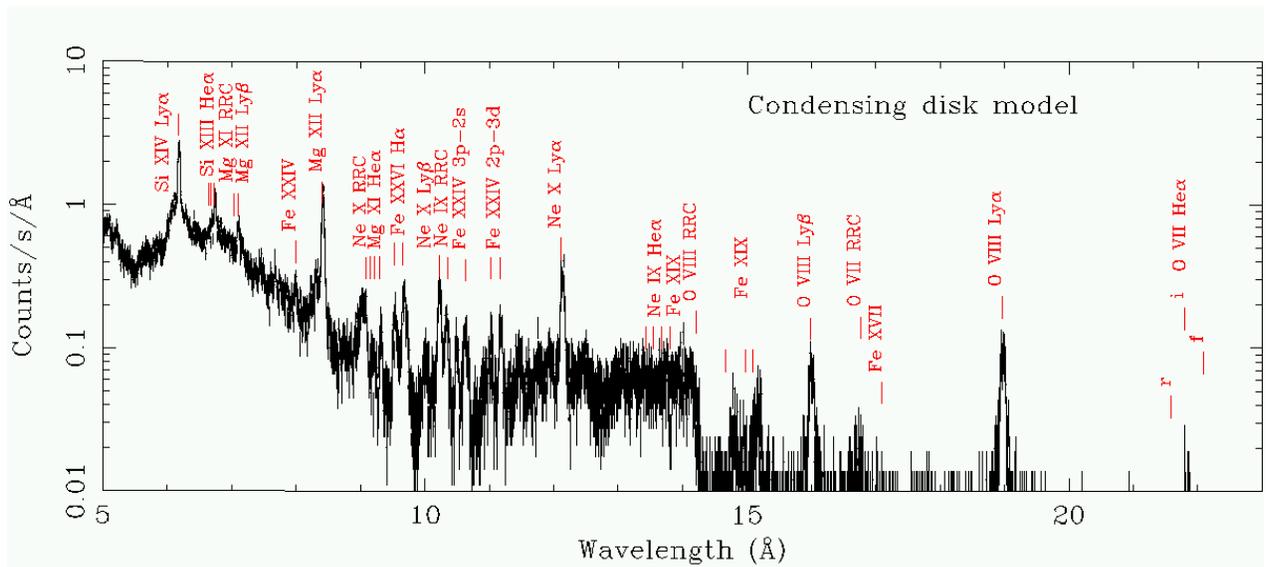}
\caption{ \label{fig:fig3}
Condensing disk spectrum (as in Fig.  \ref{fig:compevap}). Simulated
50~ks observation with \it Chandra \rm $MEG$, +1 order. }
\end{figure}

\begin{figure}
\plotone{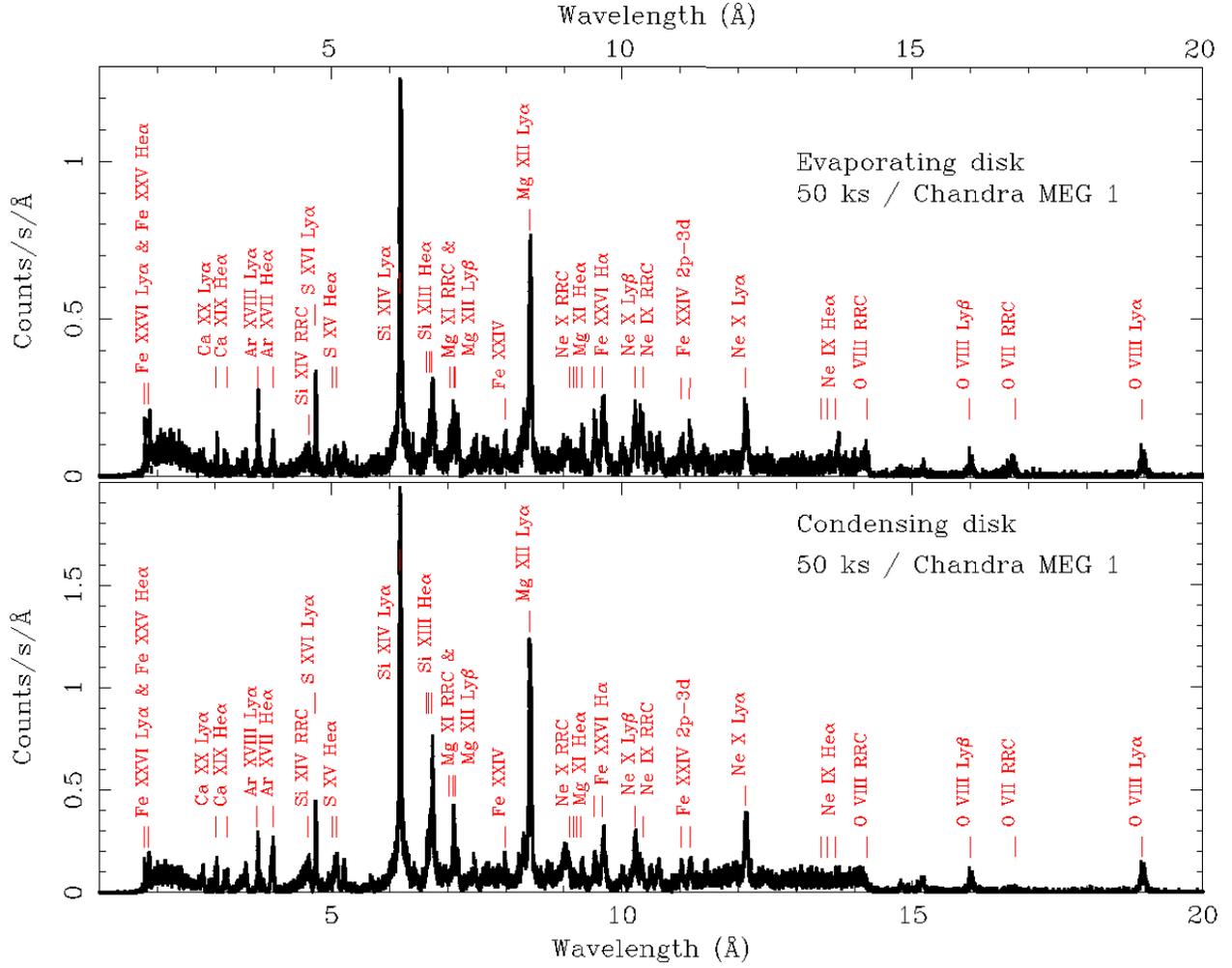}
\caption{ \label{fig:megevap}
Evaporating and condensing disk spectra with occulted neutron star. 
Simulated 50~ks observations with the \it
Chandra \rm medium energy grating $MEG$, +1 order. The evaporating disk
spectrum shown corresponds to the model in
Fig. \ref{fig:lineunobs}(c),
with the neutron star practically occulted by
$N_{H}^{*} = 10^{24}$~cm$^{-2}$.  }
\end{figure}

\begin{figure}
\plotone{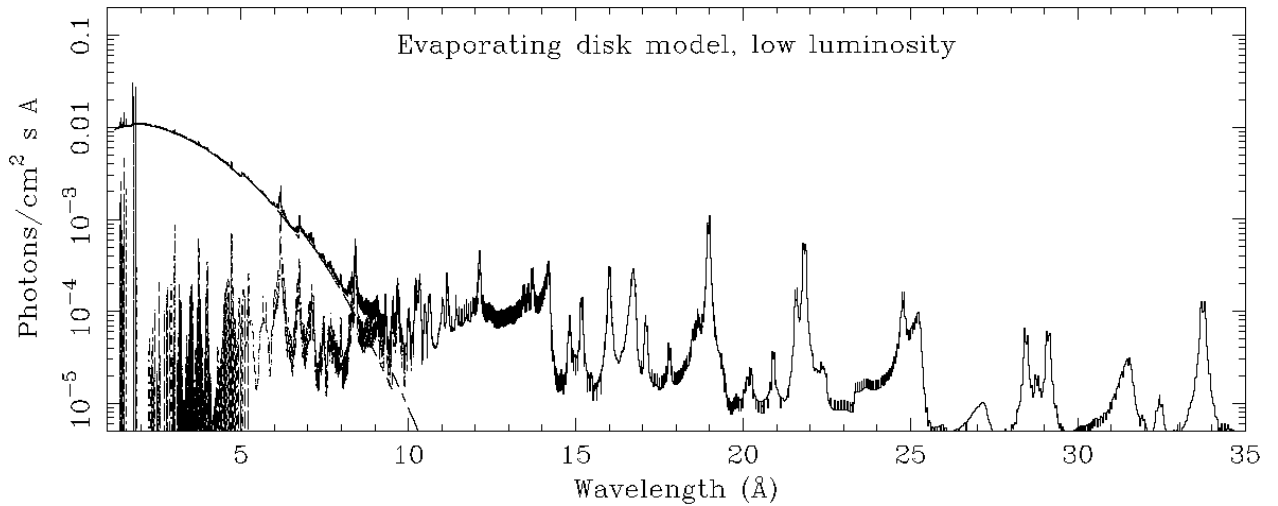}
\caption{ \label{fig:lowlumspec}	
			Model spectrum of an LMXB with $L=0.1 L_{\rm Edd}$,
			with $\Delta E_{\rm x} = 2$~eV bins.
			Doppler shifts from the projected disk orbital velocity is included
			at an inclination of $75 \degree$.
			The disk emission has $N_{H} = 10^{21}$~cm$^{-2}$, and
			the continuum emission has 
			$N_{H}^{*} = 5 \times 10^{22}$~cm$^{-2}$.
			Compare to the $L_{\rm Edd}$ case in 
			Fig. \ref{fig:lineunobs}(b). }	
\end{figure}

\begin{figure}
\epsscale{.7}
\plotone{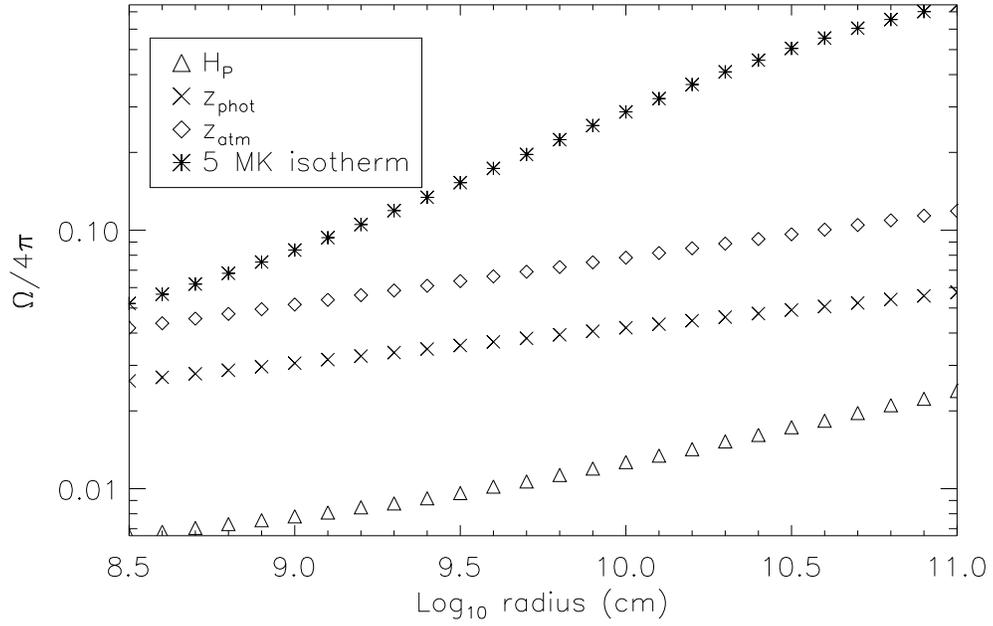}
\caption{ \label{fig:omega_low}
			Total solid angle subtended by the disk photosphere
			($z_{\rm phot}$), atmosphere ($z_{\rm atm}$), and corona
			($T = 5 \times 10^6$~K isotherm).
			The disk structure for the low-luminosity
			($L = 0.1 L_{\rm Edd}$) case is shown, whose atmosphere
			subtends a solid angle 30 to 40 percent smaller than the
			$L = L_{\rm Edd}$ case on Fig. \ref{fig:omega}.}
\end{figure}

\begin{figure}
\plotone{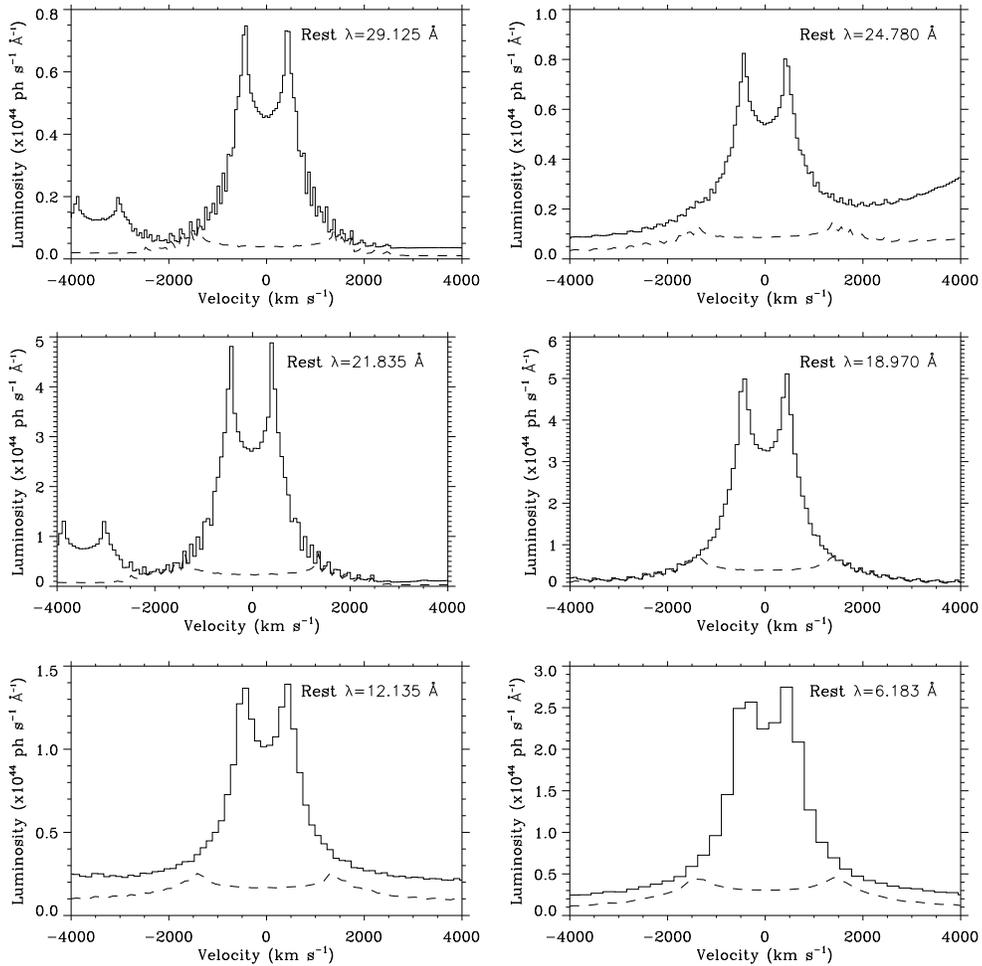}
\caption{ \label{fig:lineprof}
			From top to bottom and left to right: line profiles
			for \ion{N}{6} He$\alpha$, \ion{N}{7} Ly$\alpha$, 
			\ion{O}{7} He$\alpha$, \ion{O}{8} Ly$\alpha$,
			\ion{Ne}{10} Ly$\alpha$, and \ion{Si}{14} Ly$\alpha$.
			The bin size is $\Delta \lambda = 0.005$ \ \AA.
			The solid line corresponds to a disk with
			maximum radius of $r = 10^{11}$~cm, while the
			dashed line is the weaker and broader line profile produced
			by a disk with $r = 10^{10}$~cm.}
\end{figure}

%% If you are not including electronic art with your submission, you may
%% mark up your captions using the \figcaption command. See the 
%% User Guide for details.
%%
%% No more than seven \figcaption commands are allowed per page, 
%% so if you have more than seven captions, insert a \clearpage 
%% after every seventh one. 

%% Tables should be submitted one per page, so put a \clearpage before
%% each one.

%% Two options are available to the author for producing tables:  the
%% deluxetable environment provided by the AASTeX package or the LaTeX
%% table environment.  Use of deluxetable is preferred.
%%

%% Three table samples follow, two marked up in the deluxetable environment,
%% one marked up as a LaTeX table.

%% In this first example, note that the \tabletypesize{}
%% command has been used to reduce the font size of the table.
%% Note also that the \label command needs to be placed 
%% inside the \tablecaption.

\begin{table}
\caption[Line fluxes for disk models]{Line fluxes for disk models.
Line fluxes in units of $10^{-12}$~erg~cm$^{-2}$~s$^{-1}$,
typically over an interval $E/\Delta E \sim 100$. Estimated 
systematic normalization error limit of $\sim 50$ \%, due to the 1-D transfer
calculation. These line fluxes do not include any interstellar absorption
effects, unlike those from Table 1 in \citet{jimenez}.  }
\label{tbl-1}
\begin{center}
\begin{tabular}{crrrr}
\\
\hline \hline
 & $L = 0.1 L_{\rm Edd}$ & $L = L_{\rm Edd}$ & $L = L_{\rm Edd}$ & $L = L_{\rm Edd}$  \\
 & $r = 10^{11}$~cm   &  $r = 10^{11}$~cm &  $r = 10^{10}$~cm &  $r = 10^{11}$~cm  \\
Line(s) & Evaporating & Evaporating & Evaporating  & Condensing \\
\hline \hline
\ion{C}{6} Ly$\alpha$ 		& 0.11 & 2.0 & 0.49 & 1.7 \\
\ion{N}{6} He$\alpha$ 		& 0.063 & 1.3 & 0.27 & 0.98 \\
\ion{N}{7} Ly$\alpha$ 		& 0.061 & 1.1 & 0.29 &  1.1 \\
\ion{O}{7} He$\alpha$ 		& 0.26 & 5.1 & 1.0 & 2.4 \\
\ion{O}{8} Ly$\alpha$ 		& 0.36 & 6.1 & 1.9 & 8.4 \\
\ion{O}{7} RRC 			& 0.12 & 2.1 & 0.43 & 1.0 \\
\ion{O}{8} Ly$\beta$ 		& 0.087 & 1.47 & 0.46 & 2.3 \\
\ion{Fe}{19} L (0.815~keV) 	& 0.038 & 0.67 & 0.22 & 1.4 \\
\ion{Ne}{10} Ly$\alpha$ 		& 0.099 & 1.7 & 0.54 & 2.8 \\
\ion{Fe}{26} H$\alpha$ (1.28~keV) & 0.040 & 0.84 & 0.25 & 0.98 \\
\ion{Mg}{12} Ly$\alpha$ 		& 0.072 & 1.3 & 0.35 & 2.1 \\
\ion{Si}{13} He$\alpha$  	& 0.080 & 1.3 & 0.44 & 2.5 \\
\ion{Si}{14} Ly$\alpha$ 		& 0.024 & 4.2 & 1.4 & 6.2 \\
\ion{S}{15} He$\alpha$ 		& 0.066 & 1.3 & 0.47 & 2.5 \\
\ion{S}{16} Ly$\alpha$ 		& 0.12  & 2.3 & 0.72 & 3.0 \\
\ion{Ar}{17} He$\alpha$ 		& 0.061 & 1.3 & 0.46 & 2.5 \\
\ion{Ar}{18} Ly$\alpha$ 		& 0.067 & 1.3 & 0.35 & 1.7 \\
\ion{Ca}{19} He$\alpha$ 		& 0.025 & 0.53 & 0.19 & 0.90 \\
\ion{Ca}{20} Ly$\alpha$ 		& 0.027 & 0.52 & 0.17 & 0.64 \\
\ion{Fe}{25} He$\alpha$ 		& 0.91 & 16 & 4.6 & 16 \\
\ion{Fe}{26} Ly$\alpha$ 		& 0.57 & 12 & 4.0 & 12 \\
\hline \\
\end{tabular}
\end{center}
\end{table}

\clearpage

\appendix

\section{Radiative Recombination Emission}
\label{sec:radrec}

We describe the numerical calculation of the radiative recombination
emission, including both recombination lines and
radiative recombination continua.  
A similar method for an optically thin gas in the photoionized
wind of a High Mass X-ray Binary $HMXB$ was described by 
\citet{sako}.

Consider an infinitesimal volume $dV$ at which a single ionization
parameter $\xi$, temperature $T$, electron density $n_{\rm e}$, and
elemental abundances $A_z$ describe the state of a gas.
The $\xi(T)$ function is found from thermal balance and ionization
equilibrium, for a given ionizing spectrum $F_\nu$ (section
\ref{sec:ionpar}).
In the radiative recombination process 
\begin{equation}
\label{eq:recformula}
 Z^{+(i+1)} + e^- \rightarrow  Z^{+i*} + h\nu_{RRC}
\end{equation}
an electron recombines with an ion with
net charge +$(i+1)$, assumed in its ground state, producing a new ion 
with net charge +$i$, which might be excited.
The radiative recombination continuum photon has energy
\begin{equation}
\label{eq:rrcen}
E_{\rm x} = h\nu_{RRC} = \chi + KE_e,
\end{equation}
where $\chi$ is the ionization energy of ion $Z^{+i*}$, and $KE_e$
is the initial kinetic energy of the electron, assumed to be on a
Maxwell distribution with temperature $T$. 
The radiative recombination rate to $Z^{+i}$ (where $Z^{+i}$
can be in any quantum state) in units of s$^{-1}$ is 
\begin{equation}
\label{eq:recrate}
\Gamma_{RR} = n_{\rm e} n_{z,i+1} \alpha_{RR} dV
\end{equation}
where $n_{z,i+1}$ is the $Z^{+(i+1)}$ ion number density,
equation (\ref{eq:recrate}) defines $\alpha_{RR}$, the total radiative
recombination rate coefficient in units of cm$^3$~s$^{-1}$.
Note $\alpha_{RR}$ depends on $Z$, $i$, and $T$.

\subsection{Recombination Lines}

After recombination, a fraction $\eta_{u \rightarrow l}$ of the $Z^{+i}$ ions
produce a radiative cascade photon by an electronic transition from
upper level $u$ to lower level $l$. 
The line luminosity of photons from this transition in units of erg~s$^{-1}$ is
\begin{equation}
\label{eq:linelum}
dL_{u \rightarrow l}  = 
	n_{\rm e} n_{z,i+1} E_{u \rightarrow l} \eta_{u \rightarrow l} \alpha_{RR} dV
\end{equation}
where $E_{u \rightarrow l}$ is the transition energy in ergs.
To create a synthetic spectrum, the line luminosities 
$dL_{u \rightarrow l}$ at energies $E_{u \rightarrow l}$ 
in the X-ray band are added for all the levels $u,l$, and
all the ions $Z^{+(i+1)}$ which are abundant in the gas. 
Notice the recombination emission of the $Z^{+i}$ ion depends on
the number density of the $Z^{+(i+1)}$ ion. 

For computational purposes, various quantities from equation
(\ref{eq:linelum}) are defined. The specific line power
\begin{equation}
S_{u \rightarrow l} \equiv \eta_{u \rightarrow l} \alpha_{RR},
\end{equation}
in units of cm$^3$~s$^{-1}$, is the photon emission rate 
per $Z^{+(i+1)}$ ion, per unit electron density.
The $Z^{+(i+1)}$ ion was assumed to be in its ground state
before recombining into $Z^{+i}$.
The population fraction $f_u$ of each level
of the $Z^{+i}$ ion is computed explicitly, and $S_{u \rightarrow l}$
is obtained by equating the matrix of the photon emission rates per ion,
\begin{equation}
n_{\rm e} S_{u \rightarrow l} = f_u A_{u \rightarrow l}
\end{equation}
where $A_{u \rightarrow l}$ is the rate of spontaneous decay
for $Z^{+i}$.
After solving  $S_{u \rightarrow l}$ for a grid of temperatures,
typically in the 10 to 80~eV range, 
it is fit to a power law
\begin{equation}
S_{u \rightarrow l}  = C_{u \rightarrow l} T^{- \gamma_{u \rightarrow l}}
\end{equation}
where the exponent $\gamma_{u \rightarrow l}$ is typically 0.6--0.8. 
The number density of $Z^{+(i+1)}$ is calculated with
\begin{equation}
n_{z,i+1} = n_H A_z f_{z,i+1}
\end{equation}
where $n_H$ is the proton density, $A_z$ is the fractional abundance of
element $Z$ relative to $H$, and $f_{z,i+1}$ is the fractional
abundance of the $Z^{+(i+1)}$ ion relative to all the $Z$ ions.
The differential emission measure for $Z^{+(i+1)}$ is defined as
\begin{equation}
\label{eq:emme}
d(EM_{z,i+1}) \equiv n_{\rm e} n_{z,i+1} dV,
\end{equation}
in units of cm$^{-3}$. 
The line luminosity in equation (\ref{eq:linelum}) can therefore
be re-written as 
\begin{equation}
\label{eq:lumdem}
dL_{u \rightarrow l}  = 
	E_{u \rightarrow l} S_{u \rightarrow l} d(EM_{z,i+1}) .
\end{equation}
If the emission measure is defined as $d(EM) \equiv n_{\rm e}^2 dV$, then
$dL_{u \rightarrow l} = P_{u \rightarrow l} d(EM)$, where
$P_{u \rightarrow l}$ is defined as the line power,
with units of erg~cm$^3$~s$^{-1}$. The emission measure is
useful for calculating the luminosity of an optically thin gas.
Since the accretion disk atmosphere does have some optical depth,
$d(EM)$ and $d(EM_{z,i+1})$ will only be used to track the regions where
the emission originates. The 
radiative recombination line list includes
transitions from levels with principal quantum number
$n \leq 4$ or 5 typically, although in some cases
levels of up to $n=7$ are included.

\subsection{Radiative Recombination Continuum}

To calculate the shape and luminosity of the RRC, 
a Maxwell thermal distribution, the photoionization cross sections, 
and the Milne relation were used.
The monochromatic version of the RR coefficient in equation (\ref{eq:recrate}),
for electrons with velocities between $v$ and $v+dv$ is
\begin{equation}
\alpha_{RR,\nu} = \sigma_{RR,\nu} v f_v dv,
\end{equation}
where $\sigma_{RR,\nu}$ is the RR cross section of ion $Z^{+(i+1)}$,
and the number of electrons in that velocity range is $f_v dv$,
which is assumed to be given by the Maxwellian distribution
\begin{equation}
\label{eq:maxwell}
f_v = \biggr( \frac{2}{\pi}\biggl)^{1/2} \biggr( \frac{m}{kT}\biggl)^{3/2} 
		v^2 e^{-mv^2 / 2 kT},
\end{equation}
where $m$ is the electron mass.
Thus, the monochromatic RRC emissivity of $Z^{+i}$ for thermal electrons is
\begin{equation}
\label{eq:rrcem}
j_\nu = n_{\rm e} n_{z,i+1} E_{\rm x} \sigma_{RR,\nu} v f_v \frac{dv}{dE_{\rm x}},
\end{equation}
in units of erg~cm$^{-3}$~s$^{-1}$~erg$^{-1}$.
Because radiative recombination is the inverse process of
photoionization, a relationship between their cross sections is
derived by equating their transition rates obtained from Fermi's golden
rule \citep{salz}. Detailed balance yields a cross section ratio proportional
to the ratio of the density of final states for each reaction. For
recombination and photoionization, this is the Milne relation 
\begin{equation}
\label{eq:milne}
\sigma_{RR,\nu} = \frac{g_i}{g_{i+1}} \biggr( \frac{E_{\rm x}}{mcv} \biggl)^2 \sigma_{PE,\nu},
\end{equation}
where $\sigma_{PE,\nu}$ is the photoionization cross section
for the valence electron of $Z^{+i}$, 
and $g_i,g_{i+1}$ are the statistical weights of the energy levels of ions
$Z^{+i}$ and $Z^{+(i+1)}$, respectively. Note $g=2J +1$, for total
angular momentum quantum number $J$.
From equation (\ref{eq:rrcen}), and equations (\ref{eq:maxwell})--(\ref{eq:milne}), one
can derive the RRC emissivity
\begin{equation}
\label{eq:emrrc}
j_\nu = \biggr( \frac{2}{\pi}\biggl)^{1/2} n_{\rm e} n_{z,i+1} \frac{g_i}{g_{i+1}} 
	c \sigma_{PE,\nu} 
	\biggr( \frac{E_{\rm x}^2}{mc^2 k T} \biggl)^{3/2} e^{-(E_{\rm x} - \chi) / kT}	
\end{equation}
which is in the same units as equation (\ref{eq:rrcem}).
The ground state photoionization cross sections are taken from
\citet{salom}, and the code based upon that paper is used to calculate
the cross sections from excited levels, such that
\begin{equation}
\label{eq:photoxs}
\sigma_{PE,\nu} = 10^{-18} n^\prime \frac{Ry}{\chi^\prime}
			\exp{ \biggl[ \sum_{q=0}^{3} a_q \biggr( \ln{ \frac{E_{\rm x}}{\chi^\prime} \biggl)^q  } \biggr]}
\end{equation}
in units of cm$^2$, where the four-element $a_q$ vector and
$\chi^\prime$ are fitting parameters, and $Ry \equiv 13.6$~eV. 
Note $\chi^\prime \sim \chi$. The
constant $n^\prime$ is a function of various occupancy numbers and
statistical weights. 

\section{Continuum Opacity}
\label{sec:opacity}
In the disk atmosphere, the recombination emission is partially
absorbed by the ionized gas above it. 
Each ionization zone in the gas column in Figure \ref{fig:column}
is denoted by an index $j=1...N$, starting from the top zone.
If a recombination emission net flux $F_{\nu,j}$ is produced in each zone $j$
of height $h_j$, then the total flux for the column is
\begin{equation}
\label{eq:opacity}
F_\nu = \sum_{j=1}^{N} F_{\nu,j} 
\exp{\biggl[  - \frac{1}{\cos{i}} \sum_{m<j} h_m \kappa_{\nu,m} \biggr]} 
\end{equation}
where $i$ is the inclination angle of the observer in reference to
the disk midplane normal, and $\kappa_{\nu,m}$ is the 
continuum opacity of the $m$th zone, 
\begin{equation}
\label{eq:opacity2}
\kappa_{\nu,m} = \sigma_T n_{e,m} + \sum_{z,k} \sigma_{\nu,z,k} n_{z,k,m} \ ,
\end{equation}
where $n_{z,k,m}$ is the number density
of each ion $Z^{+k}$ in the $m$th zone, 
$n_{{\rm e},m}$ is the electron density, $\sigma_T$ is the Thomson cross section,
and $\sigma_{\nu,z,k}$ is the photo-electric
absorption cross section of ion $Z^{+k}$, given by
\begin{equation}
\sigma_{\nu,z,k} = \sum_{e=1}^{z-k} \sigma_{PE, \nu,z, k, e} \  
\end{equation}
where the photoionization cross section $\sigma_{PE, \nu, z, k, e}$
for each electron $e$ in the ion $Z^{+k}$ is given by equation (\ref{eq:photoxs}),
and the cross sections for all
$Z-k$ electrons were added. This contrasts with the case of
recombination, in which only the $\sigma_{PE,\nu}$ for the
valence electron was needed. The model 
atmosphere is optically thin; i.e., the continuum optical
depth $\tau_\nu = \sum_{j=1}^{N} h_j \kappa_{\nu,j}  \ll 1 $. 

The flaring geometry of the disk atmosphere (see section \ref{sec:diskstruct}) is
not taken into account in equation (\ref{eq:opacity}). Disk flaring will
result in opacities that are larger than those in equation (\ref{eq:opacity})
at inclinations $i> 75$--$80 \degree$,
since the disk atmosphere subtends an 
angle of $\arctan{(z_{\rm atm}/r)} = 10$--$15 \degree$.


\begin{thebibliography}{}

\bibitem[Allen(1973)]{allen} Allen, C.\ W.\ 1973, Astrophysical
Quantities (3rd ed.; London: Athlone Press)  
\bibitem[Angelini et al.(1995)]{angelini1995} Angelini, L., White, 
N.\ E., Nagase, F., Kallman, T.\ R., Yoshida, A., Takeshima, T., Becker, 
C.\ \& Paerels, F.\ 1995, \apjl, 449, L41 
\bibitem[Asai, Dotani, Nagase, \& Mitsuda(2000)]{asai2000} Asai, 
K., Dotani, T., Nagase, F., \& Mitsuda, K.\ 2000, \apjs, 131, 571 
\bibitem[Balbus \& Hawley(1998)]{bh98} Balbus, S. A., \& Hawley, J. F. 1998,
	Rev. Mod. Phys., 70, 1
\bibitem[Ballantyne, Ross, \& Fabian(2001)]{balla} 
Ballantyne, D.~R., Ross, R.~R., \& Fabian, A.~C.\ 2001, \mnras, 327, 10 
\bibitem[Begelman, McKee \& Shields(1983)]{bms1983} Begelman, 
M.\ C., McKee, C.\ F.\ \& Shields, G.\ A.\ 1983, \apj, 271, 70 
\bibitem[Begelman \& McKee(1990)]{bemc} Begelman, M. C., \& McKee, C. 1990,
	\apj, 358, 375
\bibitem[Blumenthal, Drake, \& Tucker(1972)]{blum} 
Blumenthal, G.~R., Drake, G.~W.~F., \& Tucker, W.~H.\ 1972, \apj, 172, 205
\bibitem[Brandt \& Schulz(2000)]{cirx1} Brandt, W.~N.~\& 
Schulz, N.~S.\ 2000, \apjl, 544, L123 
\bibitem[Branduardi-Raymont et al.(2001)]{brandu} 
Branduardi-Raymont, G., Sako, M., Kahn, S.\ M., Brinkman, A.\ C., Kaastra, 
J.\ S., \& Page, M.\ J.\ 2001, \aap, 365, L140 
\bibitem[Buff \& McCray(1974)]{buff}Buff, J.\ \& McCray, R.\ 1974, \apj, 189, 147
\bibitem[Church \& Balucinska-Church(1993)]{church1993} Church, 
M.\ J.\ \& Balucinska-Church, M.\ 1993, \mnras, 260, 59 
\bibitem[Church \& Balucinska-Church(1995)]{church1995} Church, 
M.\ J.\ \& Balucinska-Church, M.\ 1995, \aap, 300, 441 
\bibitem[Church et al.(1997)]{church1997} Church, M.\ J., Dotani, 
T., Balucinska-Church, M., Mitsuda, K., Takahashi, T., Inoue, H.\ \& 
Yoshida, K.\ 1997, \apj, 491, 388 
%church 1998 also church
\bibitem[Church et al.(1998)]{church1998} Church, M. J., Balucinska-Church, M.,
	Dotani, T., \& Asai, K. 1998, \apj, 504, 516
\bibitem[Church(2001)]{church2000} Church, M.~J.\ 2001, Advances 
in Space Research, 28, 323 
\bibitem[Cottam et al.(2001a)]{exo0748} Cottam, J., Kahn, S.~M., 
Brinkman, A.~C., den Herder, J.~W., \& Erd, C.\ 2001a, \aap, 365, L277 
\bibitem[Cottam et al.(2001b)]{4u1822} Cottam, J., Sako, M., 
Kahn, S.~M., Paerels, F., \& Liedahl, D.~A.\ 2001b, \apjl, 557, L101 
\bibitem[Czerny \& King(1989)]{czerny}Czerny, M.\ \& King, A.R.\ 1989, \mnras, 236, 843
\bibitem[Davidson \& Netzer(1979)]{david1979} Davidson, K.\ \& Netzer, H.\ 1979, Rev.\ Mod.\
Phys., 51, 715
\bibitem[de Jong, van Paradijs, \& Augusteijn(1996)]{dejong} de Jong, J. A., 
	van Paradijs, J., \& Augusteijn, T. 1996, \aap, 314, 484
\bibitem[di Matteo(1998)]{matteo} di Matteo, T.\ 1998, \mnras, 
299, L15 
\bibitem[Dubus et al.(1999)]{dub99} Dubus, G., Lasota, J. P., Hameury, J. M.,
	\& Charles, P. 1999, \mnras, 303, 139
\bibitem[Ferland et al.(1998)]{ferland1998} Ferland, G.\ J., 
Korista, K.\ T., Verner, D.\ A., Ferguson, J.\ W., Kingdon, J.\ B.\ \& 
Verner, E.\ M.\ 1998, PASP, 110, 761 
\bibitem[Field(1965)]{field1965} Field, G.\ B.\ 1965, \apj, 142, 
531 
\bibitem[Field(1969)]{field1969}Field, G.B., Goldsmith, D.W., \& Habing, H.J. 1969,
\apj, 155, L149
\bibitem[Frank, King, \& Lasota(1987)]{frank1987} Frank, J., 
King, A.\ R., \& Lasota, J.\ P.\ 1987, \aap, 178, 137 
\bibitem[Frank, King \& Raine(1992)]{fkr1985} Frank, J., King, 
A.\ R.\ \& Raine, D.\ J.\ 1992, Accretion Power in Astrophysics (2d ed.; New York: Cambridge 
University Press)
\bibitem[Gabriel \& Jordan(1969)]{gabriel1969} Gabriel, A.\ H.\ \& 
Jordan, C.\ 1969, \mnras, 145, 241 
\bibitem[Halpern \& Grindlay(1980)]{halpern} Halpern, J.\ P.\ 
\& Grindlay, J.\ E.\ 1980, \apj, 242, 1041 
\bibitem[Hawley, Balbus, \& Stone(2001)]{3disk} Hawley, 
J.~F., Balbus, S.~A., \& Stone, J.~M.\ 2001, \apjl, 554, L49 
\bibitem[Hawley \& Balbus(2002)]{lowacc} Hawley, J.~F.~\& 
Balbus, S.~A.\ 2002, \apj, 573, 738 
\bibitem[Hess, Kahn, \& Paerels(1997)]{hess} Hess, C. J., Kahn, S. M., \& Paerels, F. B. S.
	1997, \apj, 478, 94
\bibitem[Jimenez-Garate, Raymond, Liedahl, \& 
Hailey(2001)]{jimenez} Jimenez-Garate, M.~A., Raymond, J.~C., 
Liedahl, D.~A., \& Hailey, C.~J.\ 2001, \apj, 558, 448 
\bibitem[Jimenez-Garate et al.(2002)]{herx1} Jimenez-Garate, M.~A.,
Hailey, C.~J., den Herder, J.~W., Zane, S., Ramsay, G.\ 2002, \apj,
578, in press (astro-ph/0206181)
\bibitem[Kallman \& McCray(1982)]{kallman1982} Kallman, T.\ R.\ \& 
McCray, R.\ 1982, \apjs, 50, 263 
\bibitem[Klapisch et al.(1977)]{hullac} Klapisch, M., Schwab, J. L.,
	Fraenkel, J. S., \& Oreg, J. 1977, Opt. Soc. Am., 61, 148 
\bibitem[Ko \& Kallman(1991)]{ko1991} Ko, Y.\ \& Kallman, T.\ 
R.\ 1991, \apj, 374, 721 
%also ko94
\bibitem[Ko \& Kallman(1994)]{ko1994} Ko, Y.\ \& Kallman, T.\ 
R.\ 1994, \apj, 431, 273 
\bibitem[Krolik, McKee \& Tarter(1981)]{kmt} Krolik, J. H., McKee, C. F., \&
 Tarter, C. B. 1981, \apj, 249, 422
\bibitem[Li, Gu, \& Kahn(2001)]{li} Li, Y., Gu, M.~F., \& 
Kahn, S.~M.\ 2001, \apj, 560, 644
\bibitem[Liedahl et al.(1992)]{lied1992}Liedahl, D.A., Kahn, S.M., Osterheld, A.L., and Goldstein, W.H.\ 
1992, \apj, 391, 306
\bibitem[Liedahl \& Paerels(1996)]{lied}Liedahl, D. A., \& Paerels F. 1996,
		\apj, 468, 33
\bibitem[Liedahl(1999)]{lied99}Liedahl, D. A. 1999, in X-ray Spectroscopy in
Astrophysics, EADN School proceedings, ed. J. A. van Paradijs, \&
J. A. M. Bleeker (Amsterdam: Springer), 189  
\bibitem[Liu, Mineshige, \& Shibata(2002)]{liu} Liu, B.~F., 
Mineshige, S., \& Shibata, K.\ 2002, \apjl, 572, L173
\bibitem[Machida, Hayashi, \& Matsumoto(2000)]{machida} 
Machida, M., Hayashi, M.~R., \& Matsumoto, R.\ 2000, \apjl, 532, L67 
\bibitem[Matt, Fabian \& Ross(1993)]{matt1993} Matt, G., Fabian, 
A.\ C.\ \& Ross, R.\ R.\ 1993, \mnras, 262, 179 
\bibitem[cf., Mauche, Liedahl, \& Fournier(2001)]{mauche} Mauche, 
C.~W., Liedahl, D.~A., \& Fournier, K.~B.\ 2001, \apj, 560, 992
\bibitem[McClintock, London, Bond \& Grauer(1982)]{mcclintock1982} 
McClintock, J.\ E., London, R.\ A., Bond, H.\ E.\ \& Grauer, A.\ D.\ 1982, 
\apj, 258, 245 
\bibitem[for a detailed discussion, see McKee \& Begelman(1990)]{mckee} McKee, C.\ F.\ \& 
Begelman, M.\ C.\ 1990, \apj, 358, 392 
\bibitem[Meyer \& Meyer-Hofmeister(1982)]{meyer1982} Meyer, F.\ 
\& Meyer-Hofmeister, E.\ 1982, \aap, 106, 34
\bibitem[Mihalas(1978)]{mihalas} Mihalas, D.\ 1978, Stellar Atmospheres, (2d ed.; San 
Francisco: W.\ H.\ Freeman and Co.)  
\bibitem[Miller \& Stone(2000)]{miller2000} Miller, K.\ A.\ \& 
Stone, J.\ M.\ 2000, \apj, 534, 398 
\bibitem[Morrison \& McCammon(1983)]{abun} Morrison, R., \& McCammon, D. 
	1983, \apj, 270, 119
\bibitem[Murray \& Chiang(1997)]{murray_chiang1997} Murray, N.\ \& 
Chiang, J.\ 1997, \apj, 474, 91 
\bibitem[Nayakshin, Kazanas \& Kallman(2000)]{nay2000} 
Nayakshin, S., Kazanas, D.\ \& Kallman, T.\ R.\ 2000, \apj, 537, 833 
\bibitem[Nayakshin \& Kallman(2001)]{nay2001} Nayakshin, S.\ \& 
Kallman, T.\ R.\ 2001, \apj, 546, 406 
\bibitem[Paerels et al.(2001)]{paerels} Paerels, F.~et al.\ 
2001, \apj, 546, 338 
\bibitem[Parmar, White, Giommi \& Gottwald(1986)]{parmar1986} 
Parmar, A.\ N., White, N.\ E., Giommi, P.\ \& Gottwald, M.\ 1986, \apj, 
308, 199 
\bibitem[Parmar et al.(2000)]{parmar2000} Parmar, A.\ N., 
Oosterbroek, T., Del Sordo, S., Segreto, A., Santangelo, A., Dal Fiume, D.\ 
\& Orlandini, M.\ 2000, \aap, 356, 175 
\bibitem[Parmar, Oosterbroek, Boirin, \& Lumb(2002)]{parmar2002} 
Parmar, A.~N., Oosterbroek, T., Boirin, L., \& Lumb, D.\ 2002, \aap, 386, 
910 
\bibitem[Porquet \& Dubau(2000)]{porquet} Porquet, D., \& Dubau, J. 2000,
	\aap, 143, 495
\bibitem[Press(1994)]{numrec} Press, W. H. 1994,  Numerical Recipes
in FORTRAN : The Art of Scientific Computing (Cambridge: Cambridge
University Press)
\bibitem[Proga \& Kallman(2002)]{proga} Proga, D.~\& Kallman, 
T.~R.\ 2002, \apj, 565, 455 
% also raymond93
\bibitem[Raymond(1993)]{ray93} Raymond, J. C. 1993, \apj, 412, 267
\bibitem[Reynolds \& Fabian(1995)]{rey1995} Reynolds, C.\ S.\ 
\& Fabian, A.\ C.\ 1995, \mnras, 273, 1167 
\bibitem[Ross \& Fabian(1993)]{ross1993} Ross, R.\ R.\ \& 
Fabian, A.\ C.\ 1993, \mnras, 261, 74 
\bibitem[R{\' o}{\. z}a{\' n}ska \& Czerny(1996)]{rozanska1996} R{\' o}{\. z}a{\' n}ska, A.\ \& 
Czerny, B.\ 1996, Acta Astronomica, 46, 233 
\bibitem[R{\' o}{\. z}a{\' n}ska, Czerny, {\.Z}ycki \& 
Pojma{\'n}ski(1999)]{rozanska1999} R{\' o}{\. z}a{\' n}ska, A., Czerny, 
B., {\.Z}ycki, P.\ T.\ \& Pojma{\'n}ski, G.\ 1999, \mnras, 305, 481 
\bibitem[R{\' o}{\. z}a{\' n}ska, Dumont, Czerny, \& 
Collin(2002)]{roza2002} R{\' o}{\. z}a{\' n}ska, A., Dumont, 
A.-M., Czerny, B., \& Collin, S.\ 2002, \mnras, 332, 799 
\bibitem[Sako et al.(1999)]{sako} Sako M., Liedahl, D. A., Kahn, S. M., \& Paerels, F. 1999,
\apj, 525, 921
\bibitem[Saloman, Hubble, \& Scofield(1988)]{salom} Saloman, E. B.,
  Hubble, J. H., \& Scofield, J. H. 1988, At. Data Nucl. Data Tables,
  38, 1
\bibitem[Salzmann(1988)]{salz} Salzmann, D. 1998,
Atomic physics in hot plasmas (New York: Oxford University Press)
\bibitem[Savin et al.(1999)]{savin1999} Savin, D.\ W.\ et al.\ 
1999, \apjs, 123, 687 
\bibitem[Schulz(1999)]{schultz1999} Schulz, N.\ S.\ 1999, \apj, 
511, 304 
\bibitem[Schulz et al.(2001)]{schulz2001} Schulz, N.~S., 
Chakrabarty, D., Marshall, H.~L., Canizares, C.~R., Lee, J.~C., \& Houck, 
J.\ 2001, \apj, 563, 941 
\bibitem[Shakura \& Sunyaev(1973) or SS73]{ss73} Shakura, N. I., \& Sunyaev, R. A.
	1973, \aap, 24, 337
\bibitem[Spitzer(1962)]{spitzer} Spitzer, L.\ 1962, Physics of 
Fully Ionized Gases (New York: Interscience)  
\bibitem[Stella \& Rosner(1984)]{stella} Stella, L.\ \& 
Rosner, R.\ 1984, \apj, 277, 312 
\bibitem[Tanaka et al.(1995)]{tanaka1995} Tanaka, Y.~et al.\ 1995, 
\nat, 375, 659 
\bibitem[Tarter, Tucker \& Salpeter(1969)]{tarter}Tarter, C.B., Tucker,
W.H., \& Salpeter, E.E.\ 1969, \apj, 156, 943
\bibitem[Vrtilek et al.(1990)]{vrtilek} Vrtilek, S. D., Raymond, J. C., 
	Garcia, M. R., Verbunt, F., Hasinger, G., \& Kurster, M. 1990,
	\aap, 235, 162
\bibitem[Vrtilek et al.(1991)]{vrtilek1991} Vrtilek, S.\ D., 
McClintock, J.\ E., Seward, F.\ D., Kahn, S.\ M.\ \& Wargelin, B.\ J.\ 
1991, \apjs, 76, 1127 
\bibitem[White \& Swank(1982)]{white1982} White, N.\ E.\ \& 
Swank, J.\ H.\ 1982, \apjl, 253, L61 
\bibitem[White \& Holt(1982)]{whiteh1982} White, N.\ E.\ \& Holt, 
S.\ S.\ 1982, \apj, 257, 318 
\bibitem[White et al.(1984)]{white1984} White, N.\ E., Parmar, 
A.\ N., Sztajno, M., Zimmermann, H.\ U., Mason, K.\ O., \& Kahn, S.\ M.\ 
1984, \apjl, 283, L9 
\bibitem[White et al.(1995)]{white1995} White, N., Nagase, F. \&
Parmar, A. N. 1995, in X-ray Binaries, ed. W. G. H. Lewin, J. van
Paradijs, \& E. P. J. van den Heuvel (Cambridge: Cambridge Univ. Press), 1 
\bibitem[Zeldovich \& Pikelner (1969)]{zel}Zeldovich, Y. B., \& Pikelner, 
	S. B. 1969, Soviet Physics JETP, 29, 170
\bibitem[Zycki, Krolik, Zdziarski \& Kallman(1994)]{zycki1994} 
Zycki, P.\ T., Krolik, J.\ H., Zdziarski, A.\ A.\ \& Kallman, T.\ R.\ 1994, 
\apj, 437, 597 

\end{thebibliography}
\end{document}